\pdfoutput=1
\documentclass[aps,prd,twocolumn,groupedaddress]{revtex4-1}

\usepackage{graphicx}
\usepackage{natbib}
\usepackage{color}
\usepackage{url}

\newcommand{\beq}{\begin{equation}}
\newcommand{\beqa}{\begin{eqnarray}}
\newcommand{\eeq}{\end{equation}}
\newcommand{\eeqa}{\end{eqnarray}}

\newcommand{\simgt}{\lower.5ex\hbox{$\; \buildrel > \over \sim \;$}}
\newcommand{\simlt}{\lower.5ex\hbox{$\; \buildrel < \over \sim \;$}}

\newcommand{\bd}[1]{\mbox{\boldmath $#1$}}

\usepackage[usenames,dvipsnames]{xcolor}

\preprint{DESY 16-112}

\begin{document}
\title{
  Cosmological constraints on dark matter annihilation and decay:
  Cross-correlation analysis of the extragalactic $\gamma$-ray background and cosmic shear
}

\author{Masato Shirasaki}
\affiliation{
National Astronomical Observatory of Japan, 
Mitaka, Tokyo 181-8588, Japan}
\email{masato.shirasaki@nao.ac.jp}

\author{Oscar Macias}
\author{Shunsaku Horiuchi}
\affiliation{
Center for Neutrino Physics, Department of Physics, Virginia Tech, Blacksburg, Virginia
24061, USA
}
\email{oscar.macias@vt.edu}
\email{horiuchi@vt.edu}

\author{Satoshi Shirai}
\affiliation{
Deutsches Elektronen-Synchrotron (DESY), 22607 Hamburg, Germany
}

\author{Naoki Yoshida}
\affiliation{
Department of Physics, University of Tokyo, Tokyo 113-0033, Japan\\
Kavli Institute for the Physics and Mathematics of the Universe (WPI),
University of Tokyo, Kashiwa, Chiba 277-8583, Japan
}
\affiliation{
CREST, Japan Science and Technology Agency, 4-1-8 Honcho, Kawaguchi, Saitama, 332-0012, Japan
}

\begin{abstract}
  We derive constraints on dark matter (DM) annihilation cross section
  and decay lifetime from cross-correlation analyses of the
  data from Fermi-LAT and weak lensing surveys that cover a wide
  area of $\sim660$ squared degrees in total.
  We improve upon our previous analyses 
  by using an updated extragalactic $\gamma$-ray background data
  reprocessed with the Fermi Pass 8 pipeline,
  and by using well-calibrated shape measurements of about twelve million
  galaxies in the Canada-France-Hawaii Lensing Survey (CFHTLenS)
  and Red-Cluster-Sequence Lensing Survey (RCSLenS).
  We generate a large set of full-sky mock catalogs 
  from cosmological $N$-body simulations  
  and use them to estimate statistical errors accurately. 
  The measured cross correlation is consistent with null detection, 
  which is then used to place strong cosmological constraints 
  on annihilating and decaying DM.
  For leptophilic DM, the constraints are improved 
  by a factor of $\sim100$ in the mass range of $O(1)$ TeV
  when including contributions from secondary $\gamma$ rays
  due to the inverse-Compton upscattering of background photons.
  Annihilation cross-sections of
  $\langle \sigma v \rangle \sim 10^{-23}\, {\rm cm}^3/{\rm s}$
  are excluded for TeV-scale DM depending on channel.
  Lifetimes of $\sim 10^{25}$ sec are also excluded for 
  the decaying TeV-scale DM. 
  Finally, we apply this analysis to wino DM 
  and exclude the wino mass around 200 GeV.
  These constraints will be further tightened, 
  and all the interesting wino DM parameter region can be tested, 
  by using data from
  future wide-field cosmology surveys. 
\end{abstract}

\maketitle

\section{\label{sec:intro}INTRODUCTION}

An array of astronomical observations  
over a wide range of redshifts and length scales
consistently support the existence of cosmic dark matter (DM).
Recent observations include the statistical analysis of the
cosmic microwave background (CMB) anisotropies
(e.g., Refs.~\cite{Hinshaw:2012aka, Ade:2013zuv}), 
spatial clustering of galaxies 
(e.g., Ref.~\cite{Eisenstein:2005su}),
galaxy rotation curves
(e.g., Ref.~\cite{Persic:1995ru}),
and direct mapping of matter distribution through gravitational lensing 
(e.g., Ref.~\cite{Clowe:2006eq}).

Gravitational lensing is a direct and most promising probe
of the matter density distribution in the Universe.    
A foreground gravitational field 
causes small distortions of the images of distant background galaxies.  
The small distortions contain, collectively, 
rich information on the foreground matter distribution
and its growth over cosmic time.
In the past decades, the coherent lensing effect between galaxy pairs
with angular separation of $\sim$ degree 
has been successfully detected in wide-area surveys (e.g., Refs~\cite{Bacon:2000sy, Kilbinger:2012qz, Becker:2015ilr}).
Most importantly,
the large angular scale signals, called cosmic shear, probe the matter
distribution in an {\it unbiased} manner. 
However, cosmic shear alone does not provide, by definition,
any information on possible electromagnetic
signatures from DM, 
and thus it cannot be used to probe the particle properties of DM
such as annihilation cross section and decay lifetime.

The extragalactic $\gamma$-ray background
is thought to be a potential probe of DM, if DM annihilates or decays to
produce high-energy photons.
Weakly interacting massive particles (WIMPs) are promising
DM candidates that can naturally explain the observed abundance of cosmic DM 
if the WIMP mass ranges from 10 GeV to 10 TeV 
and the self-annihilation occurs around the weak-interaction scale \cite{1996PhR...267..195J}.
Dark matter decay lifetime remains largely unknown, and, in fact,
there are {\it not} strong cosmological and astrophysical evidences for stable DM;
the possibility of very long-lived particles with a lifetime 
longer than the age of the Universe of 13.8 Gyr remains viable. 
DM annihilation or decay produces a variety of 
cascade products and thus leaves characteristic imprints 
in, for example, the cosmic $\gamma$-ray background.
The isotropic $\gamma$-ray background (IGRB) is a promising
target to search for DM annihilation or decay~\cite{Funk:2015ena}.
Although the mean IGRB intensity can be explained 
by (extrapolating) unresolved astrophysical sources (e.g., \cite{2015ApJ...800L..27A}),
there remains substantial uncertainties and thus there is room for
contribution from other unknown sources.
The anisotropies in the diffuse $\gamma$-ray background should in principle
contain rich information about DM contributions at small and large length scales
(e.g., see Ref.~\cite{Fornasa:2015qua} for review). 

It has been proposed that the cross-correlation of the IGRB 
with large-scale structure provides 
a novel probe of the microscopic properties of DM 
\cite{Camera:2012cj, 2014FrP.....2....6F, 2014PhRvD..90b3514A, Ando:2014aoa, 2015JCAP...06..029C}.
Positive correlations with actual galaxy survey data 
\cite{2015ApJS..217...15X} have been reported, and implications for
the nature of DM have been discussed \cite{2015PhRvL.114x1301R, Cuoco:2015rfa}.

In this paper, we search indirect DM signals through
cross-correlation of the IGRB and cosmic shear.
We improve the cross-correlation measurement
over our previous analysis \cite{Shirasaki:2014noa}
by using the latest $\gamma$-ray data taken from the Fermi-LAT
and two publicly available galaxy catalogs, the Canada-France-Hawaii Lensing Survey (CFHTLenS)
and the Red-Cluster-Sequence Lensing Survey (RCSLenS),
that provide with precise galaxy shape measurement.
We apply a set of Galactic $\gamma$-ray emission models to 
characterize the foreground emission from our own Galaxy,
and also utilize full-sky simulations of cosmic shear to construct 
realistic mock galaxy catalogues specifically for CFHTLenS
and RCSLenS.
In order to make the best use of the cross-correlation signals 
over a wide range of angular separations,
we calculate the statistical uncertainties associated
with the intrinsic galaxy shapes, the Poisson photon noise, 
and the sample variance of cosmic shear. To this end,
we make use of our large mock catalogues in a manner
closely following the actual observations.
The methodology presented in this paper 
is readily applicable to cross-correlation analyses
of the IGRB and cosmic shear with ongoing and future surveys,
such as the Hyper-Suprime Cam, the Dark Energy Survey,
the Large Synoptic Survey Telescope, and the Cherenkov Telescope Array.

The rest of the paper is organized as follows.
In Section~\ref{sec:DM}, we summarize the basics of the
two observables of interest: IGRB and cosmic shear.
We also present a theoretical model of the cross-correlation of 
the IGRB and cosmic shear in annihilating or decaying DM scenarios. 
In Section~\ref{sec:data}, 
we describe the $\gamma$-ray data and the galaxy imaging survey 
for shape measurement. 
The details of the cross-correlation analysis are provided in Section~\ref{sec:cross}.
In Section~\ref{sec:res}, 
we show the result of our cross-correlation analysis,
and derive constraints on particle DM.
Concluding remarks and discussions are given in Section~\ref{sec:con}. 
Throughout the paper, we adopt the standard $\Lambda$CDM model
with the following parameters;
matter density $\Omega_{\rm m0}=0.279$, 
dark energy density $\Omega_{\Lambda}=0.721$, 
the density fluctuation amplitude
$\sigma_{8}=0.823$,
the parameter of the equation of state of dark energy $w_{0} = -1$,
Hubble parameter $h=0.700$ and 
the scalar spectral index $n_s=0.972$.
These parameters are consistent with 
the WMAP nine-year results \citep{Hinshaw:2012aka}.

\section{\label{sec:DM}A MODEL OF DARK MATTER}
\subsection{Extragalactic probe}
We consider two observables to probe 
dark matter properties: the IGRB and cosmic shear.
The former may contain $\gamma$ rays from 
DM annihilation/decay 
located outside of our galaxy,
while the latter provides unbiased information of the
DM mass distribution in the Universe. 
\subsubsection*{Isotropic $\gamma$-ray background}

The IGRB intensity $I_\gamma$ is defined by 
the number of photons per unit energy, area, time, and solid angle,
\begin{equation}\label{eq:Intensity}
E_\gamma I_\gamma = \frac{c}{4\pi} \int {\rm d}z \frac{P_\gamma (E'_\gamma,z)}{H(z)(1+z)^4} e^{-\tau(E'_\gamma,z)},
\end{equation}
where $E_\gamma$ is the observed $\gamma$-ray energy, 
$E'_\gamma = (1+z) E_\gamma$ is the energy 
of the $\gamma$ ray at redshift $z$, 
$H(z) = H_0 [\Omega_{\rm m0}(1+z)^3+\Omega_\Lambda]^{1/2}$ 
is the Hubble parameter in a flat Universe, 
and the exponential factor in the integral
takes into account the effect of $\gamma$-ray attenuation 
during propagation owing to pair creation 
on diffuse extragalactic photons. 
For the $\gamma$-ray optical depth $\tau\left(E'_\gamma, z \right)$, 
we adopt the model in Ref.~\citep{Gilmore:2011ks}.
In Eq.~(\ref{eq:Intensity}), $P_\gamma$ represents 
the volume emissivity 
(i.e., the photon energy emitted per unit volume, time, and energy range),
which is given by
\beqa\label{eq:dmEmissivity}
P_\gamma(E_\gamma, z)=
E_\gamma {\cal S}_{\rm dm}(E_\gamma, z) 
{\cal F}_{\rm dm}(\bd{r}, z),
\eeqa
where ${\cal S}_{\rm dm}$ 
is a $\gamma$-ray source function and ${\cal F}_{\rm dm}$ represents 
the relevant density field of $\gamma$-ray sources.

Here we consider two particle properties of DM,
annihilation and decay.
For annihilating DM, the two functions ${\cal S}_{\rm dm}$
and ${\cal F}_{\rm dm}$ are given by
\beqa
{\cal S}_{\rm dm} (E_\gamma, z) &=& 
\frac{\langle \sigma v \rangle}{2 m^{2}_{\rm dm}}
\left(\frac{{\rm d}N_{\gamma, a}}{{\rm d}E_\gamma}+Q_{{\rm IC},a}(E_\gamma, z)
\right) , \\
\label{eq:source_func_a}
{\cal F}_{\rm dm}(\bd{r}, z) &=&
\left[\rho_{\rm dm}(\bd{r}, z)\right]^{2},
\eeqa
where 
${\rm d}N_{\gamma, a} /{\rm d}E_\gamma$ 
is the $\gamma$-ray spectrum per annihilation,
$Q_{{\rm IC}, a}$ describes the $\gamma$-ray energy distribution through 
the inverse-Compton (IC) scattering
between $e^{\pm}$ produced by annihilation and background photons,
$\langle \sigma v \rangle$ is the annihilation cross section times the relative velocity averaged with the velocity distribution function,
and 
$m_{\rm dm}$ is the DM particle mass.
For decaying dark matter, 
\beqa
{\cal S}_{\rm dm} (E_\gamma, z) &=& 
\frac{\Gamma_{\rm d}}{m_{\rm dm}}
\left(\frac{{\rm d}N_{\gamma, d}}{{\rm d}E_\gamma}+Q_{{\rm IC}, d}(E_\gamma, z)
\label{eq:source_func_d}
\right) , \\
{\cal F}_{\rm dm}(\bd{r}, z) &=& \rho_{\rm dm}(\bd{r}, z),
\eeqa
where 
${\rm d}N_{\gamma, d} /{\rm d}E_\gamma$ 
is the $\gamma$-ray spectrum per decay,
$Q_{{\rm IC}, d}$ is the contribution from 
the inverse-Compton scattering of decay products $e^{\pm}$,
and 
$\Gamma_{\rm d}$ is the decay rate.
Note that we assume the same source distributions of the primary and secondary photons in this paper. 
The source distribution of secondary photons can differ from 
that of primary photons, because the annihilation/decay products
can propagate or diffuse while they produce secondary photons.
The diffusion time scale of electrons with $\sim$100 GeV 
is much longer than the cooling time scale due to inverse-Compton energy losses in both galaxy-sized and cluster-sized haloes 
\cite{Ishiwata:2009dk,2014IJMPD..2330007B}.
We can safely ignore the propagation effect of sources 
of secondary photons in our halo model approach.

We assume that the IGRB intensity is measured in the energy range
$E_{\gamma, {\rm min}}$ to $E_{\gamma, {\rm max}}$ 
along a given angular direction $\hat{\bd n}$.
The DM contribution in Eq.~(\ref{eq:Intensity}) 
is given by 
\beqa
I_{\gamma}(\hat{\bd n}) &=& \int {\rm d}\chi
\, W_{\rm dm}(\chi) {\cal F}_{\rm dm}(\chi \hat{\bd n}, z(\chi)), 
\label{eq:Intensity_dm}
\\
W_{\rm dm}(\chi) &=&
\int_{E_{\gamma, {\rm min}}}^{E_{\gamma, {\rm min}}}
\frac{{\rm d}E_{\gamma}}{4\pi} \, \frac{{\cal S}_{\rm dm}(E'_{\gamma}, z(\chi))}{(1+z(\chi))^3} 
e^{-\tau(E'_\gamma, z(\chi))},
\eeqa
where $\chi(z)$ is the comoving distance.
Throughout this paper, 
we set $E_{\gamma, {\rm min}} = 1\, {\rm GeV}$
and $E_{\gamma, {\rm max}} = 500\, {\rm GeV}$, respectively.

\subsubsection*{Cosmic shear}

Images of distant galaxies are distorted by the weak lensing effect 
due to foreground matter distributions.
The weak lensing effect is commonly characterized by the following 
two-dimensional matrix;
\beqa
A_{ij} = \frac{\partial \beta^{i}}{\partial \theta^{j}}
           \equiv \left(
\begin{array}{cc}
1-\kappa -\gamma_{1} & -\gamma_{2}  \\
-\gamma_{2} & 1-\kappa+\gamma_{1} \\
\end{array}
\right), \label{distortion_tensor}
\eeqa
where
$\kappa$ is the convergence, $\gamma$ is shear, and
$\mbox{\boldmath $\theta$}$ and $\mbox{\boldmath $\beta$}$
represent the observed position and 
the true position of a source object, respectively.

When considering the metric perturbation in the Newtonian gauge,
one finds that each component of $A_{ij}$ can be related 
to the second derivative of the gravitational potential 
\cite{Bartelmann:1999yn}.
Since the gravitational potential can be related to the matter density
through the Poisson equation, the convergence field is expressed as 
\beqa\label{eq:kappa}
\kappa(\hat{\bd n}) = \int {\rm d}\chi\, W_{\kappa}(\chi) 
\delta (\chi\hat{\bd n}, \chi(z)),
\eeqa
where $\delta$ is the matter overdensity field.
The weight $W_{\kappa}$ in Eq.~(\ref{eq:kappa}) is given by
\beqa
W_{\kappa}(\chi) &=& \frac{3}{2}
\left(\frac{H_{0}}{c}\right)^2 
\Omega_{\rm m0}(1+z(\chi))
f_{K}(\chi)\, \nonumber \\ 
&&
\,\,\,\,\,\,\,\,\,\,\,\,\,\,\,
\times
\int_{\chi}^{\chi_{H}} {\rm d}\chi^{\prime} \, 
p(\chi^{\prime}) \frac{f_{K}(\chi^{\prime}-\chi)}{f_{K}(\chi^{\prime})},
\eeqa
where $\chi_{H}$ is the comoving distance to the horizon,
$f_{K}(\chi)$ is the comoving angular diameter distance,
and $p(\chi)$ represents the source distribution normalized to 
$\int {\rm d}\chi\, p(\chi)=1$.

\subsection{Microscopic processes}
DM annihilation or decay generally yields a mixture
of various final states depending upon the particle interaction model.
We consider three benchmark models with 
$100$\% branching ratios to $\mu^{+}\mu^{-}$, $\tau^+\tau^-$, and $b\bar{b}$ final states;
in Section \ref{sec:dmlimits} we also explore a specific model of wino DM. 
We use the {\tt PPPC4DMID} package 
\citep{Cirelli:2010xx}
to estimate the primary $\gamma$-ray energy spectrum.
For $\tau^+\tau^-$ and $b\bar{b}$ final states, hadronic production of neutral pions are 
kinematically allowed and the $\gamma$-ray spectra ${\rm d}N_{\gamma, a} / {\rm d}E_\gamma$ 
are dominated by the decay of neutral pions. 
For decaying DM with particle mass of $m_{\rm dm}$,
we model the $\gamma$-ray spectrum per decay 
(${\rm d}N_{\gamma, d}/{\rm d}E_\gamma$)
using the $\gamma$-ray spectrum from annihilation 
(${\rm d}N_{\gamma, a}/{\rm d}E_\gamma$)
except with a particle mass of $m_{\rm dm}/2$.
These are {\it primary} $\gamma$-ray emissions, to be distinguished from {\it secondary} emission 
that results from interactions of the annihilation products with the environment. 

An example of a relevant secondary emission is the inverse-Compton 
$\gamma$-rays that are up-scattered background photons off
of the high energy $e^{\pm}$ produced through annihilation or decay or subsequent interactions.
We follow Refs.~\cite{Ando:2015qda, Ando:2016ang}
to calculate the inverse-Compton contributions from 
up-scattered photons, i.e., 
$Q_{{\rm IC},a}$ in Eq.~(3) and 
$Q_{{\rm IC},d}$ in Eq.~(\ref{eq:source_func_d}).
The energy spectrum of the inverse-Compton $\gamma$-rays 
can be expressed as
\cite{Ishiwata:2009dk, Ando:2015qda, Ando:2016ang}
\beqa
Q_{{\rm IC},i}(E_{\gamma}, z) 
&=& c \int {\rm d}E_{e}{\rm d}E_{\rm BG}
(1+z)\frac{{\rm d}\sigma_{\rm IC}}{{\rm d}E'_{\gamma}}
(E'_{\gamma}, E_{e}, E_{\rm BG}) \nonumber \\
&&
\,\,\,\,\,\,\,\,\,\,\,\,\,\,\,
\times
f_{\rm BG}(E_{\rm BG},z)
\frac{Y_{e, i}(E_{e})}{b_{\rm IC}(E_{e}, z)},
\label{eq:Q_IC}
\eeqa
with $i=a~{\rm or}~d$ for annihilation and decay, respectively.
Here, $E_{e}$ and $E_{\rm BG}$ represent the energy of 
$e^{\pm}$ and the background photons,
$f_{\rm BG}$ is the energy spectrum of the background photon field,
${\rm d}\sigma_{\rm IC}/{\rm d}E_{\gamma}$ is the differential
cross section of the inverse-Compton scattering,
and $b_{\rm IC}$ is the energy loss rate of $e^{\pm}$.
In Eq.~(\ref{eq:Q_IC}), $Y_{e, i}(E_{e})$ is defined by
\beqa
Y_{e, i}(E_{e}) =\int_{E_{e}}^{\infty} {\rm d}E 
\left(\frac{{\rm d}N_{e^{+}, i}}{{\rm d}E}+
\frac{{\rm d}N_{e^{-}, i}}{{\rm d}E}
\right),
\label{eq:Ye}
\eeqa
where ${\rm d}N_{e^{\pm}, i}/{\rm d}E$ is the energy distribution
of $e^{\pm}$ produced by DM annihilation $(i=a)$ 
or decay $(i=d)$.
We consider the CMB as the background photons and ignore 
the extragalactic background light (EBL) with other wavelengths.
Ref.~\cite{Ando:2016ang} has shown that the EBL would 
only have a $\sim5\%$ contribution to 
the energy loss rate of $e^{\pm}$ at 
$z=0$, and even less at higher $z$.
In order to calculate the energy spectrum of 
inverse-Compton $\gamma$-rays, 
we need the energy distribution of $e^{\pm}$
in annihilation/decays, which we calculate using {\tt PPPC4DMID} \citep{Cirelli:2010xx} for each of our adopted final states.

For CMB, the integrand of Eq.~(\ref{eq:Q_IC}) has a maximum 
at $E_{\rm BG}\sim10^{-3}$ eV and 
$E_{e}/E'_{\gamma}\sim200-300$ for $E'_{\gamma}=1$ GeV.
Hence heavy DM with mass
$m_{\rm dm} \simgt 200-300$ GeV would be needed for prominent secondary 
$\gamma$-ray emissions,
because the integrand is proportional to the cumulative number of 
$e^{\pm}$ given by Eq.~(\ref{eq:Ye}).
For $E'_{\gamma}=0.1$ GeV,
we find the maximum in the integrand of Eq.~(\ref{eq:Q_IC})
to be at $E_{\rm BG}\sim10^{-3}$ eV, and thus 
$E_{e}/E'_{\gamma}\sim1000$,
suggesting that DM with mass of $m_{\rm dm}\sim100$ GeV
yields a larger secondary contribution than the primary one 
at $E'_{\gamma}=0.1$ GeV.
Therefore, setting a low energy cut in $\gamma$-ray analysis 
would be important to efficiently search for cross-correlation signals
from DM annihilation or decay with $m_{\rm dm}\sim O(100)$ GeV.
In practice, we use the $\gamma$ rays with $E_{\gamma}\geq1$ GeV
in our analyses, because the smearing effect of Fermi's point
spread function (PSF) on cross-correlation analysis is significant
in the range of $E_{\gamma}\simlt1$ GeV at present.

\subsection{Cross-correlation statistics}\label{subsec:cross_corr_model}

Using Eqs.~(\ref{eq:Intensity_dm}) and (\ref{eq:kappa}),
with the Limber approximation \cite{Limber:1954zz},
we can calculate the angular cross-power spectrum 
of the convergence field and extragalactic $\gamma$ rays emitted 
through DM annihilation/decay as,
\beqa\label{eq:ang_power}
C_{\kappa, {\rm IGRB}}(\ell) 
&=& 
\int \frac{{\rm d} \chi}{f_{K}(\chi)^2} 
W_{\rm dm}(\chi) W_{\kappa}(\chi) 
\nonumber \\
&&
\,\,\,\,\,\,\,\, \,\,\,\,\,\,\,\, 
\times \, 
P_{\delta, {\cal F}}\left(k=\frac{\ell+1/2}{f_{K}(\chi)}, z(\chi)\right), 
\eeqa
where $P_{\delta, {\cal F}}(k, z)$ represents
the three-dimensional power spectrum between two fields 
of $\delta$ and ${\cal F}_{\rm dm}$
as a function of wave number $k$ and redshift $z$.

In practice, a more direct observable is the two-point cross-correlation function of the tangential shear 
$\gamma_{t}(\hat{\bd n})$ and $\gamma$-ray intensity
\cite{Shirasaki:2014noa}. 
The two-point cross-correlation function $\xi(\theta)$
can be calculated from the angular power spectrum 
by the following equation:
\beqa
\xi(\theta) = \int \frac{{\rm d}\ell \ell}{2\pi} C_{\kappa, {\rm IGRB}}(\ell) J_{2}(\ell \theta),
\eeqa
where $J_{2}(x)$ represents the second-order Bessel function
\citep{dePutter:2010jz, Oguri:2010vi}.
Furthermore, the smearing effect of Fermi's PSF on $\xi$ 
would be relevant to the $\gamma$-ray statistics on small angular scales.
Thus, we apply the same framework to take into account 
the PSF effect, as in Ref.~\cite{Shirasaki:2014noa}.

For annihilation, $P_{\delta, {\cal F}}(k, z)$ in 
Eq.~(\ref{eq:ang_power}) is equivalent 
to $\bar{\rho}_{\rm dm}^2 (z) P_{\delta, \delta^2}(k, z)$
where $\bar{\rho}_{\rm dm}$ is the mean DM density
at redshift $z$ and $P_{\delta, \delta^2}(k, z)$ represents
the cross power spectrum of matter overdensity and its squared.
We adopt the halo model approach to predict 
$P_{\delta, \delta^2}(k, z)$ as in Ref.~\cite{Shirasaki:2014noa}.
In the framework of the halo model \cite{Cooray:2002dia},
$P_{\delta, \delta^2}(k, z)$ can be expressed 
as a sum of two terms,
called the one-halo and two-halo term, respectively.
The former represents the two-point correlation within a given DM halo, 
and the latter corresponds to the  
clustering of DM halos. 
In the halo-model calculation, 
we adopt the model of halo mass function 
and linear halo bias as in Refs.~\cite{Tinker:2008ff, Tinker:2010my},
and assume the NFW density profile \cite{Navarro:1996gj}
with concentrations as given in Ref.~\cite{Prada:2011jf}.
For the boost factor $b_{sh}$ that describes the effective increase
of the amplitude of the DM density clumping, 
we consider two phenomenological models by 
Refs.~\cite{Gao:2011rf,Sanchez-Conde:2013yxa}.
In one model, subhalo properties are scaled as power laws 
and extrapolated many orders of magnitudes to 
the smallest subhalos ($10^{-6} M_\odot$)
\cite{Gao:2011rf}. 
This procedure yields a large boost factor, but it is rather 
uncertain whether the extrapolation remains valid to such small halos. 
We thus test another model as an optimistic scenario, 
where the halo concentration is given by a relation that flattens at small halo masses, 
yielding a smaller boost factor \cite{Sanchez-Conde:2013yxa}. 
We regard the latter model as a conservative case,
because the concentration-mass relation is derived from 
field halos rather than subhalos. 
For a given mass, a subhalo is expected to be more concentrated than a field halo,
and hence the boost factor in the latter model remains modest. 
Ref.~\cite{Bartels:2015uba} estimates that the net effect is 
a factor of 2--5 increase in $b_{sh}$ of Ref.~\cite{Sanchez-Conde:2013yxa}.

\begin{figure}[!t]
\begin{center}
       \includegraphics[clip, width=0.8\columnwidth, bb=0 0 516 484]
       {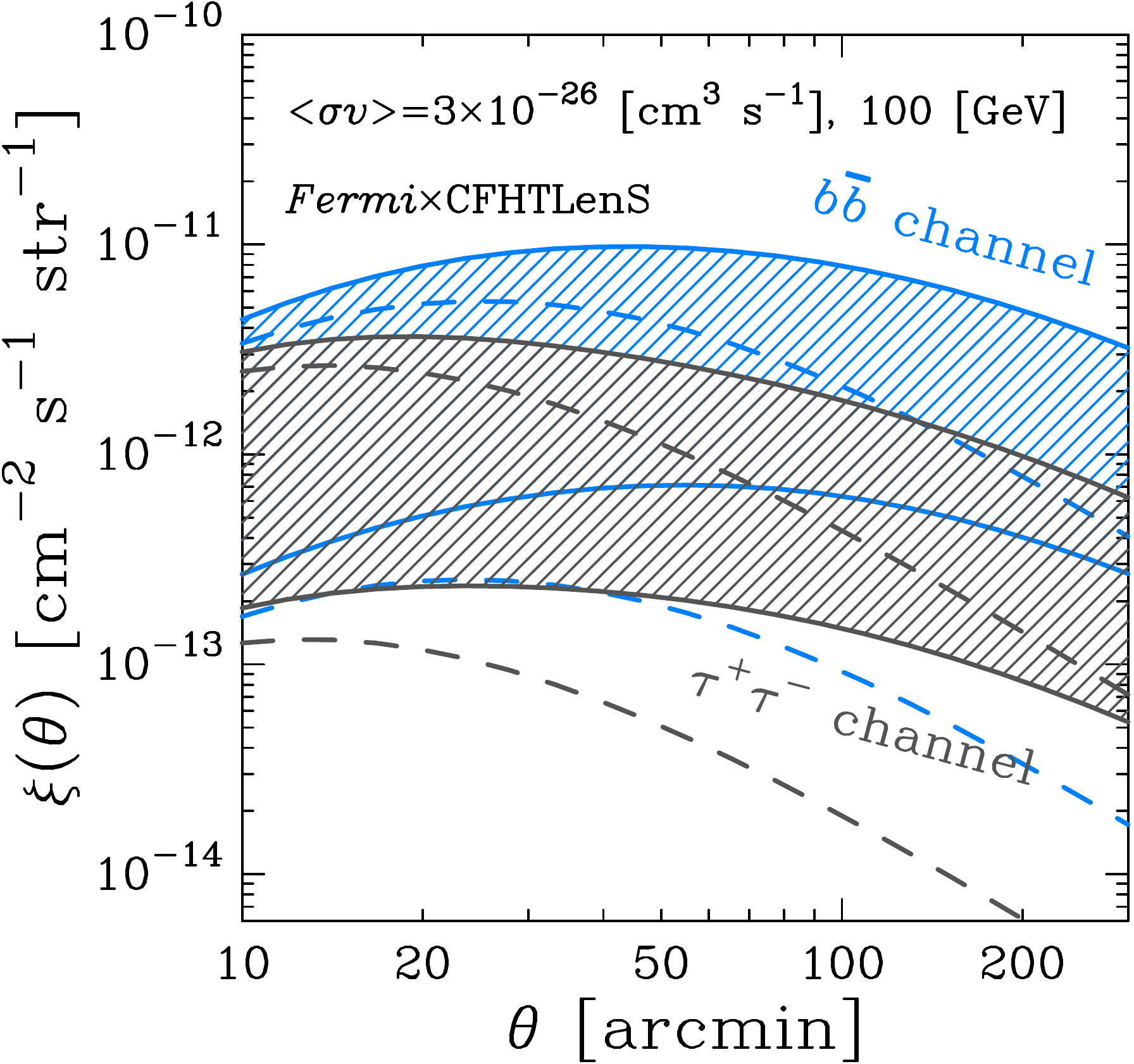}
       \includegraphics[clip, width=0.8\columnwidth, bb=0 0 516 484]
       {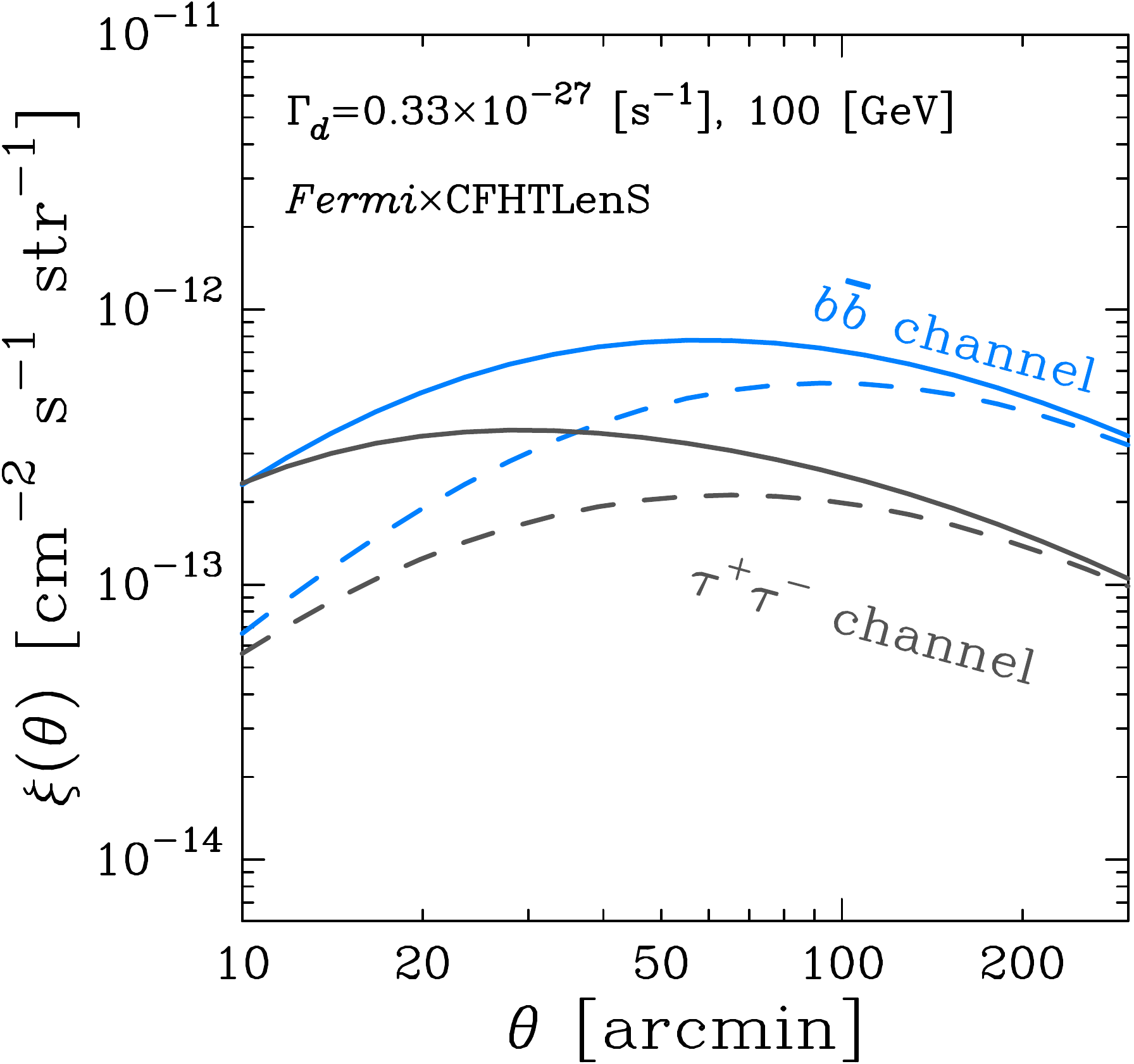}
     \caption{
     \label{fig:model_100GeV}
     We plot the model IGRB-cosmic shear cross-correlation 
     as a function of angle separation.
     We assume a DM particle mass of 100 GeV 
     and ignore secondary $\gamma$ rays
     in this figure. The different colors represent 
     different channels: 
     $b\bar{b}$ (cyan) and $\tau^{+}\tau^{-}$ channel (gray).
     {\it Top}: DM annihilation with the canonical cross section of
     $\langle \sigma v \rangle = 3\times 10^{-26}\, {\rm cm}^3/{\rm s}$.
     For each colored line, the shaded band indicates
     our conservative estimate of the theoretical uncertainty caused by 
     the subhalo boost factor model.
     The solid line shows the total correlation function
     while the dashed line represents the so-called one-halo term (see text).
     {\it Bottom}: DM decay with decay rate of 
     $3.3\times10^{-28}\, {\rm s}^{-1}$.
     The solid line shows the correlation function with proper modeling
     of nonlinear gravitational growth,
     while the dashed line represents the expected signal 
     with linear growth of density perturbations.
     Note that we assume the source galaxy distribution 
     of Ref.~\cite{Kilbinger:2012qz} and 
     the $\gamma$-ray PSF as in Ref.~\cite{2009ApJ...697.1071A}.
     }
    \end{center}
\end{figure}

Fig.~\ref{fig:model_100GeV} shows our benchmark model of $\xi(\theta)$.
Here, we consider two representative channels ($b\bar{b}$ and $\tau^{+}\tau^{-}$) for a 100 GeV DM,
and only show the primary $\gamma$-ray contribution. 
The top panel shows the case of annihilating DM 
with a thermal cross section $\langle \sigma v \rangle = 3\times 10^{-26}\, {\rm cm}^3/{\rm s}$.
We also show the model uncertainty
originating from the choice of the boost factor.
The expected correlation for each channel lies 
in the shaded region in the top panel.
The uncertainty of the boost factor causes an
uncertainty in $\xi(\theta)$ of a factor of $\sim10$ for annihilating DM.
Our model predicts that 
the contribution from inside a single DM halo (i.e., arising from subhalos) dominate 
on scale of $\simlt 30$ arcmin, 
while the clustering of DM halos  
induce significant correlations at $\simgt 1$ deg.

For decaying DM, $P_{\delta, {\cal F}}(k, z)$ in 
Eq.~(\ref{eq:ang_power}) is given by
 $\bar{\rho}_{\rm dm} (z) P_{\delta}(k, z)$,
where $P_{\delta}(k, z)$ is
the three-dimensional power spectrum of the matter overdensity.
The power spectrum $P_{\delta}(k, z)$ has been calibrated
with a set of N-body simulations 
over a wide range of $k$ and $z$ for cosmological analyses.
We adopt a fitting formula of 
$P_{\delta}(k, z)$
that reproduces the nonlinear effects on two-point statistics 
at $\simgt1\, {\rm Mpc}$ scales, consistently with the result of cosmological N-body simulations
\cite{Takahashi:2012em}.
For comparison, we also calculate the cross-correlation
adopting linear matter power spectrum.

\begin{figure}[!t]
\begin{center}
       \includegraphics[clip, width=0.8\columnwidth, bb=0 0 516 470]
       {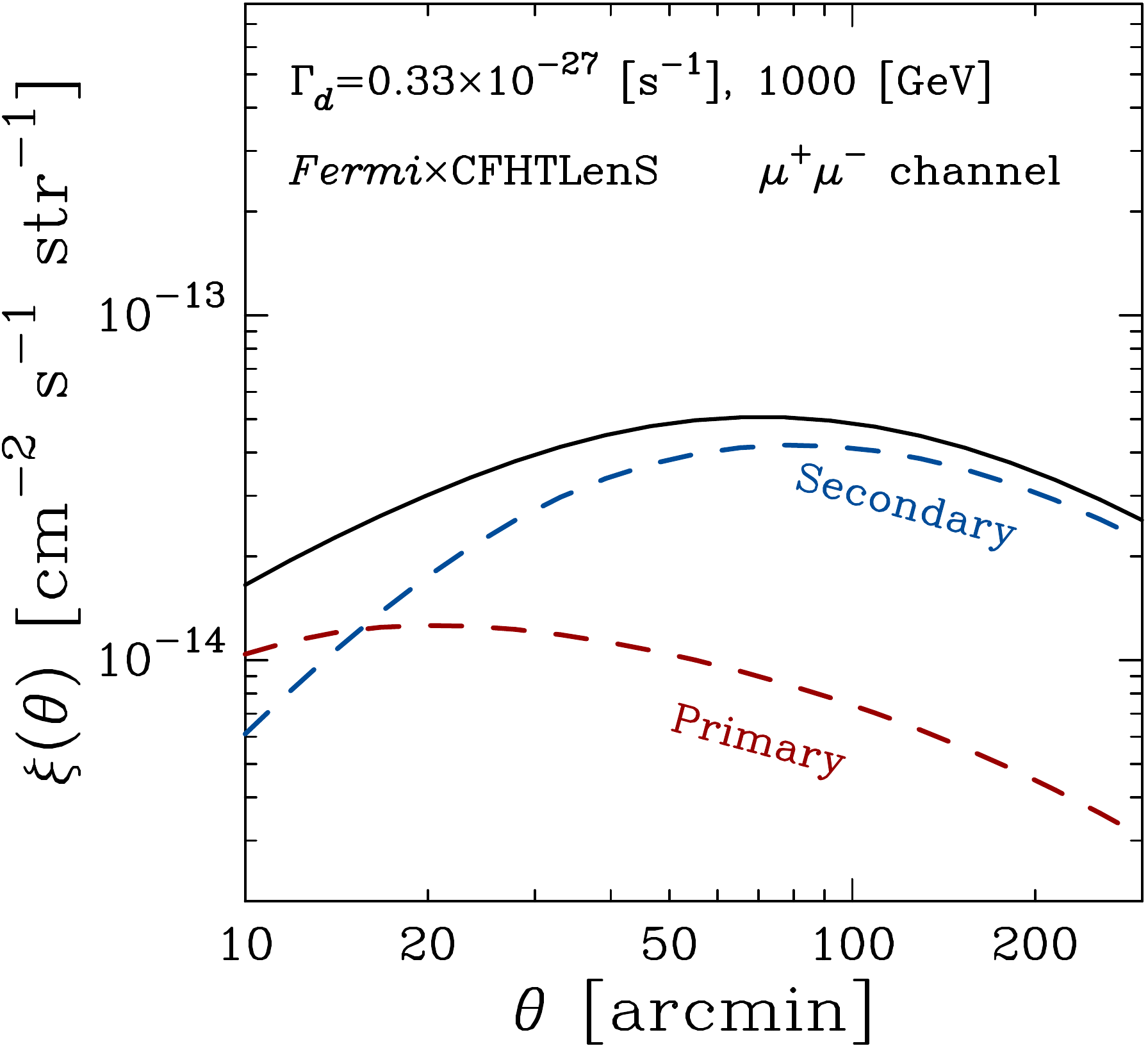}
     \caption{
     \label{fig:impact_IC_mm_decay_1000GeV}
     The contributions from primary and secondary $\gamma$ rays to 
     the cross-correlation function.
     Shown are the correlation expected from DM decay.
     We assume a DM particle mass of 1 TeV and a decay rate
     of $3.3\times10^{-28}\, {\rm s}^{-1}$ 
     for the $\mu^{+}\mu^{-}$ model.
     The red dashed line shows the contribution from primary $\gamma$ rays,
     while the blue shows secondary $\gamma$ rays induced by 
     the inverse-Compton upscattering of CMB photons by the 
     decay products $e^{\pm}$. The solid line is the sum of the
     primary and secondary contributions.
  } 
    \end{center}
\end{figure}

The bottom panel of Fig.~\ref{fig:model_100GeV}
shows the case of decaying DM 
with decay rate of $\Gamma_{\rm d} 
= 3.3\times 10^{-28}\, {\rm s}^{-1}$.
There, the solid lines represent the model with nonlinear power
spectrum $P_{\delta}$, while the dashed lines are with linear $P_{\delta}$.
We find that the nonlinear gravitational growth enhances the signal
at $\simlt30$ arcmin scales by a factor of a few.
The linear approximation 
is valid only at scales larger than a few degrees.

We next consider the effect of secondary $\gamma$ rays 
on our cross-correlation statistics $\xi$.
Fig.~\ref{fig:impact_IC_mm_decay_1000GeV}
summarizes the contributions for a 1 TeV dark matter decaying 
into $\mu^{+}\mu^{-}$ final states.
The red line shows the correlation signals originating from
primary $\gamma$ rays, while
the blue line is for the secondary contribution produced
by the inverse-Compton scattering of CMB photons by
the decay products $e^{\pm}$.
As shown in Fig.~\ref{fig:impact_IC_mm_decay_1000GeV},
the secondary $\gamma$ rays dominate the cross-correlation
statistics when large amounts of energetic leptons are generated, 
as in the case of heavy DM particles with 1-10 TeV masses
annihilating/decaying in leptonic channels. 
The different angular dependence between the red and blue lines
originates from the energy dependence of the $\gamma$-ray PSF.
The primary $\gamma$ rays have a dominant contribution to the IGRB intensity
for $E_{\gamma}>100$ GeV,
while the secondary mainly contributes to observed 
photon counts with lower energies of $E_{\gamma}\sim 1$ GeV 
\cite{Ando:2016ang}.
Since the size of the $\gamma$-ray PSF is smaller for higher energy photons, 
the secondary contribution tends to be 
affected more by the PSF smoothing
than the primary contribution.
 
\section{\label{sec:data}DATA}
\subsection{\label{subsec:fermi}Fermi-LAT data selection and methodology}

The Fermi-LAT detects $\gamma$-ray photons in the energy range 20 MeV to over 300 GeV~\citep{2009ApJ...697.1071A}. In operation since August 2008, this telescope makes all sky observations every $\sim 3$ hours, making it an excellent tool to investigate diffuse $\gamma$-ray background photons.  The LAT instrument is described in detail in Ref.~\citep{2009ApJ...697.1071A}.

\begin{figure*}[ht!]
\begin{center}

\includegraphics[width=0.9\linewidth, bb=0 0 454 574]
{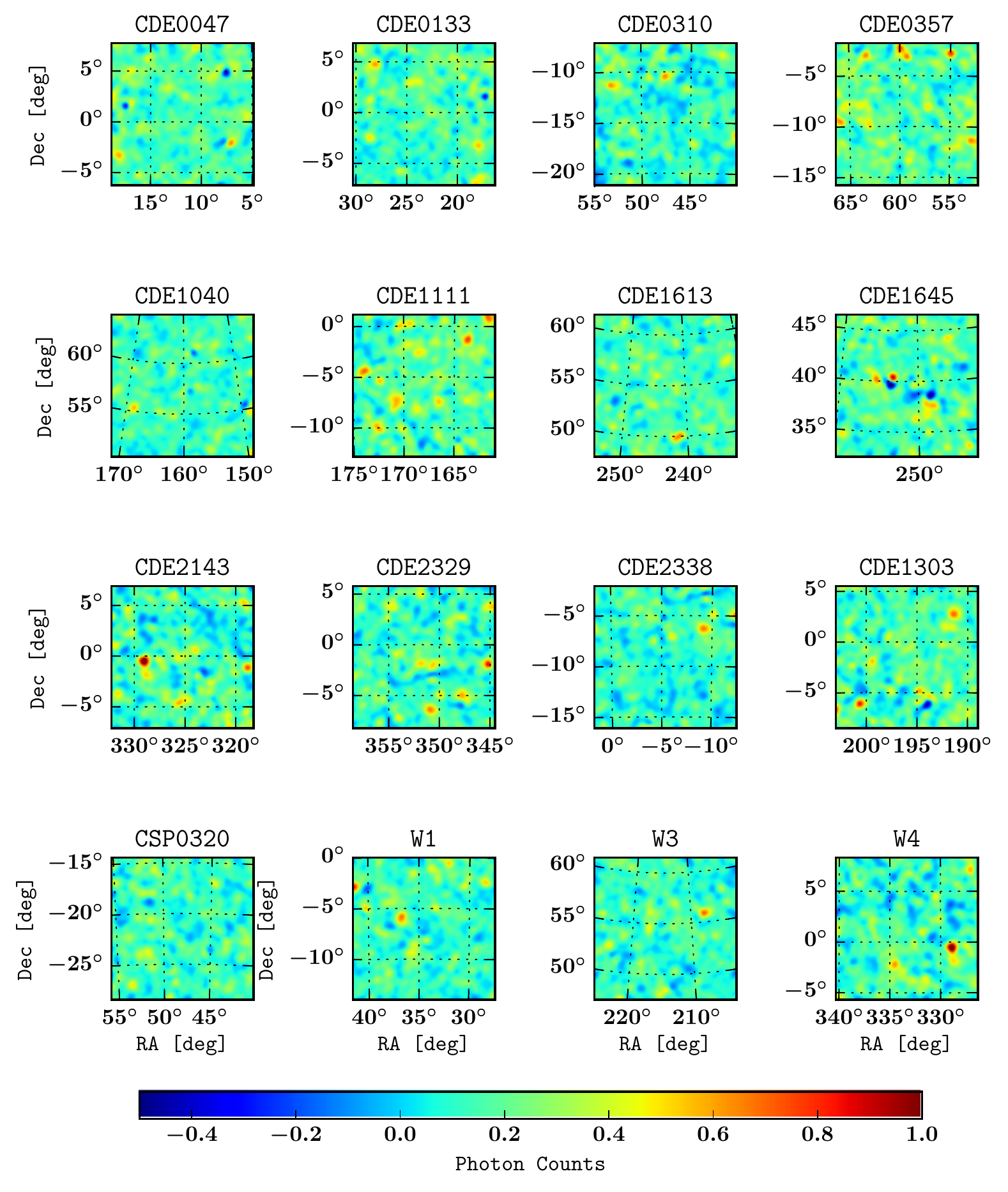} 

\caption{ \label{fig:IGRBresiduals} Residual IGRB counts for energies $>1$ GeV in the lensing survey regions. The isotropic components are calculated by subtracting the best-fit 
DGB and point sources from
the raw photon counts in each ROI (see text). All maps comprise an area of $14^{\circ}\times 14^{\circ}$ and have been passed through a Gaussian filter of 0.3$^{\circ}$.   }
\end{center}
\end{figure*}    

The analysis presented here was carried out 
with $\sim 7$ years of observations from August 4, 2008$
-$September 4, 2015 of the recently released \textsc{Pass 8} 
data.
For each CFHTLens and RCSLens patch, 
we selected events within a squared region of 
$14^{\circ}\times14^{\circ}$ (around their respective centroids) 
with energies greater than 1 GeV, and using combined 
\textit{Front} and \textit{Back} events. 
At low energies the angular resolution of the LAT 
is poor and source confusion could introduce a bias in the analysis, 
the chosen low energy limit is a compromise between 
sensitivity and statistics. 

Events  and  instrument  response  functions  (IRFs)  for  the lowest-residual  cosmic ray (CR)  background  ``ULTRACLEANVETO''~\footnote{ULTRACLEANVETO events are the recommended class of photons for investigation of the IGRB at intermediate to high latitudes. The reader is referred to the Cicerone (\protect\url{http://fermi.gsfc.nasa.gov/ssc/data/analysis/documentation/Pass8_usage.html}) for further details} from the  \textsc{Pass 8}  were used. The zenith angles were chosen to be smaller than 100$^{\circ}$ to reduce contamination from  the Earth limb. Time intervals when the rocking angle was more than 52$^{\circ}$ and when the Fermi satellite was within the South Atlantic Anomaly were also excluded.

In order to determine the IGRB photons in all of the CFHTLens and RCSLens region of interest (ROI) we performed a maximum-likelihood fit~\citep{3FGL} with the \textit{pyLikelihood} analysis tool \footnote{
\protect\url{http://fermi.gsfc.nasa.gov/ssc/data/analysis/documentation/}}. The energy binning was set to 24 logarithmic evenly spaced bins and the Fermi Science Tools \footnote{http://fermi.gsfc.nasa.gov/ssc/data/analysis/} v10r0p5 was used.

In the fit we employed all 3FGL~\citep{3FGL} $\gamma$-ray point-sources present in each ROI plus the standard diffuse Galactic background (DGB) \texttt{gll$_{-}$iem$_{-}$v06.fits} and the isotropic extra-galactic background model \texttt{iso$_{-}$P8R2$_{-}$ULTRACLEANVETO$_{-}$V6$_{-}$v06.txt}. In addition, we included specialized templates for the \textit{Sun} and \textit{Moon} in our fits. These account for diffuse $\gamma$-ray photons resulting from CR interactions with the solar radiation field, solar atmosphere,  and the lunar lithosphere. The \textit{Sun} and \textit{Moon} templates were constructed by making use of the tool \textit{gtsuntemp} as explained in the Cicerone~\footnote{
\protect\url{http://fermi.gsfc.nasa.gov/ssc/data/analysis/scitools/solar_template.html}}.   

The normalization and spectral index of all the $\gamma$-ray point sources that fell within $10^{\circ}$ of the center of each patch were left free in the fit while all the spectral parameters of sources which were within $5^{\circ}$ of the ROI perimeter were fixed to their catalog values~\citep{3FGL}. We also varied the normalizations corresponding to the isotropic and DGB components but kept the \textit{Sun} and \textit{Moon} fluxes fixed to their nominal values. Convergence was reached iteratively by following a relaxation method that has proven to be successful in a region with a high density of $\gamma$-ray point sources and large parameter degeneracies~\citep{Macias:2013vya,Abazajian:2014fta}. Lastly, the extragalactic diffuse photons were obtained by subtracting the best-fit DGB and point sources from the raw counts maps. A mosaic with the resulting IGRB residual images is shown in Fig.~\ref{fig:IGRBresiduals}. 
Note that the residual obtained in this way in principle contains also isotropic detector backgrounds. The normalization of the isotropic 
extragalactic background template is typically about 0.6--0.7. However, we are able to reproduce well the IGRB derived by the Fermi collaboration \cite{Ackermann:2014usa} as shown in Figure 5. From this we conclude that detector backgrounds are safely removed by our conservative photon selection and zenith angle cuts. 
As described in Section \ref{subsec:sys}, 
we examine multiple templates for the Galactic diffuse emission model, whose uncertainties can dominate analyses of residuals.
Nevertheless, we find that the statistical uncertainty would dominate 
the systematic uncertainties in the diffuse modeling in the current survey.
The details are found in Section \ref{subsec:sys}.


\subsection{\label{subsec:lens}CFHTLenS and RCSLenS}

We use the cosmic shear data from 
the Canada-France-Hawaii Telescope Lensing Survey (CFHTLenS) \cite{Heymans:2012gg}
and the Red Cluster Sequence Lensing Survey (RCSLenS)
\cite{Hildebrandt:2016wyi}.
Table~\ref{tab:lens_survey} summarizes the number of source galaxies
and the effective survey area for each subregion used in the present paper.

\begin{table}[!t]
\begin{center}
\begin{tabular}{|c|c|r|r|}
\tableline
Survey & Name of patch & $N_{\rm gal}$ & Effective area (${\rm deg}^2$)  \\ \tableline
& W1 & 2,570,270  & 63.6 \\
& W3 & 1,649,718  & 41.5 \\ 
& W4 & 770,356     & 21.2 \\
\tableline
CFHTLenS & & 4,990,344 & 126.3 \\
\tableline
& CDE0047 & 954,554  & 55.2 \\
& CDE0133 & 430,775  & 27.8 \\ 
& CDE0310 & 101,851  & 68.7 \\
& CDE0357 & 455,616  & 27.7 \\
& CDE1040 & 429,987  & 27.6 \\ 
& CDE1111 & 1,075,176 & 67.7 \\
& CDE1303 & 229,533  & 13.4 \\
& CDE1613 & 381,365  & 24.9 \\ 
& CDE1645 & 404,837  & 24.0 \\
& CDE2143 & 1,178,282 & 71.1 \\
& CDE2329 & 691,348   & 38.9 \\ 
& CDE2338 & 1,070,288 & 64.9 \\
& CSP0320 & 436,054   & 22.8 \\
\tableline
RCSLenS & & 7,839,666 & 534.7 \\
\tableline
TOTAL & & 12,830,010 & 661.0 \\
\tableline
\end{tabular} 
\caption{
\label{tab:lens_survey}
Summary of lensing data used in this paper.
$N_{\rm gal}$ represents the number 
of source galaxies used in our cross-correlation measurement,
and the last column shows the effective survey area evaluated in 
Ref.~\cite{Harnois-Deraps:2016huu}.
Note that the W2 field in CFHTLenS
and CDE1514 field in RCSLenS are discarded 
due to a conservative mask on the $\gamma$-ray data 
of $|b|<30$ deg about the Galactic plane. 
}
\end{center}
\end{table}

\subsubsection*{CFHTLenS}
CFHTLenS is a 154 square degree multi-color 
optical survey in five optical bands $u^{*}, g^{\prime}, r^{\prime}, i^{\prime}, z^{\prime}$. 
CFHTLenS is optimized for weak lensing analysis 
with a full multi-color depth of $i^{\prime}_{AB} = 24.7$ 
with optimal sub-arcsec seeing conditions. 
The survey consists of four separated regions called W1, W2, W3, and W4, with an area of $\sim$ 72, 30, 50, and 25 square degrees, 
respectively.

In CFHTLenS, catalog construction mainly consists of the following three processes:  
photometric redshift measurement \citep{Hildebrandt:2011hb}, weak 
lensing data processing with THELI \citep{Erben:2012zw}, and shear measurement with $lens$fit \citep{Miller:2012am}.
A detailed systematic error study of the shear 
measurements in combination with the 
photometric redshifts is presented in Ref.~\citep{Heymans:2012gg}
and additional error analyses of the photometric 
redshift measurements are presented in Ref.~\citep{Benjamin:2012qp}.

The ellipticities of the source galaxies in the data 
have been calculated using the $lens$fit algorithm. 
$lens$fit performs 
a Bayesian model fitting to the imaging data by varying a galaxy's 
ellipticity and size, and by marginalizing over the centroid position. 
It adopts a 
forward convolution process that convolves 
the galaxy model with the PSF 
to estimate the posterior probability of the model given the data. 
For each galaxy, the ellipticity $\epsilon$ 
is estimated as the mean likelihood of the model posterior probability 
after marginalizing over galaxy size, centroid position, and bulge fraction. 
An inverse variance weight $w$ is 
given by the variance of the ellipticity likelihood surface 
and the variance of the ellipticity distribution of the galaxy population 
(see Ref.~\citep{Miller:2012am} for further details).

The photometric redshifts $z_{p}$ are 
estimated by the {\tt BPZ} code \cite{Benitez:1998br}. 
The true redshift distribution of sources is 
well described by the sum of the probability distribution functions 
(PDFs) estimated from {\tt BPZ} \citep{Benjamin:2012qp}. 
The galaxy-galaxy-lensing redshift scaling analysis 
confirms that contamination is not significant for galaxies selected at $0.2 < z_{p} < 1.3$ \citep{Heymans:2012gg}. 
In this redshift range, the weighted median redshift is 
$\sim0.7$ and the effective weighted 
number density $n_{\rm eff}$ is 11 per square arcmins. 
We have used the source galaxies with $0.2 < z_{p} < 1.3$ 
to measure the cross-correlation of cosmic shear and IGRB. 

To make a reliable lensing measurement, 
we use the following criteria of source galaxies 
in addition to the selection of redshift. 
First, we discard galaxies that have the flag ${\tt MASK} > 1$,
indicating masked objects. 
We use the galaxies that have ellipticity weight $w >0$, 
and the ellipticity fitting flag
${\tt fitclass}=0$, which indicates that the shape is reliably estimated.

\subsubsection*{RCSLenS}

RCSLenS fully utilizes multiband imaging data taken
by the Red-sequence Cluster Survey 2 (RCS2) \cite{Gilbank:2010zv}.
RCS2 is a 785 square degree multi-color 
imaging survey with sub-arcsecond seeing in four bands 
to a depth of $\sim24.3$ mag in the $r$-band
and originally designed to optically select
a very large sample of galaxy clusters over a wide redshift range.
RCSLenS reanalyzes the imaging data in RCS2 with a dedicated weak lensing pipeline examined in the CFHTLenS survey.
RCSLenS consists of 14 separate regions with varying survey area 
between $\sim36-100$ square degrees. 

The lensing data in RCSLenS is obtained by reanalysis of RCS2 imaging 
data with THELI and $lens$fit, as in CFHTLenS.
The effective source number density in RCSLenS is 
$\sim5.5$ per square arcmins.
The photometric redshift estimates by {\tt BPZ}
are only derived for pointings with four-color information,
corresponding to about two-thirds of the survey area.
Ref.~\cite{Harnois-Deraps:2016huu} has estimated
the underlying source redshift distribution
by using the CFHTLenS sample 
incorporated with near-IR and GALEX near-UV data. 
This new photometric  sample, called CFHTLenS-VIPERS, 
is calibrated against 60,000 spectroscopic redshifts 
\cite{Coupon:2015rua}.
The source redshift distribution in RCSLenS then has been
constructed with stacking of the photometric redshift PDF 
over the CFHTLenS-VIPERS galaxies 
that pass the same RCSLenS selection criteria.
The resulting redshift distribution can be fitted 
by the sum of exponential functions with nine free parameters
and the median redshift is found to be $\sim0.5$.
The detailed function form and best-fitted parameters 
are summarized in Ref.~\cite{Harnois-Deraps:2016huu}.

For the RCSLenS, we perform the same selection of source galaxies
as in the case of CFHTLenS; removal with ${\tt MASK} > 1$ and selection with 
$w >0$ and ${\tt fitclass}=0$.
In addition, we adopt the selection in $r$-band magnitude 
of $18 < r < 24$, as in Ref.~\cite{Harnois-Deraps:2016huu}.

\section{\label{sec:cross}CROSS CORRELATION}
\subsection{Measurement}\label{subsec:measure}
In order to detect the cross-correlation signal of IGRB and cosmic shear, 
we introduce the following estimator \cite{Shirasaki:2014noa}, 
\beqa
\xi_{\rm IGRB-WL}(\theta)
=\frac{\sum_{ij}I^{\rm obs}_{\gamma, i}({\bd \phi}_{i}) w_{j}\epsilon_{t, j}({\bd \phi}_{i}+{\bd \theta}_{j})}{(1+K(\theta))\sum_{ij}w_{j}},
\label{eq:CCest}
\eeqa
where $I^{\rm obs}_{\gamma, i}({\bd \phi}_{i})$ is the observed IGRB intensity in pixel $i$ of the $\gamma$-ray map, 
$w_{j}$ is the weight related to the shape measurement, 
and $\epsilon_{t, j}({\bd \phi}_{i}+{\bd \theta}_{j})$ is the tangential component of the $j$-th galaxy's ellipticity, defined by
\beqa
\epsilon_{t}({\bd \phi}_{i}+{\bd \theta}_{j}) 
&=& 
-\epsilon_{1}({\bd \phi}_{i}+{\bd \theta}_{j})\cos(2\alpha_{ij})
\nonumber \\ 
&&
\,\,\,\,\,\,\,\,\,\,\,\,\,\,\,
-\epsilon_{2}({\bd \phi}_{i}+{\bd \theta}_{j})\sin(2\alpha_{ij}),
\eeqa
where $\alpha_{ij}$ is defined as the angle measured from the right ascension direction to a line 
connecting the $i$-th pixel and the $j$-th galaxy. 
The overall factor $1+K(\theta)$ in Eq.~(\ref{eq:CCest})
is used to correct for the multiplicative shear bias $m$ in 
the shape measurement with $lens$fit \citep{Miller:2012am}. 
It is given by 
\beqa
1+K(\theta)=\frac{\sum_{ij}w_{j}(1+m_{j})}{\sum_{ij}w_{j}}.
\eeqa
For binning in angular separation $\theta$, 
we employ a logarithmically equally spaced binning 
in the range of 10 -- 300 arcmin.
The number of bins is set to be 10.

In order to quantify the systematics in the shape measurement,
we consider the cross-correlation 
using another component of weak lensing 
shear that is rotated $45^\circ$ from the tangential shear component.
We denote this shear component $\gamma_{\times}$.
This is commonly used as an indicator of systematics 
in the shape measurement, 
since the correlation signal with $\gamma_{\times}$ 
should vanish statistically
in the case of $\it perfect$ shape measurement 
and no intrinsic alignment.
In addition, Eq.~(\ref{eq:CCest}) should be dependent on the
details of the subtraction of Galactic $\gamma$ rays from the observed 
photon count maps.
We take into account this systematic uncertainty 
by using different models of the Galactic $\gamma$-ray foreground emission.
The details of the analyses are summarized in Section~\ref{subsec:sys}.

\subsection{Statistical uncertainties}\label{subsec:stat}

\begin{figure*}
\begin{center}
       \includegraphics[clip, width=0.8\columnwidth, bb=0 0 438 417]
       {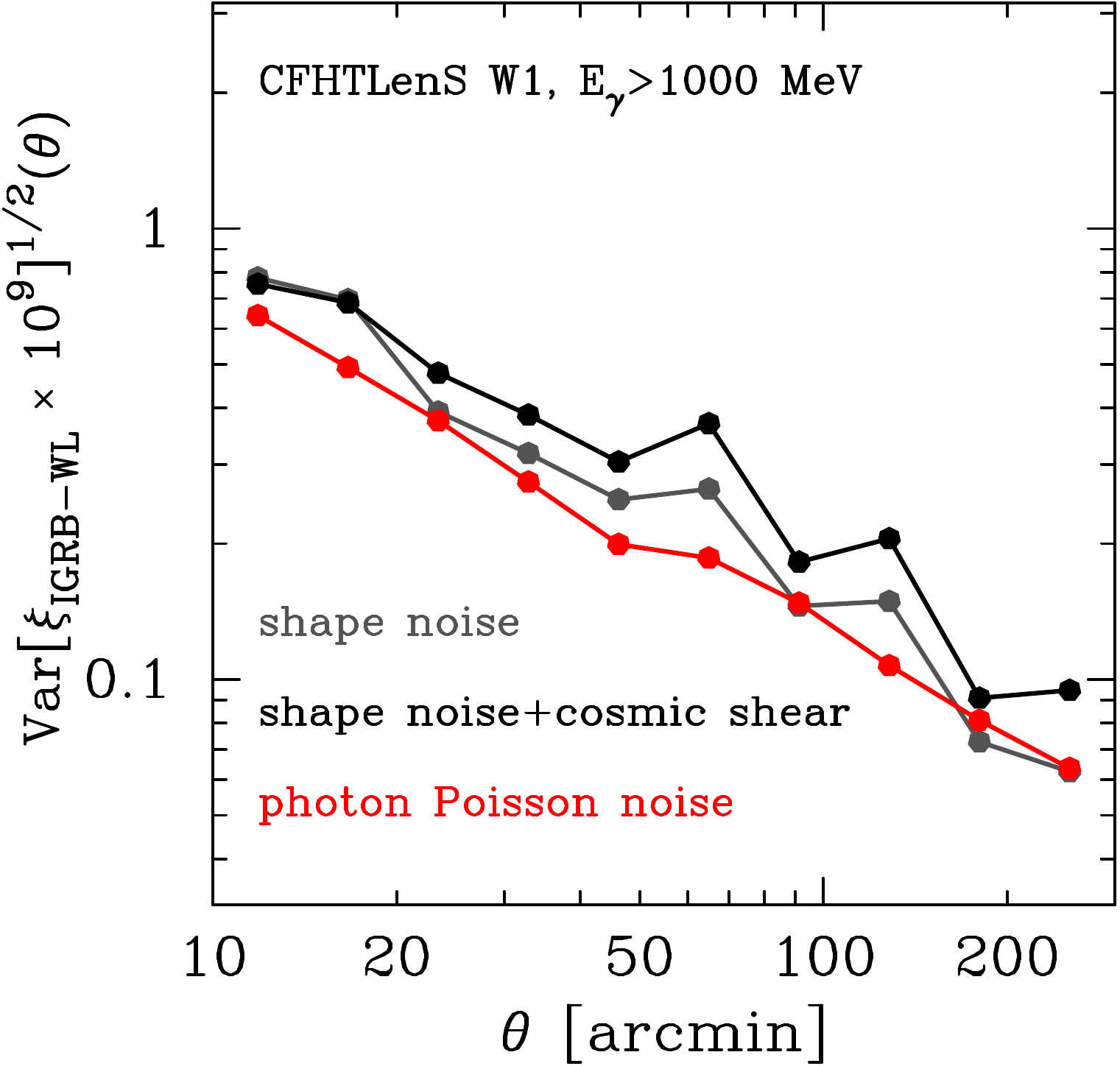}
       \includegraphics[clip, width=0.8\columnwidth, bb=0 0 438 417]
       {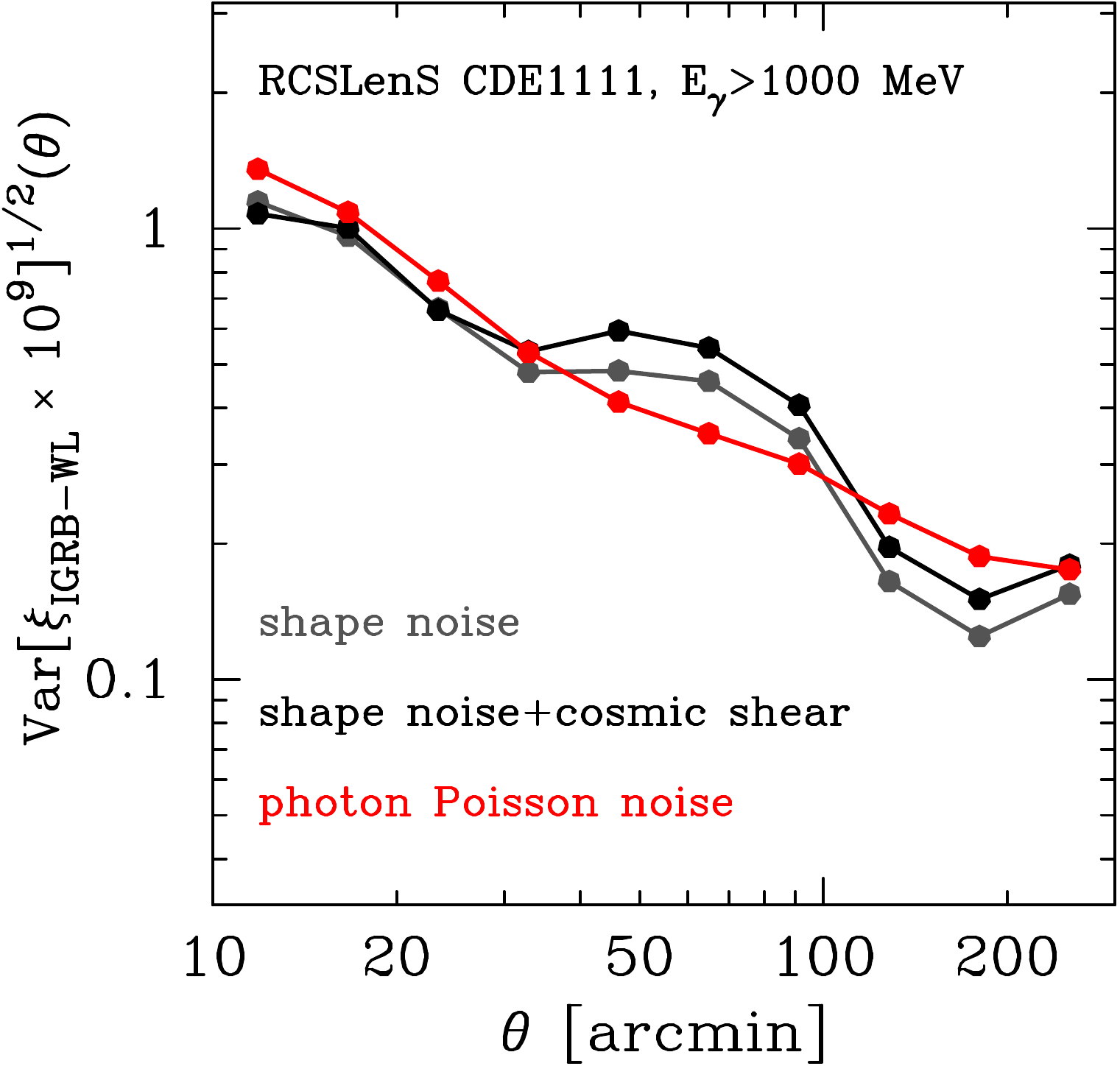}
     \caption{
     \label{fig:stat_error_exam}
     Variance of the cross-correlation 
     signals estimated from a set of 
     randomized realizations, mock catalogs and observed maps.
     The red line shows the statistical error 
     associated with the Poisson error due to the finite 
     number of $\gamma$-ray photon counts.
     The gray line shows the statistical error associated with 
     the intrinsic shape of source galaxies.
     The black line corresponds to a more realistic uncertainty
     incorporating the intrinsic shape and the sample variance
     of weak lensing.
  } 
    \end{center}
\end{figure*}

To discuss the statistical significance of 
the measured estimator from real data, 
we need to estimate the covariance of Eq.~(\ref{eq:CCest}).
According to Ref.~\cite{Shirasaki:2014noa},
the covariance in the case of $\langle\xi_{\rm IGRB-WL}\rangle =0$
can be expressed as 
\beqa\label{eq:CCcov}
&&{\rm Cov}\left[\xi_{\rm IGRB-WL}(\theta_{i}), \xi_{\rm IGRB-WL}(\theta_{j})\right]
\nonumber \\
&&
\,\,\,\,\,\,\,\,\,\,\,\,\,\,\,
\,\,\,\,\,\,\,\,\,\,\,\,\,\,\,
=C_{\rm SN+p}(\theta_i, \theta_j)
+ C_{\rm WL+p}(\theta_i, \theta_j) 
\nonumber \\
&&
\,\,\,\,\,\,\,\,\,\,\,\,\,\,\,
\,\,\,\,\,\,\,\,\,\,\,\,\,\,\,
+ C_{\rm SN+obs}(\theta_i, \theta_j)
+ C_{\rm WL+obs}(\theta_i, \theta_j),
\eeqa
where $C_{\rm SN+p}$ represents the covariance associated 
with the intrinsic shape of source galaxies and
the Poisson noise of photon counts,
$C_{\rm WL+p}$ is the contribution of 
the sample variance of cosmic shear and photon noise,
$C_{\rm SN+obs}$ stands for the covariance originated from 
the intrinsic shape and the angular correlation of observed photon counts,
and 
$C_{\rm WL+obs}$ is the covariance 
caused by the sample variance of cosmic shear 
and the angular correlation of observed photon counts.

Since the cosmic shear is $\sim10^{-2}$ time smaller 
than the intrinsic shape of sources (called shape noise) 
in the current ground-based galaxy imaging survey,
the two terms $C_{\rm SN+p}$ and $C_{\rm SN+obs}$
would dominate Eq.~(\ref{eq:CCcov}) over a wide range of 
angular separations.
Nevertheless, the sample variance of cosmic shear 
can be important at angular scale of $O(1)$ degrees
where the coherent distortion of galaxy shape caused by cosmic shear
would dominate the shape noise.
In this paper, we estimate each term in Eq.~(\ref{eq:CCcov})
by using three kinds of Monte Carlo realizations 
of the observed data set.
For a given random catalog, we estimate the covariance matrix 
of the estimator Eq.~(\ref{eq:CCest}) by,
\beqa
C_{ij} &=& \frac{1}{N_{\rm re}-1} \sum_{r}
\left(\xi^{r}_{\rm IGRB-WL}(\theta_{i})-\bar{\xi}(\theta_{i})\right)
\nonumber \\
&&
\,\,\,\,\,\,\,\,\,\,\,\,\,\,\,
\,\,\,\,\,\,\,\,\,\,\,\,\,\,\,
\times
\left(\xi^{r}_{\rm IGRB-WL}(\theta_{j})-\bar{\xi}(\theta_{j})\right), 
\label{eq:cov} \\
\bar{\xi}(\theta_{i}) &=& \frac{1}{N_{\rm re}}\sum_{r} \xi^{r}_{\rm IGRB-WL}(\theta_{i}),
\eeqa
where $\xi^{r}_{\rm IGRB-WL}(\theta_{i})$ is the estimator of the $i$-th 
angular bin obtained from the $r$-th realization,
$N_{\rm re}$ is the number of randomized catalogs.

The sum of $C_{\rm SN+p}$ and $C_{\rm WL+p}$
in Eq.~(\ref{eq:CCcov}) can 
be estimated from the observed galaxy catalogue
and randomized count maps based on a Poisson distribution.
To simulate the photon count noise, 
we generate 500 randomized count maps in the same way 
as Ref.~\cite{Shirasaki:2014noa}.
We repeat the cross-correlation analysis with the 500 randomized
count maps and the observed galaxy shear catalogue.
We then estimate the statistical error related to the photon noise 
in the same manner as shown in Eq.~(\ref{eq:cov}).

We can estimate $C_{\rm SN+obs}$ in Eq.~(\ref{eq:CCcov})
by using the observed $\gamma$-ray intensity map
and randomized shear catalogs.
In order to simulate the intrinsic shape noise, 
we generate 500 randomized shear catalogs
by rotating the direction of each galaxy ellipticity 
but with fixed amplitude as in Ref.~\cite{Shirasaki:2014noa}.
We then apply Eq.~(\ref{eq:cov}) to a set of 
randomized shear catalogs
and the observed $\gamma$-ray map.

The remaining term in Eq.~(\ref{eq:CCcov}) is $C_{\rm WL+obs}$, which
can be evaluated by cross-correlation measurements with 
the observed $\gamma$-ray map
and a set of realistic mock shape catalogs.
The mock shape catalogs of interest should 
contain appropriate information of cosmic shear 
and several observational factors such as an inverse weight $w$ and survey geometry.
To generate a set of mock shape catalogs,
we follow the approach shown in Ref.~\cite{Shirasaki:2013zpa}
and utilize numerical simulations of cosmic shear based 
on multiple N-body simulations \cite{Shirasaki:2015dga}.
The simulation in Ref.~\cite{Shirasaki:2015dga} is designed 
to simulate the weak lensing effect 
over a full sky with angular resolution of $\sim1$ arcmin.
We store the full sky weak lensing simulation data 
for 30 different source redshifts with width of 
$\Delta \chi=150\, h^{-1}{\rm Mpc}$.
The angular and redshift resolution of our simulations
are suitable to create mock catalogs for the current galaxy imaging surveys, RCSLenS and CFHTLenS.
We incorporate our simulations with observed photometric redshift
and angular position of $real$ galaxies.
We simply assume that individual galaxies in CFHTLenS
and RCSLenS would follow the estimated photometric-redshift 
distribution as in Refs.~\cite{Kilbinger:2012qz} and \cite{Harnois-Deraps:2016huu}, respectively.
In the mock CFHTLenS and RCSLenS, mock galaxies can realize 
the same angular distribution on the celestial sphere as the real galaxies,
allowing us to simulate complex observational factors, e.g., 
survey geometry.
From a single full-sky simulation,
we make 50 mock shear catalogs 
by choosing the desired sky coverage as shown in Table~\ref{tab:lens_survey}.
Since we have 10 full-sky lensing simulations, 
we obtain 500 realizations of each region of RCSLenS and CFHTLenS in total.
With these 500 mock catalogs and the observed $\gamma$-ray map,
we repeat the cross-correlation analysis and then estimate the covariance
with Eq.~(\ref{eq:cov}).

Figure~\ref{fig:stat_error_exam} shows representative examples 
of the statistical uncertainty of our cross-correlation measurements.
In both panels, we show the different contributions to the covariance 
of $\xi_{\rm IGRB-WL}$:
the red line shows the covariance associated with the Poisson noise
of photon counts ($C_{\rm SN+p}$ and $C_{\rm WL+p}$);
the gray line represents the covariance from intrinsic shape noise (expressed 
as $C_{\rm SN+obs}$);
and the black line is for the covariance associated with galaxy imaging surveys including the shape noise and the sample variance of cosmic shear
($C_{\rm SN+obs}$+$C_{\rm WL+obs}$).
The left panel corresponds to the variance in the W1 field of CFHTLenS,
while the right is for CDE1111 field of RCSLenS.
Note that both covers a similar survey area, but the CFHTLenS field contains a factor $\sim2.5$ more source galaxies.
Figure~\ref{fig:stat_error_exam} clearly shows
that the sample variance of cosmic shear can be subdominant,
but contributes at the level of $\sim20\%$ to the statistical error on large angular scales of  
$\theta\simgt0.5$ degrees.
In the following, we therefore include all contributions of the covariance 
for completeness.
The total covariance is estimated as the sum 
of the red and black lines in Figure~\ref{fig:stat_error_exam}.

\subsection{Systematic uncertainties}\label{subsec:sys}

Diffuse Galactic $\gamma$ rays are produced through the interactions of high-energy CR particles with interstellar gas via nucleon-nucleon inelastic collisions and electron Bremsstrahlung, and with the interstellar radiation field, via IC scattering. Such diffuse Galactic emission  accounts for  more  than  60\%  of  the photons detected by the Fermi-LAT. Modeling this component can thus introduce a large impact on the properties of the IGRB. 

\begin{figure}[t!]
\begin{center}

\includegraphics[width=0.9\linewidth, bb=0 0 762 730] 
{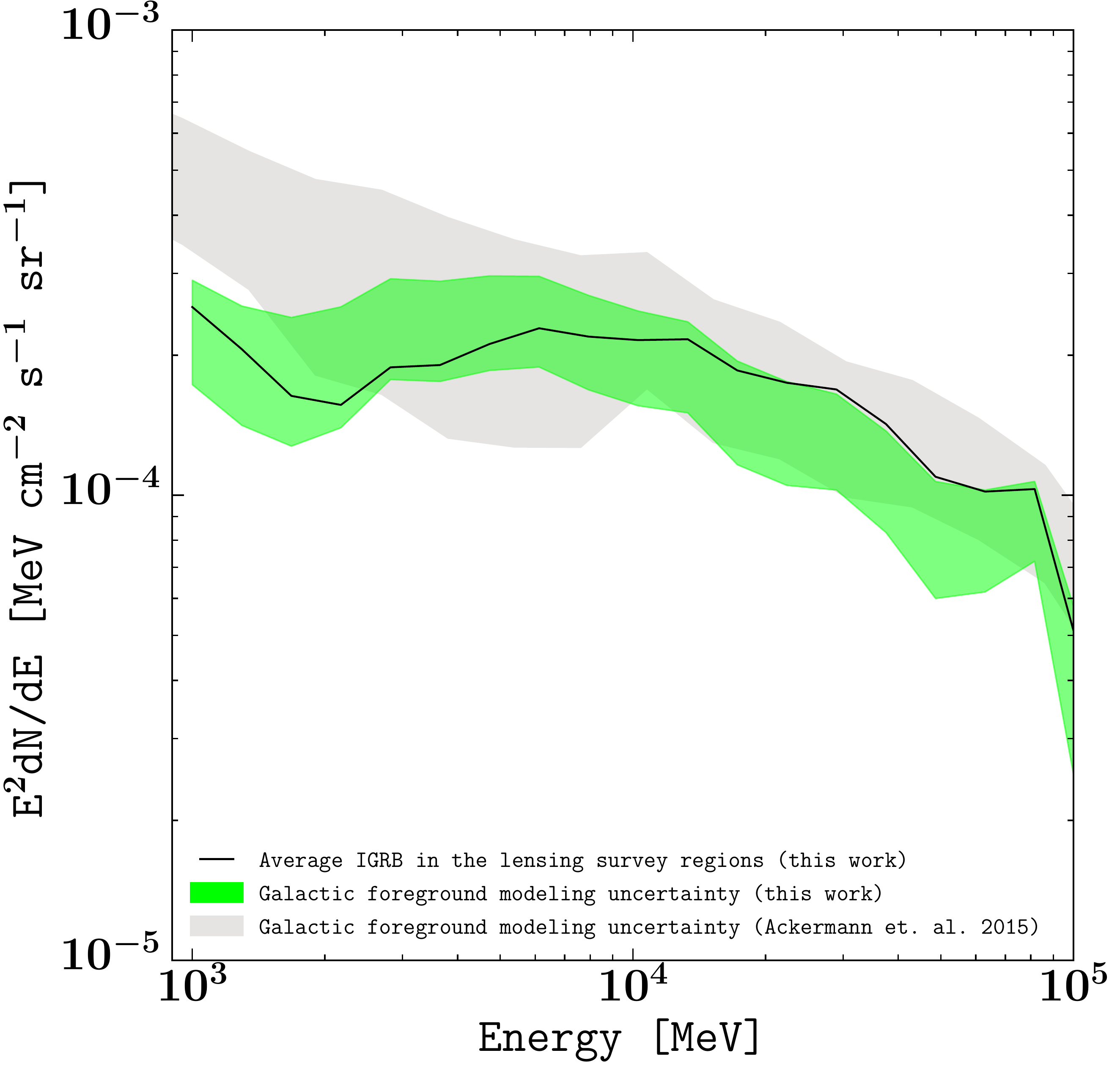}

\caption{ \label{fig:IGRBspectra}
Average spectrum of the IGRB (black solid line) obtained from a maximum-likelihood fit to the 16 different patches with the \textsc{FermiTools} analysis software. The green area represents an estimate of the systematic uncertainties as calculated with 8 different {\tt GALPROP} models of the diffuse galactic background~\citep{ackermannajelloatwood2012}. The grey band is the IGRB systematic uncertainty obtained in an all-sky study by the Fermi team~\citep{Ackermann:2014usa}. }
\end{center}
\end{figure}

We estimated the systematic uncertainties in the DGB in a similar fashion to that in Ref.~\citep{ackermannajelloatwood2012}. 
We constructed eight different diffuse emission models 
using {\tt GALPROP} \citep{Galpropsupplementary}
and each of these templates were included 
in the likelihood fit of the lensing surveys as an alternative 
to the standard DGB recommended in the 3FGL catalog~\citep{3FGL}. 

\begin{figure*}
\begin{center}
       \includegraphics[clip, width=1.5\columnwidth, bb=0 0 526 499]
       {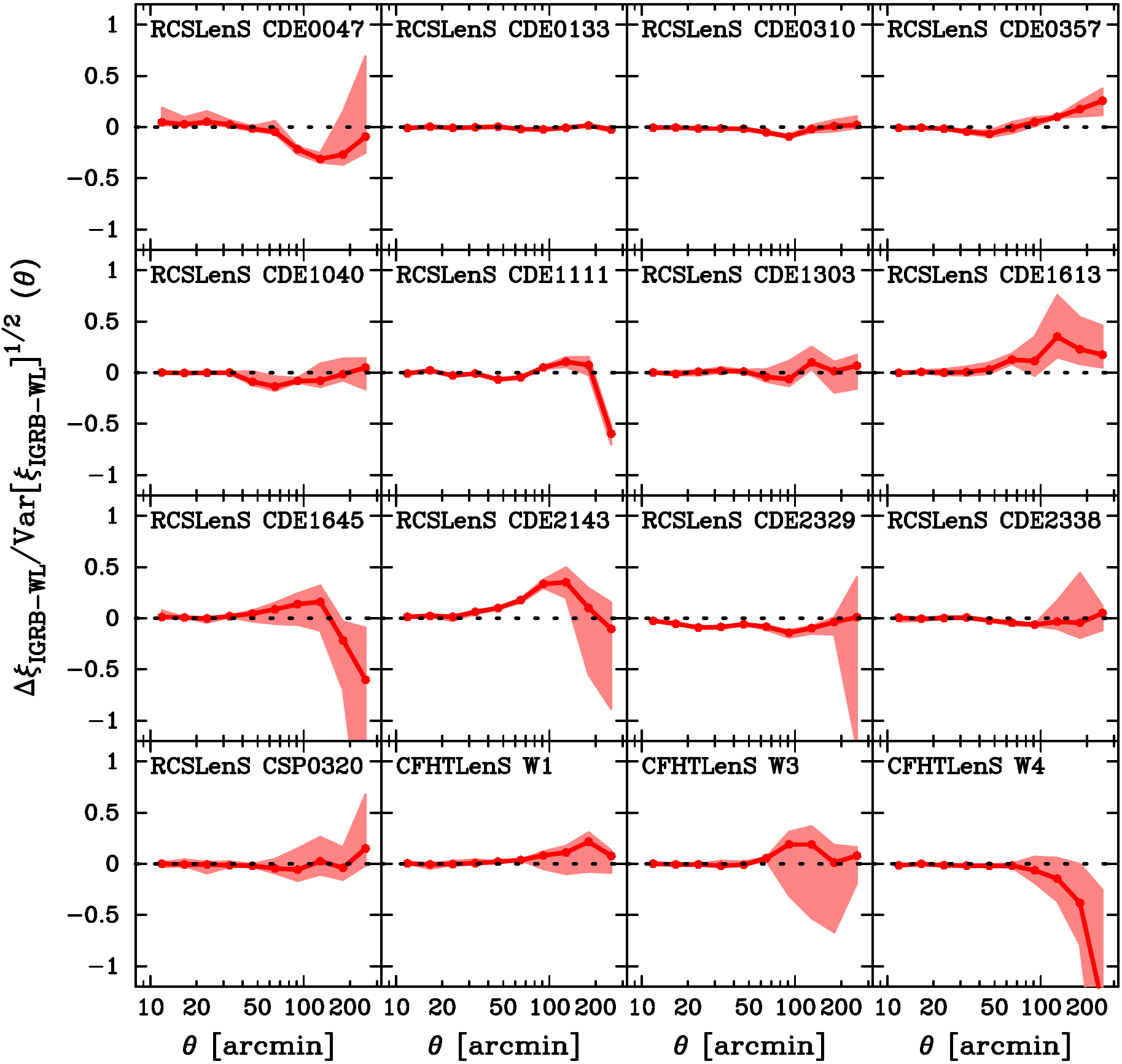}
     \caption{
     \label{fig:sys_error_table}
     Systematic uncertainties in our cross-correlation analysis. 
     Each panel shows the difference in $\xi_{\rm IGRB-WL}$ 
     between the fiducial Galactic $\gamma$-ray template
     and an alternate Galactic $\gamma$-ray template (see text).
     The differences are normalized by the statistical uncertainty.
     The red line shows the average residual over eight galactic tempts,
     while the hatched region represents the range of maximum 
     and minimum residuals for a given $\theta$.
  } 
    \end{center}
\end{figure*}

The set of alternative DGB models taken into account in this analysis consider a range of possible values for the input parameters 
that were found to affect the most the $\gamma$-ray fits 
performed by the Fermi team in Ref.~\citep{ackermannajelloatwood2012}. The parameters varied in the models are 
the CR source distribution (supernova remnants or pulsars), 
CR halo heights (4 kpc or 10 kpc),  and the atomic hydrogen spin temperature ($150$\,K or optically thin). 
In all of the {\tt GALPROP} runs an $E(B-V)$ magnitude 
cut of 5 mag was chosen. 
The impact in the resulting IGRB spectra obtained through this method is displayed as a green shaded area in Fig.~\ref{fig:IGRBspectra}.   

We use these eight model templates 
and repeat the cross-correlation analysis described in 
Sec.~\ref{subsec:measure}.
We therefore obtain eight different binned signals 
($\xi_{\rm IGRB-WL}$) as functions of angular separation
for each ROI.
We introduce the following quantity for convenience,
\beqa
\Delta \xi_{\rm IGRB-WL} &=& 
\xi_{\rm IGRB-WL} ({\tt GALPROP})
\nonumber \\
&&
\,\,\,\,\,\,\,\,\,\,\,\,\,\,\,
\,\,\,\,\,\,\,\,\,\,\,\,\,\,\,
-\xi_{\rm IGRB-WL} ({\rm Fiducial}),
\eeqa
where $\xi_{\rm IGRB-WL} ({\rm Fiducial})$ is the cross-correlation
signal measured with the IGRB intensity constructed by the fiducial
Galactic templete,
and
$\xi_{\rm IGRB-WL} ({\tt GALPROP})$ is for the IGRB intensity with 
the {\tt GALPROP} Galactic template.

Figure~\ref{fig:sys_error_table} summarizes
the systematic effect associated with the subtraction of Galactic $\gamma$ rays.
In this figure, the red line shows the average $\Delta \xi_{\rm IGRB-WL}$ 
over the eight different {\tt GALPROP} templates,
while the hatched region is for the scatter of $\Delta \xi_{\rm IGRB-WL}$.
Note that the amplitudes are normalized by the statistical uncertainty
evaluated in Sec.~\ref{subsec:stat}.
The figure shows that our measurements
are insensitive to the choice of Galactic $\gamma$-ray template,
and that the typical systematic uncertainty in each ROI
is smaller than the statistical uncertainty by a factor of $\sim0.1-0.2$.
The small effect of the Galactic $\gamma$-ray template uncertainty is
consistent with previous works \cite{Shirasaki:2015nqp},
and derives from the high latitudes of our ROIs; had we chosen 
ROIs closer to the Galactic plane, the model uncertainty would 
likely be significantly larger.  
Nevertheless, we also find that some ROIs show a relatively large difference
among adopted Galactic templates at large angular separations 
(e.g., $\theta \simgt 2 \,{\rm deg}$).
We have confirmed that our results remain robust 
when we limit the angular scale by 
$\Delta \xi_{\rm IGRB-WL}(\theta) < 0.5 \sqrt{{\rm Var}[\xi_{\rm IGRB-WL}(\theta)]}$ 
where ${\rm Var}[\xi_{\rm IGRB-WL}]$ represents the statistical variance of our cross 
correlation measurements.

\section{\label{sec:res}RESULT}
\subsection{Cross-correlation measurement}
Here we present the cross-correlation measurements of the IGRB and cosmic shear
that we use to probe the parameter space of generic DM models.

\begin{figure}[!t]
\begin{center}
       \includegraphics[clip, width=0.8\columnwidth, bb=0 0 498 495]
       {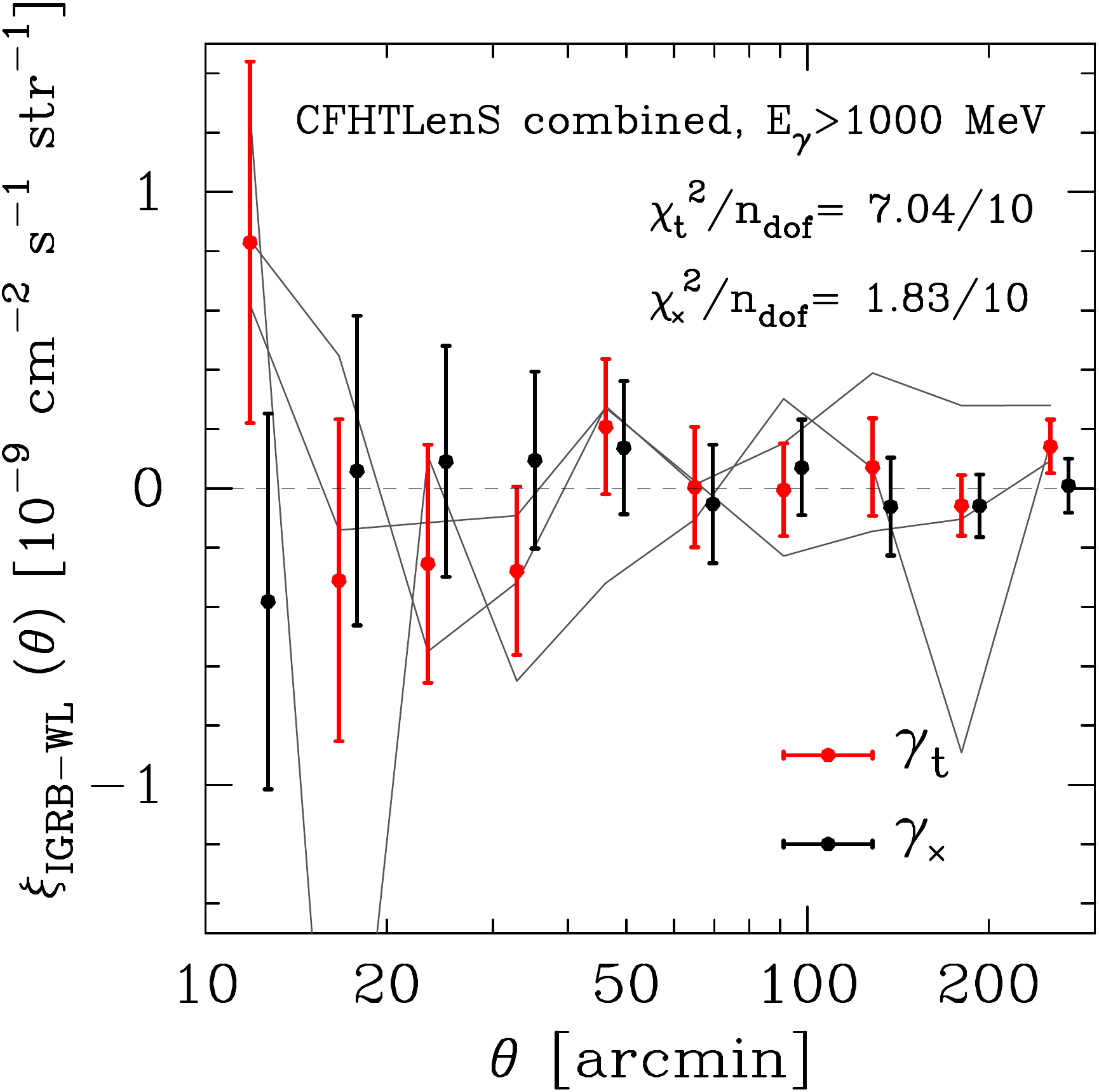}
       \includegraphics[clip, width=0.8\columnwidth, bb=0 0 498 495]
       {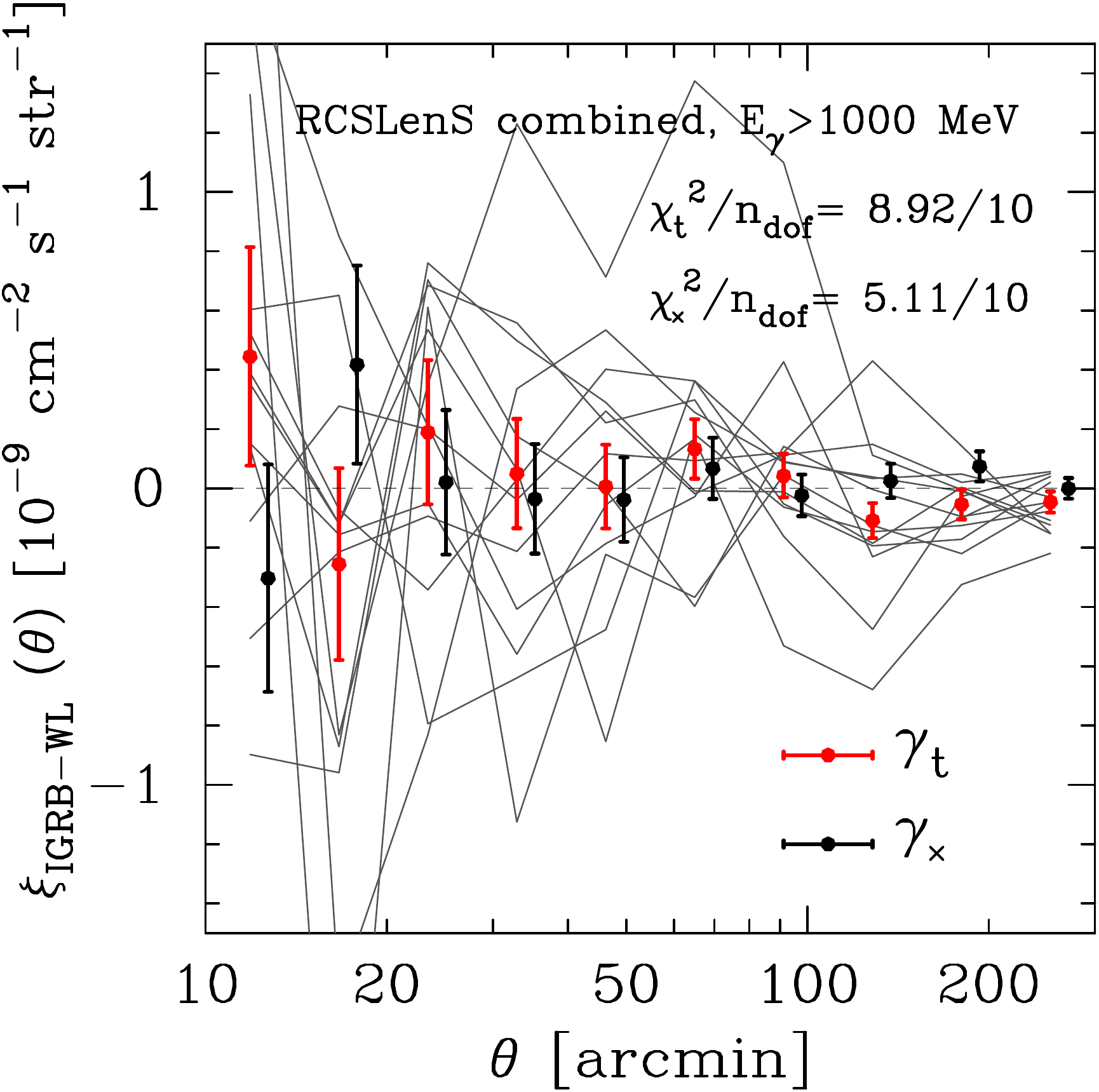}
     \caption{
     \label{fig:xi_measured}
     The cross-correlation signal of the IGRB and cosmic shear. 
     The top panel shows the ROIs within each of the CFHTLenS fields, 
     while the bottom is the equivalent for RCSLenS.
     In both panels, each gray thin line represents
     the cross-correlation signals for a single ROI,
     while the symbols with error bars are the results combining 
     several ROIs.
     The red circles show the results using tangential 
     component of shear, 
     while the black circles are for the cross component 
     of shear as an indicator of systematics in lensing data.
     The error bars are based on the statistical uncertainties
     of 500 randomized shear catalogs, mock shear catalogs,
     and randomized photon count maps.
    } 
    \end{center}
\end{figure}

Figure~\ref{fig:xi_measured} summarizes
the cross-correlation signals obtained for the CFHTLenS and RCSLenS 
fields.
The top panel shows the result obtained for CFHTLenS,
while the bottom is for RCSLenS.
In each panel, the gray thin line shows the cross-correlation signal estimated with Eq.~(\ref{eq:CCest}) for each patch in the lensing survey.
To combine the measured signal with different patches,
we introduce the weighted cross-correlation signal
over the different patches,
\beqa\label{eq:combined_xi}
\xi^{(c)}_{\rm IGRB-WL}(\theta) 
= \sum_{\alpha} {\cal W}_\alpha(\theta) \, 
\xi_{\rm IGRB-WL}(\theta | \alpha),
\eeqa
where $\xi_{\rm IGRB-WL}(\theta | \alpha)$ represents the measured signal for the $\alpha$-th ROI 
and the summation is taken over all ROIs of interest.
We determine the weight ${\cal W}_\alpha$ for a given $\theta$
by minimization of variance of $\xi^{(c)}_{\rm IGRB-WL}(\theta)$.
The minimum variance weight is then given by,
\beqa
{\cal W}_\alpha (\theta) = 
\frac{{\rm Var}^{-1}[\xi_{\rm IGRB-WL}(\theta | \alpha)]}
{\sum_{\beta}{\rm Var}^{-1}[\xi_{\rm IGRB-WL}(\theta | \beta)]},
\eeqa
where 
${\rm Var}[\xi_{\rm IGRB-WL}(\theta | \alpha)]$
represents the variance of $\xi_{\rm IGRB-WL}(\theta | \alpha)$
and 
we assume that the cross covariance between different two patches
can be ignored.

We then evaluate the significance of measured signals 
by using the signal-to-noise ratio for Eq.~(\ref{eq:combined_xi}),
\beqa
\chi^{2} &=& \sum_{i,j} 
\xi^{(c)}_{\rm IGRB-WL}(\theta_{i})
[{\bd C}^{(c)}]^{-1}_{ij}
\xi^{(c)}_{\rm IGRB-WL}(\theta_{j}), \\ \label{eq:chi2_combined}
{\bd C}^{(c)}_{ij} &=& \sum_{\alpha} 
{\cal W}_{\alpha}(\theta_{i})
{\cal W}_{\alpha}(\theta_{j})
{\bd C}_{ij}(\alpha),
\eeqa
where ${\bd C}_{ij}(\alpha)$ represents the covariance matrix
for $\xi_{\rm IGRB-WL}(\theta_{i}|\alpha)$.
The red point with error bar in Figure~\ref{fig:xi_measured}
shows the weighted signal for the tangential shear,
while the black point with error bar is for 
the cross component of shear
$\gamma_{\times}$
as an indicator of lensing systematics.
We find that the cross-correlation signals in both CFHTLenS and RCSLenS 
are consistent with null detection.
The resulting $\chi^2$ per number of degree of freedoms
is $7.04/10$ for CFHTLenS
and $8.92/10$ for RCSLenS.
When combing all fields in CFHTLenS and RCSLenS,
$\chi^2$ in Eq.~(\ref{eq:chi2_combined}) is 8.99.
Also, we confirm that the $\chi^2$ for $\gamma_{\times}$ 
is consistent with null signal, suggesting the lensing systematics in 
our measurement are under control.
We also define the significance for each ROI as
\beqa
\chi_{\alpha}^{2}= \sum_{i,j} 
\xi_{\rm IGRB-WL}(\theta_{i}|\alpha)
{\bd C}^{-1}_{ij}(\alpha)
\xi_{\rm IGRB-WL}(\theta_{j}|\alpha), \label{eq:chi2_each}
\eeqa
where we confirm that the $\chi_{\alpha}^{2}$ is still consistent 
with null detection for a given $\alpha$-th ROI.
Table~\ref{tab:chi2} summarizes the results of our cross-correlation
measurements.

\begin{table}[!t]
\begin{center}
\begin{tabular}{|c|c|r|r|}
\tableline
Survey & Name of patch & $\chi^2$ for $\gamma_{t}$ & $\chi^2$ for $\gamma_{\times}$  \\ \tableline
& W1 & 2.40/10 & 4.13/10 \\
& W3 & 5.00/10 & 2.77/10 \\ 
& W4 & 10.45/10 & 6.46/10 \\
\tableline
CFHTLenS & & 7.04/10 & 1.83/10 \\
\tableline
& CDE0047 & 6.26/10  & 4.27/10 \\
& CDE0133 & 4.91/10  & 4.11/10 \\ 
& CDE0310 & 5.22/10  & 5.10/10 \\
& CDE0357 & 7.21/10  & 8.08/10 \\
& CDE1040 & 9.40/10  & 8.53/10 \\ 
& CDE1111 & 8.67/10  & 11.92/10 \\
& CDE1303 & 1.60/10  & 4.54/10 \\
& CDE1613 & 6.50/10  & 3.26/10 \\ 
& CDE1645 & 1.40/10  & 10.04/10 \\
& CDE2143 & 2.22/10  & 5.61/10 \\
& CDE2329 & 4.40/10  & 8.72/10 \\ 
& CDE2338 & 7.84/10  & 6.40/10 \\
& CSP0320 & 4.03/10  & 6.46/10 \\
\tableline
RCSLenS & & 8.92/10 & 5.11/10 \\
\tableline
TOTAL & & 8.99/10 & 3.74/10 \\
\tableline
\end{tabular} 
\caption{
	\label{tab:chi2}
	Summary of the significance of our cross-correlation measurements.
	When combining with several fields,
	we employ the minimum variance weighting (see the text for detail).
}
\end{center}
\end{table}

\subsection{Constraints on particle dark matter}\label{sec:dmlimits}

In order to use the null detection of the cross-correlation to place 
constraints on particle DM, 
we use a maximum likelihood analysis.
We assume that the data vector ${\bd D}$ is well approximated 
by the multivariate Gaussian distribution with covariance ${\bd C}$.
In this case, the $\chi^2$ statistics (log-likelihood) is given by,
\beqa
\chi^2({\bd p}) = \sum_{i,j}(D_{i}-\mu_{i}({\bd p}))C^{-1}_{ij}(D_{j}-\mu_{j}({\bd p})), \label{eq:logL}
\eeqa
where ${\bd \mu}({\bd p})$ is the theoretical template 
for the set of parameters of interest. The theoretical prediction is computed 
as in Section~\ref{subsec:cross_corr_model}.
The parameters of interest ${\bd p}$ are  
the DM particle mass $m_{\rm dm}$ and the annihilation 
cross section $\langle \sigma v \rangle$ for annihilating DM; as for decaying
DM, we consider $m_{\rm dm}$ and the decay rate $\Gamma_{\rm d}$.

The data vector ${\bd D}$ consists of the ten measured cross-correlation 
amplitudes in the range of $\theta=[10, 300]$ arcmin as,
\beqa
D_{i} = \{ \xi(\theta_{1}), \xi(\theta_{2}),..., \xi(\theta_{10}) \},
\eeqa
where $\theta_{i}$ is the $i$-th angular separation bin.
The inverse covariance matrix ${\bd C}^{-1}$ includes 
the statistical error owing to the intrinsic shape of galaxies,
the sample variance of cosmic shear,
and the photon Poisson error (see Section \ref{subsec:stat}).
In our likelihood analysis, we assume that the 16 patches of
CFHTLenS and RCSLenS are independent of each other.  
With this assumption, the total log-likelihood 
is given by the sum of Eq.~(\ref{eq:logL})
in each patch of CFHTLenS and RCSLenS. 
We consider the 95\% confidence level 
posterior distribution function of parameters.
This is given by the contour line in the two dimensional space 
($m_{\rm dm}$ and $\langle \sigma v \rangle$, 
or $m_{\rm dm}$ and $\Gamma_{\rm d}$), 
defined as 
\beqa
\Delta \chi^2({\bd p}) = \chi^2({\bd p})-\chi^2({\bd \mu}=0)=6.17.
\eeqa

Note that the constraints in the following are based on all the measurements presented in this paper.
Compared to the constraints using CFHTLenS alone \cite{Shirasaki:2014noa}, 
the present analysis combines CFHTLenS and RCSLenS,
resulting in a factor of $\sim5$ improvement.
This derives from several factors, namely, 
(i) better $\gamma$-ray photon data quality and higher statistics,
(ii) the increase in the number of source galaxies,
and 
(iii) the use of large angular separations, i.e., $\theta \simgt 100$ arcmin.
When we limit the angular range to $\theta<100$ arcmin,
the constraints are degraded by a factor of $\sim4$,
even if both CFHTLenS and RCSLenS are used.
This indicates that the 
main cosmological information comes from large angular scales, 
making large sky coverage essential
for maximizing the information content of 
our cross-correlation statistics.
Note that our limit in this paper corresponds to the 95\% confidence level,
while our previous study using CFHTLenS in Ref~\cite{Shirasaki:2014noa}
derived constraints based on the 68\% confidence level.

\subsubsection*{Dark matter annihilation}
\begin{figure*}
\begin{center}
       \includegraphics[clip, width=0.65\columnwidth, bb=0 0 533 500]
       {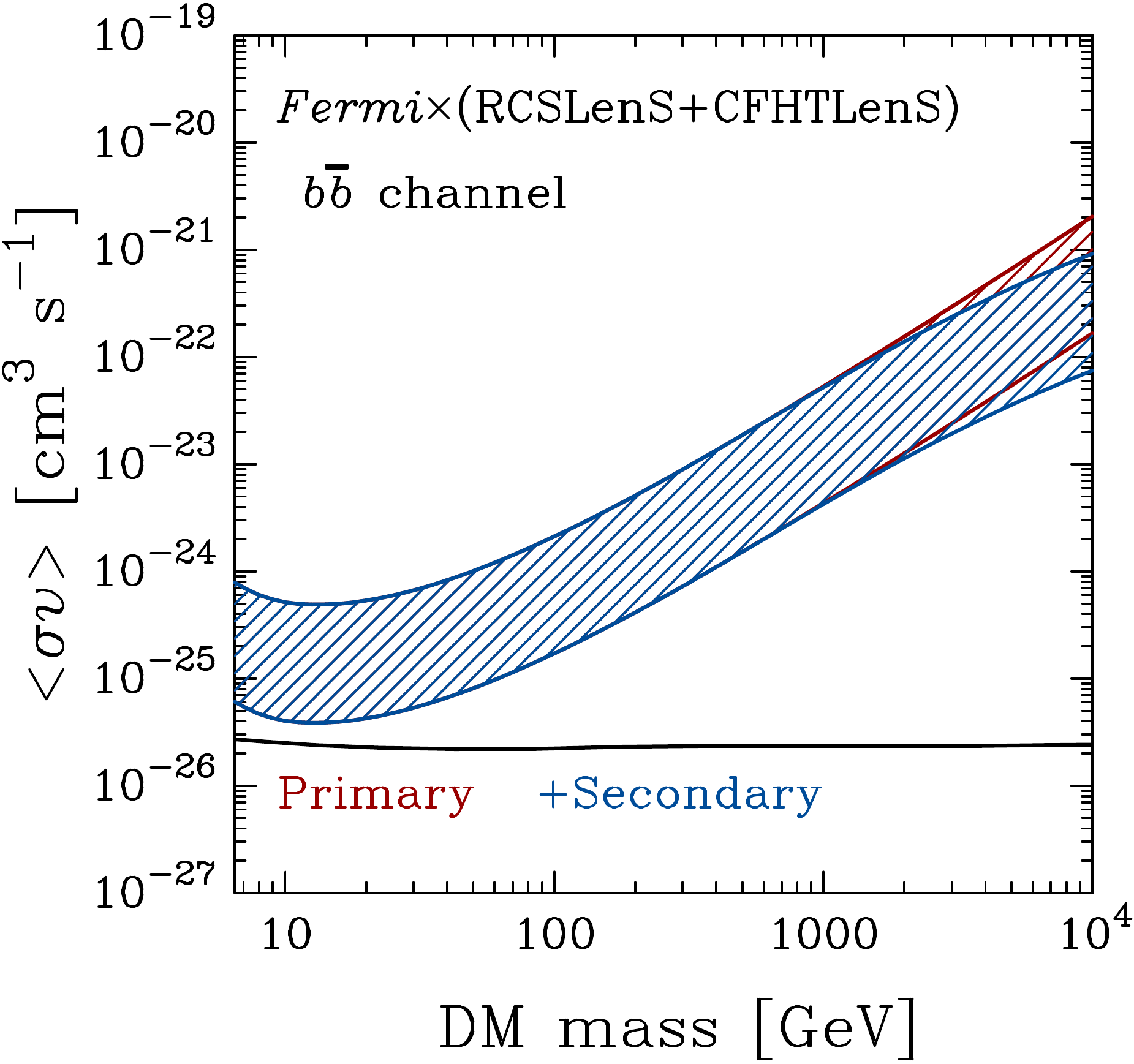}
       \includegraphics[clip, width=0.65\columnwidth, bb=0 0 533 500]
       {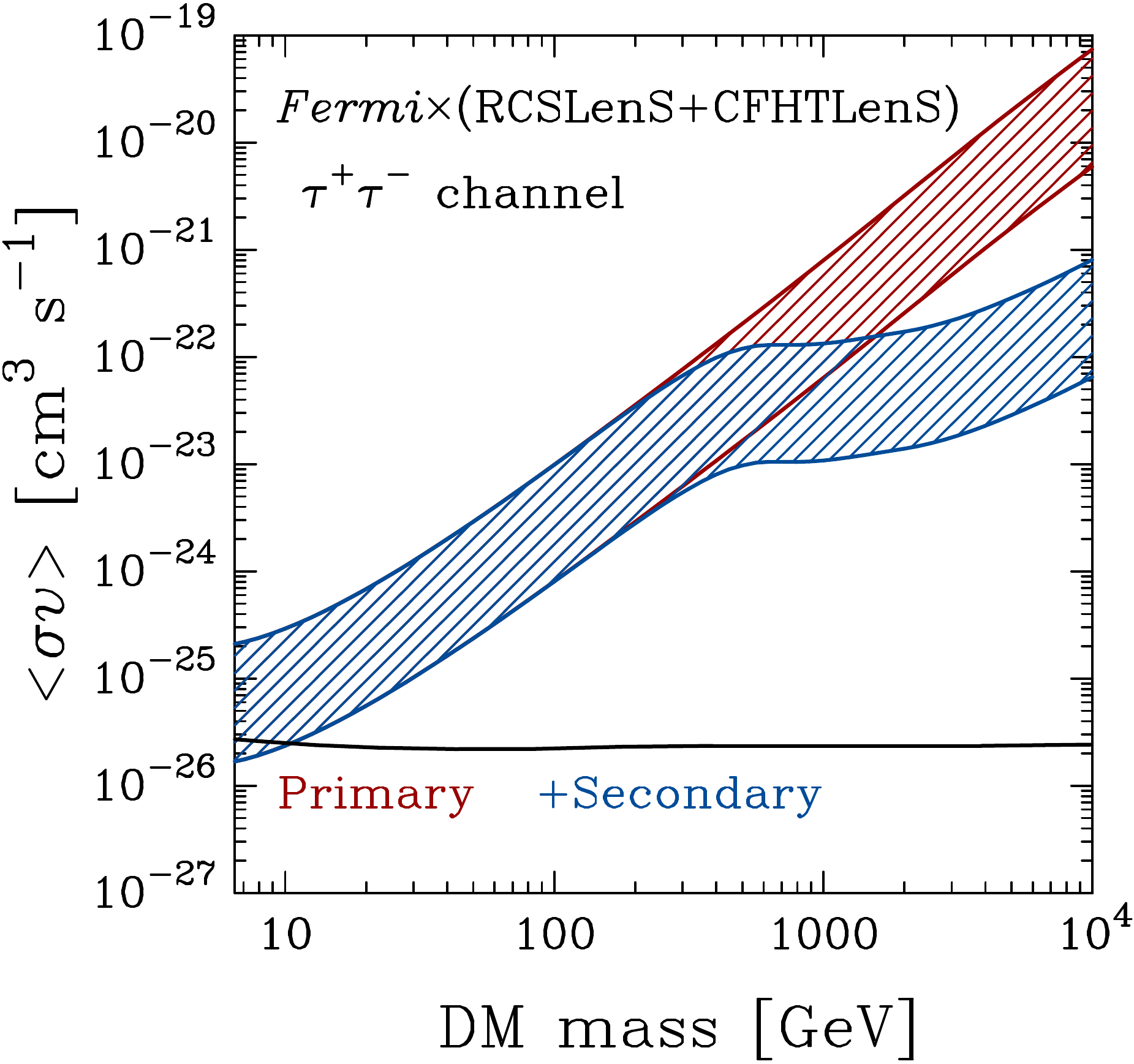}
       \includegraphics[clip, width=0.65\columnwidth, bb=0 0 533 500]
       {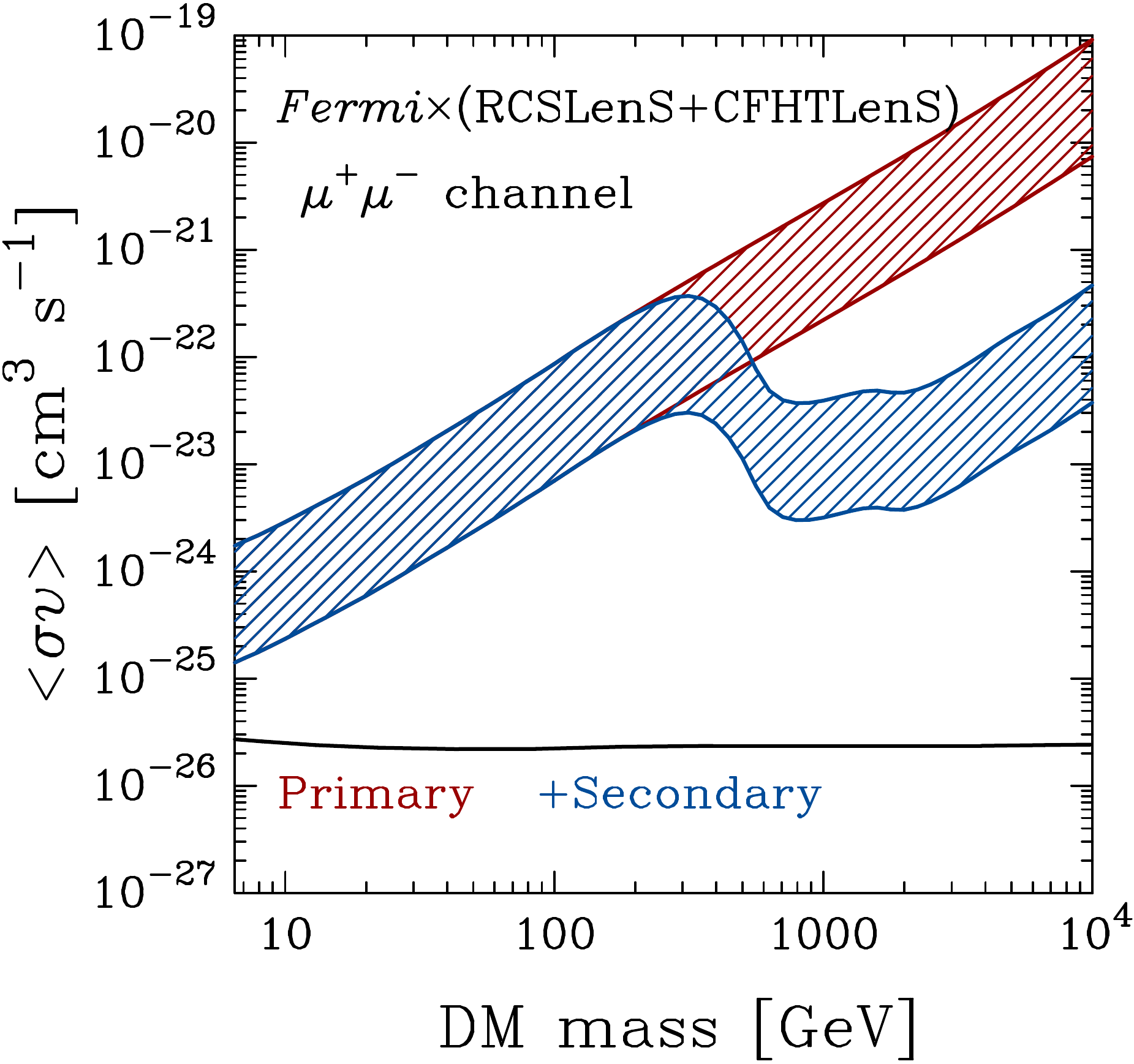}       
     \caption{
     \label{fig:ann_constraints}
     $95\%$ confidence level upper limits 
     on the DM annihilation cross section as a function 
     of DM particle mass.
     The three panels correspond to three different annihilation channels
     ($b\bar{b}$, $\tau^{+}\tau^{-}$, and $\mu^{+}\mu^{-}$).
     In each panel, constraints with (blue lines) and without (red)
     consideration of secondary $\gamma$ rays from inverse-Compton
     upscattering of CMB photons are shown. 
     For each, the shaded band represents 
     a conservative estimate of the theoretical uncertainty 
     due to DM substructure.
     The black line shows the canonical cross section expected for thermal relic DM \cite{2012PhRvD..86b3506S}.
  } 
    \end{center}
\end{figure*}

As discussed in Section \ref{subsec:cross_corr_model}, 
the choice of the boost factor $b_{sh}$
affects the theoretical predictions significantly$-$by a factor of $\sim 10$. 
We therefore derive constraints based on both; an optimistic 
scenario as in Ref.~\cite{Gao:2011rf}, and a conservative 
one following Ref.~\cite{Sanchez-Conde:2013yxa}. 

Figure~\ref{fig:ann_constraints} summarizes the constraints 
on DM annihilation with our cross-correlation measurements.
These were obtained for three 
representative channels: 
$b\bar{b}$, $\tau^{+}\tau^{-}$ and $\mu^{+}\mu^{-}$.
In Figure~\ref{fig:ann_constraints}, the red line shows the constraints
obtained by considering primary $\gamma$-ray emissions alone,
while the blue line includes the secondary emission 
due to IC scattering between secondary
$e^{\pm}$ and the CMB.
As expected, heavier DM particles have a larger
contribution of $\gamma$ rays induced by the IC process.
Thus, the constraints on heavier DM particles can be significantly
improved when the secondary emission is taken into account$-$the typical improvement is of order $\sim$100.
Notice that leptonic channels are further enhanced by
the secondary emission since more energetic 
$e^{\pm}$ tend to be produced.
For very high DM masses $m_{\rm dm}\simgt2-3\, {\rm TeV}$, we find 
that the constraints are degraded by a factor of $m^{-2}_{\rm dm}$ 
because the peak energy 
in ${\rm d}N_{\gamma, a} /{\rm d}E_\gamma$ is found to be outside our 
$\gamma$-ray energy range of 1-500 GeV.

\subsubsection*{Dark matter decay}

\begin{figure*}
\begin{center}
       \includegraphics[clip, width=0.65\columnwidth, bb=0 0 515 485]
       {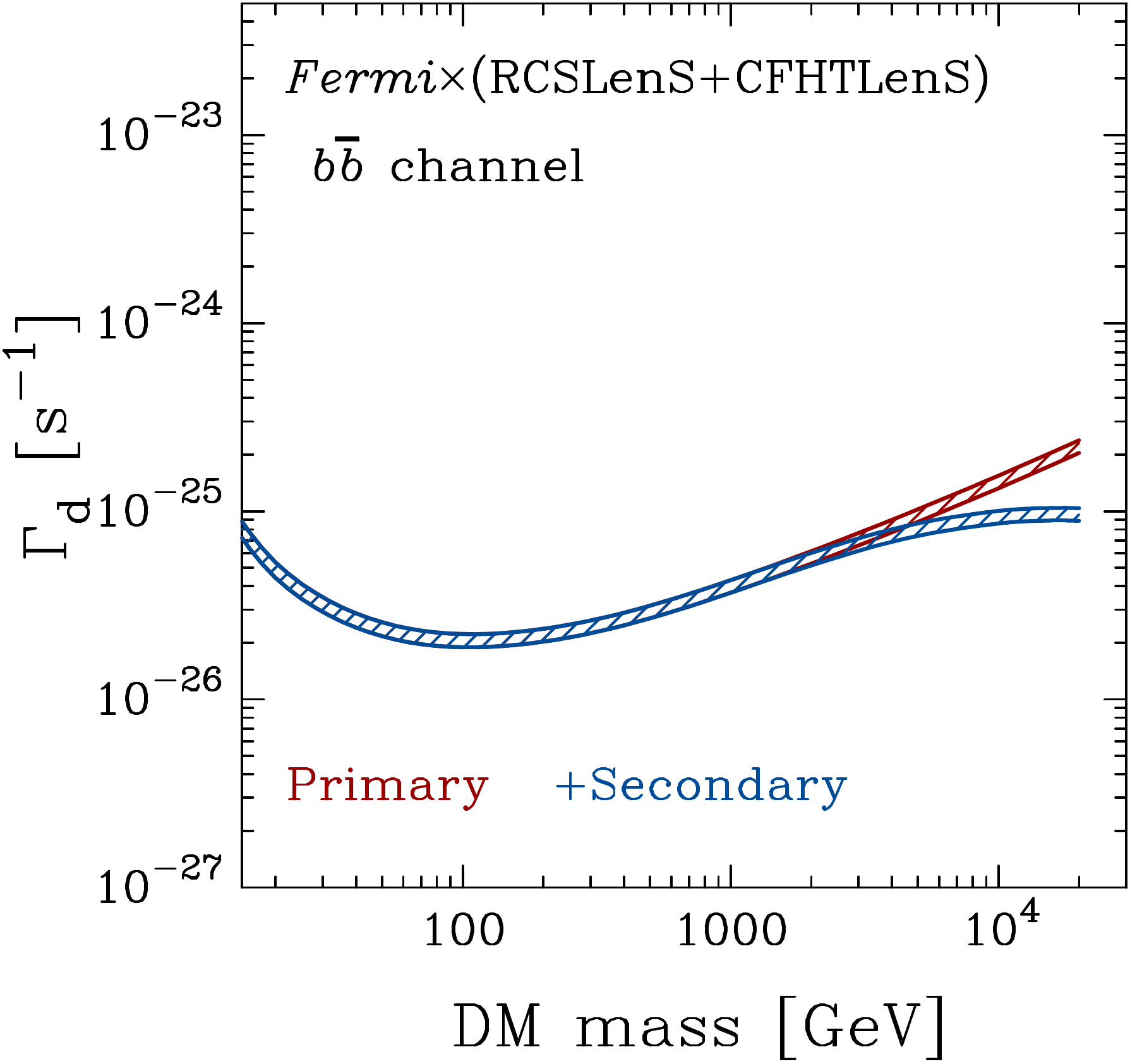}
       \includegraphics[clip, width=0.65\columnwidth, bb=0 0 515 485]
       {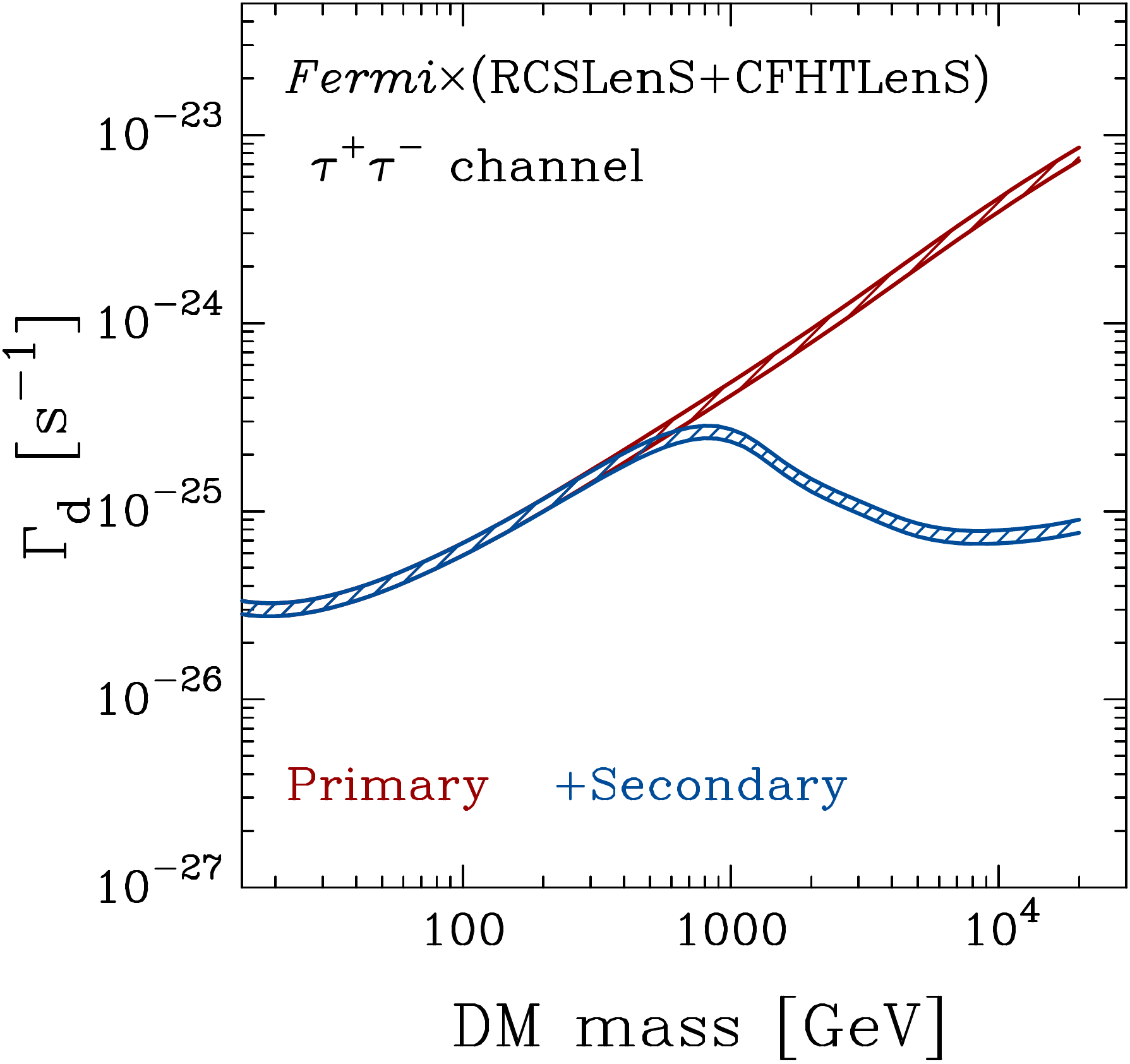}
       \includegraphics[clip, width=0.65\columnwidth, bb=0 0 515 485]
       {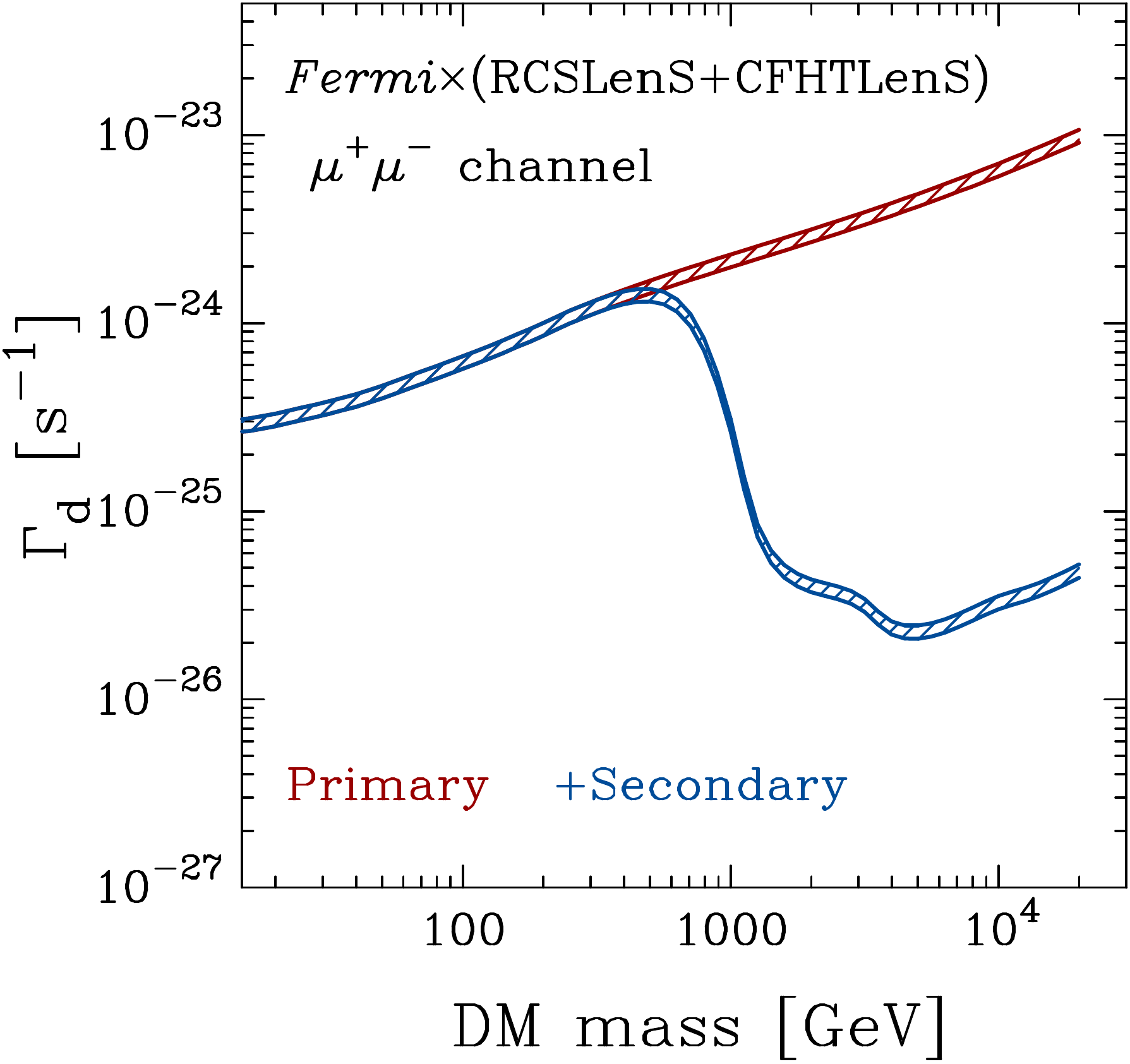}
     \caption{
     \label{fig:decay_constraints}
     The same as Figure~\ref{fig:ann_constraints}, but for
     the $95\%$ confidence level upper limit 
     on the DM decay rate. 
     The shaded band represents 
     a conservative estimate of the theoretical uncertainty 
     due to three-dimensional power spectrum of the DM distribution.
  } 
    \end{center}
\end{figure*}

We next consider decaying DM.
As discussed in Section \ref{subsec:cross_corr_model}, 
the cross-correlation signal for DM decay
can be predicted by the three-dimensional clustering 
of the DM density distribution.
Conservatively, we consider two models of $\xi_{\rm IGRB-WL}$:
one is based on the fitting formula of $P_{\delta}$ calibrated 
with a set of cosmological N-body simulations \cite{Takahashi:2012em},
and the second one is an even more conservative scenario that uses only the linear matter 
power spectrum.
Despite this rather extreme treatment of model uncertainty, 
we find that our results are fairly robust to the choice of the model 
of $P_{\delta}$, introducing only some $\sim10\%$ differences.

Our constraints on decaying DM are shown in Fig.~\ref{fig:decay_constraints}.
Similarly to the annihilating DM case,
the constraints on the decay rate for heavy DM particles
are improved by a factor of $\sim100$ when secondary emission
is accounted for. This trend is more prominent for the leptonic decay channels than the hadronic channel.

In part due to the accurate models of $P_{\delta}$, 
the uncertainty from cosmic structure formation models
in the DM decay constraint is significantly small compared to the uncertainty in 
DM annihilation.
Although the resulting constraints are $\sim$10 times weaker
than the recent constraints with the spectrum of IGRB intensity 
\cite{Ando:2015qda}
and the cross-correlation of the IGRB and local galaxy distributions
\cite{Ando:2016ang},
they are the first constraints obtained from {\it unbiased} matter distributions {\it outside} the local Universe.
Note that the previous constraints \cite{Ando:2015qda, Ando:2016ang}
have been derived by the full-sky analysis of $\gamma$ rays, while
our constraints are based on the cross-correlation analysis with 
smaller sky fraction.
When considering upcoming imaging surveys with a sky coverage of 
20,000 squared degrees, we find that the expected constraints would be
very similar to the constraints in the literature (see Section~\ref{sec:con}).

\subsubsection*{Wino dark matter}

\begin{figure}[!t]
\begin{center}
       \includegraphics[clip, width=0.85\columnwidth, bb=0 0 522 499]
       {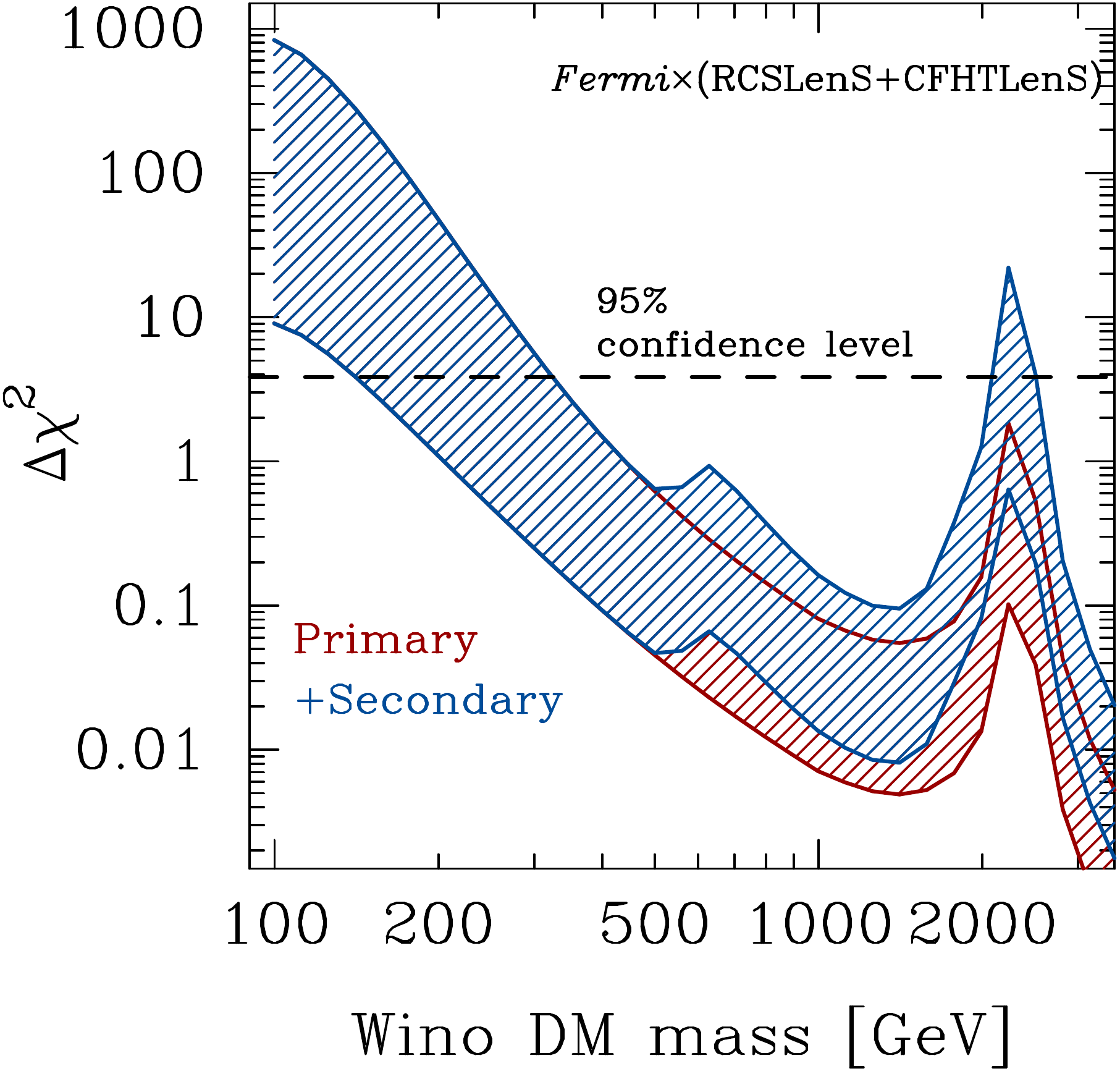}
     \caption{
     \label{fig:wino_constraints}
     The confidence level derived from 
     our cross-correlation analysis for a wino-like DM model
     \cite{Hryczuk:2011vi}.
     We consider the annihilation channel of $W^{+}W^{-}$.
     The red line shows the constraint in absence of secondary emission,
     while
     the blue is with including the primary and secondary emissions.
     Note that the shaded regions represent the model uncertainty
     associated with DM substructures.
  } 
    \end{center}
\end{figure}

Finally, we explore the parameter space of particular wino DM models
that are well motivated in particle physics.
The wino DM is the lightest supersymmetric particle (LSP)
of $\textrm{SU}(2)_{L}$ triplet Majorana fermion and realized in a wide
range of supersymmetric standard models in which the anomaly mediation
effect dominantly generates the gaugino masses \cite{Randall:1998uk, Giudice:1998xp}.

After the discovery of the Higgs boson with a mass of around 125 GeV
\cite{Aad:2012tfa, Chatrchyan:2012xdj}, 
models with the anomaly mediation and high-scale supersymmetry breaking is
one of the most attractive candidates for physics beyond the standard model~\cite{Hall:2011jd,
Hall:2012zp, Ibe:2011aa, Ibe:2012hu, Arvanitaki:2012ps,
ArkaniHamed:2012gw,Nomura:2014asa},
since this framework is compatible with the observed Higgs boson mass
\cite{Okada:1990vk,Okada:1990gg,Ellis:1990nz,Haber:1990aw,Ellis:1991zd},
and predicts the wino LSP as a good DM candidate.
The wino dark matter of a mass less than around 3 TeV is consistent
with the observed DM abundance \cite{Hisano:2006nn}.
The current LHC bound on the wino mass is around 270 GeV \cite{Aad:2013yna,CMS:2014gxa}.
At present there is no constraint on the wino DM from direct detection experiments
and 10 ton class experiments may test the wino DM
\cite{Hisano:2010fy,Hisano:2012wm,Hisano:2015rsa}.

A prominent feature of the wino DM is considerably enhanced annihilation
cross section by the Sommerfeld effects \cite{Hisano:2003ec, Hisano:2004ds}.
Due to such a large annihilation cross section, indirect detection of the wino DM
through cosmic-ray signals is promising.
In fact, the $\gamma$-ray searches such as observation of the dwarf spheroidal galaxies
by Fermi-LAT, and the galactic center by H.E.S.S. and antiproton searches
by AMS-02 \cite{Cohen:2013ama, Fan:2013faa, Hryczuk:2014hpa, Bhattacherjee:2014dya, Ibe:2015tma} 
can severely constrain the wino DM.
These constraints may, however, suffer from potentially large systematic
and astrophysical uncertainties such as the distribution of DM and that of other
astrophysical sources.
Therefore, it is important to examine the wino DM with our cross-correlation
statistics, which has totally different systematics.

In order to derive constraints on the wino mass, 
we adopt the annihilation cross-section including the one-loop and Sommerfeld
corrections~\cite{Hryczuk:2011vi}.
We assume that wino DM annihilates mainly into $W^{+}W^{-}$.
The primary and secondary $\gamma$-ray emissions for $W^{+}W^{-}$ channel are evaluated with {\tt PPPC4DMID} package 
\cite{Cirelli:2010xx} and 
the method shown in Ref.~\cite{Ando:2015qda, Ando:2016ang}.

Figure~\ref{fig:wino_constraints} shows the confidence level 
of wino DM model derived from our cross-correlation analysis.
Note that the single parameter of wino mass is of interest in the case of
wino DM model.
In this figure, the red and blue lines represent
$\Delta\chi^2$ as a function of wino mass 
in absence and presence of secondary emissions,
respectively.
The hatched regions show the model uncertainty 
of boost factor in our benchmark model (see Section~\ref{subsec:cross_corr_model} in detail).
Our result excludes the wino mass with 
$m_{\rm dm} \simlt 320$ GeV and
$2050\, {\rm GeV} \simlt m_{\rm dm} \simlt 2500 \, {\rm GeV}$ 
(95\% C.L.) in the case of the optimistic boost factor model 
and $m_{\rm dm} \simlt 140$ GeV in the conservative case.
The constraint is already close to that derived 
by the LHC \cite{Aad:2013yna}
and can be further improved with ongoing and upcoming 
imaging surveys that cover a wider area 
than $\sim660$ squared degrees considered here.

Note also that the secondary $\gamma$-ray emission originated 
from heavy DM particle tend to have lower energies.
Thus, it would be possible to make the constraint tighter 
by using $\gamma$-ray photons with lower energy.
On the other hand, the angular resolution becomes worse,
or the photon noise increases 
when we use the low-energy $\gamma$ rays.
It would also be necessary to examine carefully
systematic uncertainties associated 
with Galactic $\gamma$-ray emissions.

\section{\label{sec:con}CONCLUSION AND DISCUSSION}

We have performed cross-correlation analyses
of the IGRB and cosmic shear using 
data from the {\it Fermi-LAT} and two galaxy 
imaging surveys, CFHTLenS and RCSLenS.
Compared with our earlier
study \cite{Shirasaki:2014noa},
the present study benefits not only from the
increased survey area, but also from the following three aspects:

\begin{enumerate}

\item The use of reprocessed $\gamma$-ray data with the Pass 8 pipeline.

The improved photon event characterization decreases the size of point spread function (PSF)
in the energy range of $\simgt$1 GeV to $\sim30$ arcmin, allowing us to 
access the cross-correlation signals at small angular scales
originating from $\gamma$-ray emission within single DM halos.
The effective area is also increased, which when coupled with the improved PSF
also allows an improved characterization of the IGRB. 

\item Reduction of the statistical uncertainty by using twelve million galaxies.

Our cross-correlation analysis is based on the ensemble average 
of galaxy-photon pairs. Hence the statistical uncertainty is largely determined
by the number of galaxy-photon pairs. With 
about twelve million galaxies, we have reduced the statistical uncertainties
by factor of $\sim2.3$ compared to our previous study.

\item An accurate model of the sample variance of cosmic shear.

  There are substantial sample variances of cosmic shear
at large angular separations.  
We properly take this into account by using full-sky lensing simulations 
\cite{Shirasaki:2015dga}
incorporating observational effects in a direct manner
\cite{Shirasaki:2013zpa}.

\end{enumerate}

Our cross-correlation measurement using 661 squared-degrees sky coverage 
is still consistent with null detection.
With our accurate treatment on both statistical 
and systematic uncertainties, 
we fully utilized the cross-correlation measurement 
in the range of 10-300 arcmin, 
allowing us to improve the cosmological constraint 
by a factor of a few \footnote{
If we adopted the Planck results \cite{Ade:2015xua}, 
we would have derived slightly tighter constraints
because of the difference in $\Omega_{\rm m0}$ and $\sigma_{8}$}.
We investigate a generic class of particle DM assuming 
three representative annihilation/decay channels.
Because we include the contribution from 
secondary $\gamma$ rays produced through 
inverse-Compton scattering of CMB photons, we are able to
improve the DM constraints by a factor of $\sim100$ 
for TeV-scale DM with leptonic channels.

Using {\tt GALPROP} models, we have shown explicitly that the systematic 
uncertainties due to imperfect knowledge of Galactic $\gamma$-ray emission 
are unimportant compared to the statistical 
error in our measurement at present: 
we find that the typical systematic uncertainty is $\sim 10\%$ 
of the statistical error when using $\sim$12 million galaxies.
This conclusion is consistent with the analysis of correlation with the spatial 
distribution of galaxies \cite{Shirasaki:2015nqp}. However, the uncertainty
due to uncertain Galactic emission will likely become an issue in the future. 
First, future wide-area imaging surveys, such as the Large Synoptic Survey 
Telescope (LSST), will have a sky coverage of some 20,000 square degrees
and source number density of 30 ${\rm arcmin}^{-2}$.
We expect to use 2.16 billion galaxies in the cross-correlation measurement,
improving potentially the statistical uncertainty by a factor of $\sqrt{170}\sim13$.
Thus, the error on cross-correlation measurement would be 
dominated by systematic uncertainties due to the subtraction 
of Galactic $\gamma$-rays from the observed photon counts. Second,
recent evidence for significant large-scale time-dependent contributions 
to the cosmic rays, e.g., the Fermi bubble \cite{Su:2010qj}, are starting to
motivate studies beyond the stationary-state {\tt GALPROP} treatment.

If we can reduce the statistical error to the level
where the measurement is limited by the current systematic
uncertainties, a wide range of annihilating DM models 
with a thermal cross section can be excluded:
$m_{\rm dm} < 56-357 \, {\rm GeV}$ ($b\bar{b}$ channels),
$m_{\rm dm} < 26-96 \, {\rm GeV}$ ($\tau^{+}\tau^{-}$ channels),
and 
$m_{\rm dm} < 5-25\, {\rm GeV}$ ($\mu^{+}\mu^{-}$ channels).
Such tight constraints can actually test various hints of DM signals
reported so far.
For example, the Galactic center 
excess \cite{Goodenough:2009gk,Hooper:2010mq,Boyarsky:2010dr,
Hooper:2011ti,Abazajian:2012pn,Gordon:2013vta,Macias:2013vya,Calore:2014nla,
Calore:2014xka,Abazajian:2014fta,Daylan:2014rsa,Macias:2014sta,TheFermi-LAT:2015kwa,Lacroix:2015wfx, Horiuchi:2016zwu}, 
when interpreted as owing to DM annihilation, requires a cross section
$\langle \sigma v \rangle \sim (0.5$--$5) \times 10^{-26}\, {\rm cm}^3/{\rm s}$
over DM mass $\sim 10$--$100$ GeV, 
depending on the precise annihilation channel. 
Moreover, in the specific model of wino DM, 
all the viable wino DM mass region ($< 3$ TeV) can be tested
in the case of the optimistic boost factor model.
With the conservative assumption, the wino mass 
$m_{\rm dm} \simlt 740\, {\rm GeV}$ and 
$1830\, {\rm GeV} \simlt m_{\rm dm} \simlt  2690 \, {\rm GeV}$ 
may be excluded. 
Similarly, we find decaying DM models
in the wide range of $m_{\rm dm}=$10 GeV--20 TeV
with 
$\Gamma_{\rm d} > 1.8\times10^{-27} \, {\rm s}^{-1}$ ($b\bar{b}$ channels),
$\Gamma_{\rm d} > 5.5\times10^{-27} \, {\rm s}^{-1}$ ($\tau^{+}\tau^{-}$ channels),
and 
$\Gamma_{\rm d} > 2.8\times10^{-26} \, {\rm s}^{-1}$ ($\mu^{+}\mu^{-}$ channels)
are excluded
when the measurement would be limited 
by the current systematic uncertainties.
For specific DM models, these constraints can be further combined
with constraints from additional probes such as direct detection and collider 
searches, e.g., \citep{Horiuchi:2016tqw}.

We have shown that the cross-correlation analysis with 
the IGRB intensity and cosmic shear opens a new window
to indirect detection of DM. 
Similar analysis using data from future surveys will provide
competitive and independent constraints on the nature of DM
from the local dwarf galaxy constraints. Remaining important
issues include,
(i) reducing the uncertainty due to Galactic $\gamma$-ray subtraction,
(ii) direct measurement or constraints on the abundance of substructures
within DM halos,
(iii) optimization of detection of the cross-correlation signals
between the IGRB intensity and cosmic shear,
and 
(iv) joint analysis with Galactic and extragalactic measurements.
There appear to be a few ways to
increase the signal-to-noise ratio of the cross-correlation.
For example, detection of DM halos
in cosmic shear analyses \cite{Maturi:2004rn, Hennawi:2004ai}
can help us with reconstructing the DM distribution
more accurately.
In the next decades, a wealth of astronomical survey data
in multi-wavelengths will be 
steadily collected.
Precise statistical analysis of such big data will shed light 
on the long-standing mystery of the nature of DM.

\begin{acknowledgements}
M.S. is supported by Research Fellowships of the Japan Society for 
the Promotion of Science (JSPS) for Young Scientists.
N.Y. acknowledges financial support from JST CREST.
Numerical computations presented in this paper were in part carried out
on the general-purpose PC farm at Center for Computational Astrophysics,
CfCA, of National Astronomical Observatory of Japan.
We thank the Fermi Collaboration for the use of Fermi public data and the Fermi Science Tools.
This work is based on observations obtained with MegaPrime/MegaCam, a joint project of CFHT and CEA/IRFU, at the Canada-France-Hawaii Telescope (CFHT) which is operated by the National Research Council (NRC) of Canada, the Institut National des Sciences de l'Univers of the Centre National de la Recherche Scientifique (CNRS) of France, and the University of Hawaii. This research used the facilities of the Canadian Astronomy Data Centre operated by the National Research Council of Canada with the support of the Canadian Space Agency. 
In CFHTLenS and RCSLenS, data processing was made possible thanks to significant computing support from the NSERC Research Tools and Instruments grant program.
\end{acknowledgements}

\bibliography{ref_prd_v2}

\begin{thebibliography}{114}%
\makeatletter
\providecommand \@ifxundefined [1]{%
 \@ifx{#1\undefined}
}%
\providecommand \@ifnum [1]{%
 \ifnum #1\expandafter \@firstoftwo
 \else \expandafter \@secondoftwo
 \fi
}%
\providecommand \@ifx [1]{%
 \ifx #1\expandafter \@firstoftwo
 \else \expandafter \@secondoftwo
 \fi
}%
\providecommand \natexlab [1]{#1}%
\providecommand \enquote  [1]{``#1''}%
\providecommand \bibnamefont  [1]{#1}%
\providecommand \bibfnamefont [1]{#1}%
\providecommand \citenamefont [1]{#1}%
\providecommand \href@noop [0]{\@secondoftwo}%
\providecommand \href [0]{\begingroup \@sanitize@url \@href}%
\providecommand \@href[1]{\@@startlink{#1}\@@href}%
\providecommand \@@href[1]{\endgroup#1\@@endlink}%
\providecommand \@sanitize@url [0]{\catcode `\\12\catcode `\$12\catcode
  `\&12\catcode `\#12\catcode `\^12\catcode `\_12\catcode `\%12\relax}%
\providecommand \@@startlink[1]{}%
\providecommand \@@endlink[0]{}%
\providecommand \url  [0]{\begingroup\@sanitize@url \@url }%
\providecommand \@url [1]{\endgroup\@href {#1}{\urlprefix }}%
\providecommand \urlprefix  [0]{URL }%
\providecommand \Eprint [0]{\href }%
\providecommand \doibase [0]{http://dx.doi.org/}%
\providecommand \selectlanguage [0]{\@gobble}%
\providecommand \bibinfo  [0]{\@secondoftwo}%
\providecommand \bibfield  [0]{\@secondoftwo}%
\providecommand \translation [1]{[#1]}%
\providecommand \BibitemOpen [0]{}%
\providecommand \bibitemStop [0]{}%
\providecommand \bibitemNoStop [0]{.\EOS\space}%
\providecommand \EOS [0]{\spacefactor3000\relax}%
\providecommand \BibitemShut  [1]{\csname bibitem#1\endcsname}%
\let\auto@bib@innerbib\@empty
\bibitem [{\citenamefont {Hinshaw}\ \emph {et~al.}(2013)\citenamefont {Hinshaw}
  \emph {et~al.}}]{Hinshaw:2012aka}%
  \BibitemOpen
  \bibfield  {author} {\bibinfo {author} {\bibfnamefont {G.}~\bibnamefont
  {Hinshaw}} \emph {et~al.} (\bibinfo {collaboration} {WMAP}),\ }\href
  {\doibase 10.1088/0067-0049/208/2/19} {\bibfield  {journal} {\bibinfo
  {journal} {Astrophys. J. Suppl.}\ }\textbf {\bibinfo {volume} {208}},\
  \bibinfo {pages} {19} (\bibinfo {year} {2013})},\ \Eprint
  {http://arxiv.org/abs/1212.5226} {arXiv:1212.5226 [astro-ph.CO]} \BibitemShut
  {NoStop}%
\bibitem [{\citenamefont {Ade}\ \emph {et~al.}(2014)\citenamefont {Ade} \emph
  {et~al.}}]{Ade:2013zuv}%
  \BibitemOpen
  \bibfield  {author} {\bibinfo {author} {\bibfnamefont {P.~A.~R.}\
  \bibnamefont {Ade}} \emph {et~al.} (\bibinfo {collaboration} {Planck}),\
  }\href {\doibase 10.1051/0004-6361/201321591} {\bibfield  {journal} {\bibinfo
   {journal} {Astron. Astrophys.}\ }\textbf {\bibinfo {volume} {571}},\
  \bibinfo {pages} {A16} (\bibinfo {year} {2014})},\ \Eprint
  {http://arxiv.org/abs/1303.5076} {arXiv:1303.5076 [astro-ph.CO]} \BibitemShut
  {NoStop}%
\bibitem [{\citenamefont {Eisenstein}\ \emph {et~al.}(2005)\citenamefont
  {Eisenstein} \emph {et~al.}}]{Eisenstein:2005su}%
  \BibitemOpen
  \bibfield  {author} {\bibinfo {author} {\bibfnamefont {D.~J.}\ \bibnamefont
  {Eisenstein}} \emph {et~al.} (\bibinfo {collaboration} {SDSS}),\ }\href
  {\doibase 10.1086/466512} {\bibfield  {journal} {\bibinfo  {journal}
  {Astrophys. J.}\ }\textbf {\bibinfo {volume} {633}},\ \bibinfo {pages} {560}
  (\bibinfo {year} {2005})},\ \Eprint {http://arxiv.org/abs/astro-ph/0501171}
  {arXiv:astro-ph/0501171 [astro-ph]} \BibitemShut {NoStop}%
\bibitem [{\citenamefont {Persic}\ \emph {et~al.}(1996)\citenamefont {Persic},
  \citenamefont {Salucci},\ and\ \citenamefont {Stel}}]{Persic:1995ru}%
  \BibitemOpen
  \bibfield  {author} {\bibinfo {author} {\bibfnamefont {M.}~\bibnamefont
  {Persic}}, \bibinfo {author} {\bibfnamefont {P.}~\bibnamefont {Salucci}}, \
  and\ \bibinfo {author} {\bibfnamefont {F.}~\bibnamefont {Stel}},\ }\href
  {\doibase 10.1093/mnras/278.1.27} {\bibfield  {journal} {\bibinfo  {journal}
  {Mon.Not.Roy.Astron.Soc.}\ }\textbf {\bibinfo {volume} {281}},\ \bibinfo
  {pages} {27} (\bibinfo {year} {1996})},\ \Eprint
  {http://arxiv.org/abs/astro-ph/9506004} {arXiv:astro-ph/9506004 [astro-ph]}
  \BibitemShut {NoStop}%
\bibitem [{\citenamefont {Clowe}\ \emph {et~al.}(2006)\citenamefont {Clowe},
  \citenamefont {Bradac}, \citenamefont {Gonzalez}, \citenamefont {Markevitch},
  \citenamefont {Randall}, \citenamefont {Jones},\ and\ \citenamefont
  {Zaritsky}}]{Clowe:2006eq}%
  \BibitemOpen
  \bibfield  {author} {\bibinfo {author} {\bibfnamefont {D.}~\bibnamefont
  {Clowe}}, \bibinfo {author} {\bibfnamefont {M.}~\bibnamefont {Bradac}},
  \bibinfo {author} {\bibfnamefont {A.~H.}\ \bibnamefont {Gonzalez}}, \bibinfo
  {author} {\bibfnamefont {M.}~\bibnamefont {Markevitch}}, \bibinfo {author}
  {\bibfnamefont {S.~W.}\ \bibnamefont {Randall}}, \bibinfo {author}
  {\bibfnamefont {C.}~\bibnamefont {Jones}}, \ and\ \bibinfo {author}
  {\bibfnamefont {D.}~\bibnamefont {Zaritsky}},\ }\href {\doibase
  10.1086/508162} {\bibfield  {journal} {\bibinfo  {journal} {Astrophys. J.}\
  }\textbf {\bibinfo {volume} {648}},\ \bibinfo {pages} {L109} (\bibinfo {year}
  {2006})},\ \Eprint {http://arxiv.org/abs/astro-ph/0608407}
  {arXiv:astro-ph/0608407 [astro-ph]} \BibitemShut {NoStop}%
\bibitem [{\citenamefont {Bacon}\ \emph {et~al.}(2000)\citenamefont {Bacon},
  \citenamefont {Refregier},\ and\ \citenamefont {Ellis}}]{Bacon:2000sy}%
  \BibitemOpen
  \bibfield  {author} {\bibinfo {author} {\bibfnamefont {D.~J.}\ \bibnamefont
  {Bacon}}, \bibinfo {author} {\bibfnamefont {A.~R.}\ \bibnamefont
  {Refregier}}, \ and\ \bibinfo {author} {\bibfnamefont {R.~S.}\ \bibnamefont
  {Ellis}},\ }\href {\doibase 10.1046/j.1365-8711.2000.03851.x} {\bibfield
  {journal} {\bibinfo  {journal} {Mon. Not. Roy. Astron. Soc.}\ }\textbf
  {\bibinfo {volume} {318}},\ \bibinfo {pages} {625} (\bibinfo {year}
  {2000})},\ \Eprint {http://arxiv.org/abs/astro-ph/0003008}
  {arXiv:astro-ph/0003008 [astro-ph]} \BibitemShut {NoStop}%
\bibitem [{\citenamefont {Kilbinger}\ \emph {et~al.}(2013)\citenamefont
  {Kilbinger}, \citenamefont {Fu}, \citenamefont {Heymans}, \citenamefont
  {Simpson}, \citenamefont {Benjamin} \emph {et~al.}}]{Kilbinger:2012qz}%
  \BibitemOpen
  \bibfield  {author} {\bibinfo {author} {\bibfnamefont {M.}~\bibnamefont
  {Kilbinger}}, \bibinfo {author} {\bibfnamefont {L.}~\bibnamefont {Fu}},
  \bibinfo {author} {\bibfnamefont {C.}~\bibnamefont {Heymans}}, \bibinfo
  {author} {\bibfnamefont {F.}~\bibnamefont {Simpson}}, \bibinfo {author}
  {\bibfnamefont {J.}~\bibnamefont {Benjamin}},  \emph {et~al.},\ }\href
  {\doibase 10.1093/mnras/stt041} {\bibfield  {journal} {\bibinfo  {journal}
  {Mon.Not.Roy.Astron.Soc.}\ }\textbf {\bibinfo {volume} {430}},\ \bibinfo
  {pages} {2200} (\bibinfo {year} {2013})},\ \Eprint
  {http://arxiv.org/abs/1212.3338} {arXiv:1212.3338 [astro-ph.CO]} \BibitemShut
  {NoStop}%
\bibitem [{\citenamefont {Becker}\ \emph {et~al.}(2015)\citenamefont {Becker}
  \emph {et~al.}}]{Becker:2015ilr}%
  \BibitemOpen
  \bibfield  {author} {\bibinfo {author} {\bibfnamefont {M.~R.}\ \bibnamefont
  {Becker}} \emph {et~al.} (\bibinfo {collaboration} {DES}),\ }\href@noop {} {\
   (\bibinfo {year} {2015})},\ \Eprint {http://arxiv.org/abs/1507.05598}
  {arXiv:1507.05598 [astro-ph.CO]} \BibitemShut {NoStop}%
\bibitem [{\citenamefont {{Jungman}}\ \emph {et~al.}(1996)\citenamefont
  {{Jungman}}, \citenamefont {{Kamionkowski}},\ and\ \citenamefont
  {{Griest}}}]{1996PhR...267..195J}%
  \BibitemOpen
  \bibfield  {author} {\bibinfo {author} {\bibfnamefont {G.}~\bibnamefont
  {{Jungman}}}, \bibinfo {author} {\bibfnamefont {M.}~\bibnamefont
  {{Kamionkowski}}}, \ and\ \bibinfo {author} {\bibfnamefont {K.}~\bibnamefont
  {{Griest}}},\ }\href {\doibase 10.1016/0370-1573(95)00058-5} {\bibfield
  {journal} {\bibinfo  {journal} {Phys.Rept.}\ }\textbf {\bibinfo {volume}
  {267}},\ \bibinfo {pages} {195} (\bibinfo {year} {1996})},\ \Eprint
  {http://arxiv.org/abs/hep-ph/9506380} {hep-ph/9506380} \BibitemShut {NoStop}%
\bibitem [{\citenamefont {Funk}(2015)}]{Funk:2015ena}%
  \BibitemOpen
  \bibfield  {author} {\bibinfo {author} {\bibfnamefont {S.}~\bibnamefont
  {Funk}},\ }\href@noop {} {\  (\bibinfo {year} {2015})},\ \Eprint
  {http://arxiv.org/abs/1508.05190} {arXiv:1508.05190 [astro-ph.HE]}
  \BibitemShut {NoStop}%
\bibitem [{\citenamefont {{Ajello}}\ \emph {et~al.}(2015)\citenamefont
  {{Ajello}}, \citenamefont {{Gasparrini}}, \citenamefont
  {{S{\'a}nchez-Conde}}, \citenamefont {{Zaharijas}}, \citenamefont
  {{Gustafsson}}, \citenamefont {{Cohen-Tanugi}}, \citenamefont {{Dermer}},
  \citenamefont {{Inoue}}, \citenamefont {{Hartmann}}, \citenamefont
  {{Ackermann}}, \citenamefont {{Bechtol}}, \citenamefont {{Franckowiak}},
  \citenamefont {{Reimer}}, \citenamefont {{Romani}},\ and\ \citenamefont
  {{Strong}}}]{2015ApJ...800L..27A}%
  \BibitemOpen
  \bibfield  {author} {\bibinfo {author} {\bibfnamefont {M.}~\bibnamefont
  {{Ajello}}}, \bibinfo {author} {\bibfnamefont {D.}~\bibnamefont
  {{Gasparrini}}}, \bibinfo {author} {\bibfnamefont {M.}~\bibnamefont
  {{S{\'a}nchez-Conde}}}, \bibinfo {author} {\bibfnamefont {G.}~\bibnamefont
  {{Zaharijas}}}, \bibinfo {author} {\bibfnamefont {M.}~\bibnamefont
  {{Gustafsson}}}, \bibinfo {author} {\bibfnamefont {J.}~\bibnamefont
  {{Cohen-Tanugi}}}, \bibinfo {author} {\bibfnamefont {C.~D.}\ \bibnamefont
  {{Dermer}}}, \bibinfo {author} {\bibfnamefont {Y.}~\bibnamefont {{Inoue}}},
  \bibinfo {author} {\bibfnamefont {D.}~\bibnamefont {{Hartmann}}}, \bibinfo
  {author} {\bibfnamefont {M.}~\bibnamefont {{Ackermann}}}, \bibinfo {author}
  {\bibfnamefont {K.}~\bibnamefont {{Bechtol}}}, \bibinfo {author}
  {\bibfnamefont {A.}~\bibnamefont {{Franckowiak}}}, \bibinfo {author}
  {\bibfnamefont {A.}~\bibnamefont {{Reimer}}}, \bibinfo {author}
  {\bibfnamefont {R.~W.}\ \bibnamefont {{Romani}}}, \ and\ \bibinfo {author}
  {\bibfnamefont {A.~W.}\ \bibnamefont {{Strong}}},\ }\href {\doibase
  10.1088/2041-8205/800/2/L27} {\bibfield  {journal} {\bibinfo  {journal}
  {Astrophys.J.Lett.}\ }\textbf {\bibinfo {volume} {800}},\ \bibinfo {eid}
  {L27} (\bibinfo {year} {2015})},\ \Eprint {http://arxiv.org/abs/1501.05301}
  {arXiv:1501.05301 [astro-ph.HE]} \BibitemShut {NoStop}%
\bibitem [{\citenamefont {Fornasa}\ and\ \citenamefont
  {S{\'a}nchez-Conde}(2015)}]{Fornasa:2015qua}%
  \BibitemOpen
  \bibfield  {author} {\bibinfo {author} {\bibfnamefont {M.}~\bibnamefont
  {Fornasa}}\ and\ \bibinfo {author} {\bibfnamefont {M.~A.}\ \bibnamefont
  {S{\'a}nchez-Conde}},\ }\href {\doibase 10.1016/j.physrep.2015.09.002}
  {\bibfield  {journal} {\bibinfo  {journal} {Phys. Rept.}\ }\textbf {\bibinfo
  {volume} {598}},\ \bibinfo {pages} {1} (\bibinfo {year} {2015})},\ \Eprint
  {http://arxiv.org/abs/1502.02866} {arXiv:1502.02866 [astro-ph.CO]}
  \BibitemShut {NoStop}%
\bibitem [{\citenamefont {Camera}\ \emph {et~al.}(2013)\citenamefont {Camera},
  \citenamefont {Fornasa}, \citenamefont {Fornengo},\ and\ \citenamefont
  {Regis}}]{Camera:2012cj}%
  \BibitemOpen
  \bibfield  {author} {\bibinfo {author} {\bibfnamefont {S.}~\bibnamefont
  {Camera}}, \bibinfo {author} {\bibfnamefont {M.}~\bibnamefont {Fornasa}},
  \bibinfo {author} {\bibfnamefont {N.}~\bibnamefont {Fornengo}}, \ and\
  \bibinfo {author} {\bibfnamefont {M.}~\bibnamefont {Regis}},\ }\href
  {\doibase 10.1088/2041-8205/771/1/L5} {\bibfield  {journal} {\bibinfo
  {journal} {Astrophys.J.}\ }\textbf {\bibinfo {volume} {771}},\ \bibinfo
  {pages} {L5} (\bibinfo {year} {2013})},\ \Eprint
  {http://arxiv.org/abs/1212.5018} {arXiv:1212.5018 [astro-ph.CO]} \BibitemShut
  {NoStop}%
\bibitem [{\citenamefont {{Fornengo}}\ and\ \citenamefont
  {{Regis}}(2014)}]{2014FrP.....2....6F}%
  \BibitemOpen
  \bibfield  {author} {\bibinfo {author} {\bibfnamefont {N.}~\bibnamefont
  {{Fornengo}}}\ and\ \bibinfo {author} {\bibfnamefont {M.}~\bibnamefont
  {{Regis}}},\ }\href {\doibase 10.3389/fphy.2014.00006} {\bibfield  {journal}
  {\bibinfo  {journal} {Frontiers in Physics}\ }\textbf {\bibinfo {volume}
  {2}},\ \bibinfo {eid} {6} (\bibinfo {year} {2014})},\ \Eprint
  {http://arxiv.org/abs/1312.4835} {arXiv:1312.4835 [astro-ph.CO]} \BibitemShut
  {NoStop}%
\bibitem [{\citenamefont {{Ando}}\ \emph {et~al.}(2014)\citenamefont {{Ando}},
  \citenamefont {{Benoit-L{\'e}vy}},\ and\ \citenamefont
  {{Komatsu}}}]{2014PhRvD..90b3514A}%
  \BibitemOpen
  \bibfield  {author} {\bibinfo {author} {\bibfnamefont {S.}~\bibnamefont
  {{Ando}}}, \bibinfo {author} {\bibfnamefont {A.}~\bibnamefont
  {{Benoit-L{\'e}vy}}}, \ and\ \bibinfo {author} {\bibfnamefont
  {E.}~\bibnamefont {{Komatsu}}},\ }\href {\doibase 10.1103/PhysRevD.90.023514}
  {\bibfield  {journal} {\bibinfo  {journal} {Phys.Rev.}\ ,\ \bibinfo {eid}
  {023514}} (\bibinfo {year} {2014})},\ \Eprint
  {http://arxiv.org/abs/1312.4403} {arXiv:1312.4403} \BibitemShut {NoStop}%
\bibitem [{\citenamefont {Ando}(2014)}]{Ando:2014aoa}%
  \BibitemOpen
  \bibfield  {author} {\bibinfo {author} {\bibfnamefont {S.}~\bibnamefont
  {Ando}},\ }\href {\doibase 10.1088/1475-7516/2014/10/061} {\bibfield
  {journal} {\bibinfo  {journal} {JCAP}\ }\textbf {\bibinfo {volume} {1410}},\
  \bibinfo {pages} {061} (\bibinfo {year} {2014})},\ \Eprint
  {http://arxiv.org/abs/1407.8502} {arXiv:1407.8502 [astro-ph.CO]} \BibitemShut
  {NoStop}%
\bibitem [{\citenamefont {{Camera}}\ \emph {et~al.}(2015)\citenamefont
  {{Camera}}, \citenamefont {{Fornasa}}, \citenamefont {{Fornengo}},\ and\
  \citenamefont {{Regis}}}]{2015JCAP...06..029C}%
  \BibitemOpen
  \bibfield  {author} {\bibinfo {author} {\bibfnamefont {S.}~\bibnamefont
  {{Camera}}}, \bibinfo {author} {\bibfnamefont {M.}~\bibnamefont {{Fornasa}}},
  \bibinfo {author} {\bibfnamefont {N.}~\bibnamefont {{Fornengo}}}, \ and\
  \bibinfo {author} {\bibfnamefont {M.}~\bibnamefont {{Regis}}},\ }\href
  {\doibase 10.1088/1475-7516/2015/06/029} {\bibfield  {journal} {\bibinfo
  {journal} {JCAP}\ }\textbf {\bibinfo {volume} {6}},\ \bibinfo {eid} {029}
  (\bibinfo {year} {2015})},\ \Eprint {http://arxiv.org/abs/1411.4651}
  {arXiv:1411.4651} \BibitemShut {NoStop}%
\bibitem [{\citenamefont {{Xia}}\ \emph {et~al.}(2015)\citenamefont {{Xia}},
  \citenamefont {{Cuoco}}, \citenamefont {{Branchini}},\ and\ \citenamefont
  {{Viel}}}]{2015ApJS..217...15X}%
  \BibitemOpen
  \bibfield  {author} {\bibinfo {author} {\bibfnamefont {J.-Q.}\ \bibnamefont
  {{Xia}}}, \bibinfo {author} {\bibfnamefont {A.}~\bibnamefont {{Cuoco}}},
  \bibinfo {author} {\bibfnamefont {E.}~\bibnamefont {{Branchini}}}, \ and\
  \bibinfo {author} {\bibfnamefont {M.}~\bibnamefont {{Viel}}},\ }\href
  {\doibase 10.1088/0067-0049/217/1/15} {\bibfield  {journal} {\bibinfo
  {journal} {Astrophys.J.Suppl.}\ }\textbf {\bibinfo {volume} {217}},\ \bibinfo
  {eid} {15} (\bibinfo {year} {2015})},\ \Eprint
  {http://arxiv.org/abs/1503.05918} {arXiv:1503.05918} \BibitemShut {NoStop}%
\bibitem [{\citenamefont {{Regis}}\ \emph {et~al.}(2015)\citenamefont
  {{Regis}}, \citenamefont {{Xia}}, \citenamefont {{Cuoco}}, \citenamefont
  {{Branchini}}, \citenamefont {{Fornengo}},\ and\ \citenamefont
  {{Viel}}}]{2015PhRvL.114x1301R}%
  \BibitemOpen
  \bibfield  {author} {\bibinfo {author} {\bibfnamefont {M.}~\bibnamefont
  {{Regis}}}, \bibinfo {author} {\bibfnamefont {J.-Q.}\ \bibnamefont {{Xia}}},
  \bibinfo {author} {\bibfnamefont {A.}~\bibnamefont {{Cuoco}}}, \bibinfo
  {author} {\bibfnamefont {E.}~\bibnamefont {{Branchini}}}, \bibinfo {author}
  {\bibfnamefont {N.}~\bibnamefont {{Fornengo}}}, \ and\ \bibinfo {author}
  {\bibfnamefont {M.}~\bibnamefont {{Viel}}},\ }\href {\doibase
  10.1103/PhysRevLett.114.241301} {\bibfield  {journal} {\bibinfo  {journal}
  {Physical Review Letters}\ }\textbf {\bibinfo {volume} {114}},\ \bibinfo
  {eid} {241301} (\bibinfo {year} {2015})},\ \Eprint
  {http://arxiv.org/abs/1503.05922} {arXiv:1503.05922} \BibitemShut {NoStop}%
\bibitem [{\citenamefont {Cuoco}\ \emph {et~al.}(2015)\citenamefont {Cuoco},
  \citenamefont {Xia}, \citenamefont {Regis}, \citenamefont {Branchini},
  \citenamefont {Fornengo},\ and\ \citenamefont {Viel}}]{Cuoco:2015rfa}%
  \BibitemOpen
  \bibfield  {author} {\bibinfo {author} {\bibfnamefont {A.}~\bibnamefont
  {Cuoco}}, \bibinfo {author} {\bibfnamefont {J.-Q.}\ \bibnamefont {Xia}},
  \bibinfo {author} {\bibfnamefont {M.}~\bibnamefont {Regis}}, \bibinfo
  {author} {\bibfnamefont {E.}~\bibnamefont {Branchini}}, \bibinfo {author}
  {\bibfnamefont {N.}~\bibnamefont {Fornengo}}, \ and\ \bibinfo {author}
  {\bibfnamefont {M.}~\bibnamefont {Viel}},\ }\href@noop {} {\  (\bibinfo
  {year} {2015})},\ \Eprint {http://arxiv.org/abs/1506.01030} {arXiv:1506.01030
  [astro-ph.HE]} \BibitemShut {NoStop}%
\bibitem [{\citenamefont {Shirasaki}\ \emph {et~al.}(2014)\citenamefont
  {Shirasaki}, \citenamefont {Horiuchi},\ and\ \citenamefont
  {Yoshida}}]{Shirasaki:2014noa}%
  \BibitemOpen
  \bibfield  {author} {\bibinfo {author} {\bibfnamefont {M.}~\bibnamefont
  {Shirasaki}}, \bibinfo {author} {\bibfnamefont {S.}~\bibnamefont {Horiuchi}},
  \ and\ \bibinfo {author} {\bibfnamefont {N.}~\bibnamefont {Yoshida}},\ }\href
  {\doibase 10.1103/PhysRevD.90.063502} {\bibfield  {journal} {\bibinfo
  {journal} {Phys.Rev.}\ }\textbf {\bibinfo {volume} {D90}},\ \bibinfo {pages}
  {063502} (\bibinfo {year} {2014})},\ \Eprint {http://arxiv.org/abs/1404.5503}
  {arXiv:1404.5503 [astro-ph.CO]} \BibitemShut {NoStop}%
\bibitem [{\citenamefont {Gilmore}\ \emph {et~al.}(2012)\citenamefont
  {Gilmore}, \citenamefont {Somerville}, \citenamefont {Primack},\ and\
  \citenamefont {Dominguez}}]{Gilmore:2011ks}%
  \BibitemOpen
  \bibfield  {author} {\bibinfo {author} {\bibfnamefont {R.}~\bibnamefont
  {Gilmore}}, \bibinfo {author} {\bibfnamefont {R.}~\bibnamefont {Somerville}},
  \bibinfo {author} {\bibfnamefont {J.}~\bibnamefont {Primack}}, \ and\
  \bibinfo {author} {\bibfnamefont {A.}~\bibnamefont {Dominguez}},\ }\href@noop
  {} {\bibfield  {journal} {\bibinfo  {journal} {Mon.Not.Roy.Astron.Soc.}\
  }\textbf {\bibinfo {volume} {422}},\ \bibinfo {pages} {3189} (\bibinfo {year}
  {2012})},\ \Eprint {http://arxiv.org/abs/1104.0671} {arXiv:1104.0671
  [astro-ph.CO]} \BibitemShut {NoStop}%
\bibitem [{\citenamefont {Ishiwata}\ \emph {et~al.}(2009)\citenamefont
  {Ishiwata}, \citenamefont {Matsumoto},\ and\ \citenamefont
  {Moroi}}]{Ishiwata:2009dk}%
  \BibitemOpen
  \bibfield  {author} {\bibinfo {author} {\bibfnamefont {K.}~\bibnamefont
  {Ishiwata}}, \bibinfo {author} {\bibfnamefont {S.}~\bibnamefont {Matsumoto}},
  \ and\ \bibinfo {author} {\bibfnamefont {T.}~\bibnamefont {Moroi}},\ }\href
  {\doibase 10.1016/j.physletb.2009.07.004} {\bibfield  {journal} {\bibinfo
  {journal} {Phys. Lett.}\ }\textbf {\bibinfo {volume} {B679}},\ \bibinfo
  {pages} {1} (\bibinfo {year} {2009})},\ \Eprint
  {http://arxiv.org/abs/0905.4593} {arXiv:0905.4593 [astro-ph.CO]} \BibitemShut
  {NoStop}%
\bibitem [{\citenamefont {{Brunetti}}\ and\ \citenamefont
  {{Jones}}(2014)}]{2014IJMPD..2330007B}%
  \BibitemOpen
  \bibfield  {author} {\bibinfo {author} {\bibfnamefont {G.}~\bibnamefont
  {{Brunetti}}}\ and\ \bibinfo {author} {\bibfnamefont {T.~W.}\ \bibnamefont
  {{Jones}}},\ }\href {\doibase 10.1142/S0218271814300079} {\bibfield
  {journal} {\bibinfo  {journal} {International Journal of Modern Physics D}\
  }\textbf {\bibinfo {volume} {23}},\ \bibinfo {eid} {1430007-98} (\bibinfo
  {year} {2014})},\ \Eprint {http://arxiv.org/abs/1401.7519} {arXiv:1401.7519}
  \BibitemShut {NoStop}%
\bibitem [{\citenamefont {Bartelmann}\ and\ \citenamefont
  {Schneider}(2001)}]{Bartelmann:1999yn}%
  \BibitemOpen
  \bibfield  {author} {\bibinfo {author} {\bibfnamefont {M.}~\bibnamefont
  {Bartelmann}}\ and\ \bibinfo {author} {\bibfnamefont {P.}~\bibnamefont
  {Schneider}},\ }\href {\doibase 10.1016/S0370-1573(00)00082-X} {\bibfield
  {journal} {\bibinfo  {journal} {Phys.Rept.}\ }\textbf {\bibinfo {volume}
  {340}},\ \bibinfo {pages} {291} (\bibinfo {year} {2001})},\ \Eprint
  {http://arxiv.org/abs/astro-ph/9912508} {arXiv:astro-ph/9912508 [astro-ph]}
  \BibitemShut {NoStop}%
\bibitem [{\citenamefont {Cirelli}\ \emph {et~al.}(2011)\citenamefont
  {Cirelli}, \citenamefont {Corcella}, \citenamefont {Hektor}, \citenamefont
  {Hutsi}, \citenamefont {Kadastik}, \citenamefont {Panci}, \citenamefont
  {Raidal}, \citenamefont {Sala},\ and\ \citenamefont
  {Strumia}}]{Cirelli:2010xx}%
  \BibitemOpen
  \bibfield  {author} {\bibinfo {author} {\bibfnamefont {M.}~\bibnamefont
  {Cirelli}}, \bibinfo {author} {\bibfnamefont {G.}~\bibnamefont {Corcella}},
  \bibinfo {author} {\bibfnamefont {A.}~\bibnamefont {Hektor}}, \bibinfo
  {author} {\bibfnamefont {G.}~\bibnamefont {Hutsi}}, \bibinfo {author}
  {\bibfnamefont {M.}~\bibnamefont {Kadastik}}, \bibinfo {author}
  {\bibfnamefont {P.}~\bibnamefont {Panci}}, \bibinfo {author} {\bibfnamefont
  {M.}~\bibnamefont {Raidal}}, \bibinfo {author} {\bibfnamefont
  {F.}~\bibnamefont {Sala}}, \ and\ \bibinfo {author} {\bibfnamefont
  {A.}~\bibnamefont {Strumia}},\ }\href {\doibase
  10.1088/1475-7516/2012/10/E01, 10.1088/1475-7516/2011/03/051} {\bibfield
  {journal} {\bibinfo  {journal} {JCAP}\ }\textbf {\bibinfo {volume} {1103}},\
  \bibinfo {pages} {051} (\bibinfo {year} {2011})},\ \bibinfo {note} {[Erratum:
  JCAP1210,E01(2012)]},\ \Eprint {http://arxiv.org/abs/1012.4515}
  {arXiv:1012.4515 [hep-ph]} \BibitemShut {NoStop}%
\bibitem [{\citenamefont {Ando}\ and\ \citenamefont
  {Ishiwata}(2015)}]{Ando:2015qda}%
  \BibitemOpen
  \bibfield  {author} {\bibinfo {author} {\bibfnamefont {S.}~\bibnamefont
  {Ando}}\ and\ \bibinfo {author} {\bibfnamefont {K.}~\bibnamefont
  {Ishiwata}},\ }\href {\doibase 10.1088/1475-7516/2015/05/024} {\bibfield
  {journal} {\bibinfo  {journal} {JCAP}\ }\textbf {\bibinfo {volume} {1505}},\
  \bibinfo {pages} {024} (\bibinfo {year} {2015})},\ \Eprint
  {http://arxiv.org/abs/1502.02007} {arXiv:1502.02007 [astro-ph.CO]}
  \BibitemShut {NoStop}%
\bibitem [{\citenamefont {Ando}\ and\ \citenamefont
  {Ishiwata}(2016)}]{Ando:2016ang}%
  \BibitemOpen
  \bibfield  {author} {\bibinfo {author} {\bibfnamefont {S.}~\bibnamefont
  {Ando}}\ and\ \bibinfo {author} {\bibfnamefont {K.}~\bibnamefont
  {Ishiwata}},\ }\href@noop {} {\  (\bibinfo {year} {2016})},\ \Eprint
  {http://arxiv.org/abs/1604.02263} {arXiv:1604.02263 [hep-ph]} \BibitemShut
  {NoStop}%
\bibitem [{\citenamefont {Limber}(1954)}]{Limber:1954zz}%
  \BibitemOpen
  \bibfield  {author} {\bibinfo {author} {\bibfnamefont {D.~N.}\ \bibnamefont
  {Limber}},\ }\href@noop {} {\bibfield  {journal} {\bibinfo  {journal}
  {Astrophys.J.}\ }\textbf {\bibinfo {volume} {119}},\ \bibinfo {pages} {655}
  (\bibinfo {year} {1954})}\BibitemShut {NoStop}%
\bibitem [{\citenamefont {de~Putter}\ and\ \citenamefont
  {Takada}(2010)}]{dePutter:2010jz}%
  \BibitemOpen
  \bibfield  {author} {\bibinfo {author} {\bibfnamefont {R.}~\bibnamefont
  {de~Putter}}\ and\ \bibinfo {author} {\bibfnamefont {M.}~\bibnamefont
  {Takada}},\ }\href {\doibase 10.1103/PhysRevD.82.103522} {\bibfield
  {journal} {\bibinfo  {journal} {Phys.Rev.}\ }\textbf {\bibinfo {volume}
  {D82}},\ \bibinfo {pages} {103522} (\bibinfo {year} {2010})},\ \Eprint
  {http://arxiv.org/abs/1007.4809} {arXiv:1007.4809 [astro-ph.CO]} \BibitemShut
  {NoStop}%
\bibitem [{\citenamefont {Oguri}\ and\ \citenamefont
  {Takada}(2011)}]{Oguri:2010vi}%
  \BibitemOpen
  \bibfield  {author} {\bibinfo {author} {\bibfnamefont {M.}~\bibnamefont
  {Oguri}}\ and\ \bibinfo {author} {\bibfnamefont {M.}~\bibnamefont {Takada}},\
  }\href {\doibase 10.1103/PhysRevD.83.023008} {\bibfield  {journal} {\bibinfo
  {journal} {Phys.Rev.}\ }\textbf {\bibinfo {volume} {D83}},\ \bibinfo {pages}
  {023008} (\bibinfo {year} {2011})},\ \Eprint {http://arxiv.org/abs/1010.0744}
  {arXiv:1010.0744 [astro-ph.CO]} \BibitemShut {NoStop}%
\bibitem [{\citenamefont {Cooray}\ and\ \citenamefont
  {Sheth}(2002)}]{Cooray:2002dia}%
  \BibitemOpen
  \bibfield  {author} {\bibinfo {author} {\bibfnamefont {A.}~\bibnamefont
  {Cooray}}\ and\ \bibinfo {author} {\bibfnamefont {R.~K.}\ \bibnamefont
  {Sheth}},\ }\href {\doibase 10.1016/S0370-1573(02)00276-4} {\bibfield
  {journal} {\bibinfo  {journal} {Phys.Rept.}\ }\textbf {\bibinfo {volume}
  {372}},\ \bibinfo {pages} {1} (\bibinfo {year} {2002})},\ \Eprint
  {http://arxiv.org/abs/astro-ph/0206508} {arXiv:astro-ph/0206508 [astro-ph]}
  \BibitemShut {NoStop}%
\bibitem [{\citenamefont {Tinker}\ \emph {et~al.}(2008)\citenamefont {Tinker},
  \citenamefont {Kravtsov}, \citenamefont {Klypin}, \citenamefont {Abazajian},
  \citenamefont {Warren} \emph {et~al.}}]{Tinker:2008ff}%
  \BibitemOpen
  \bibfield  {author} {\bibinfo {author} {\bibfnamefont {J.~L.}\ \bibnamefont
  {Tinker}}, \bibinfo {author} {\bibfnamefont {A.~V.}\ \bibnamefont
  {Kravtsov}}, \bibinfo {author} {\bibfnamefont {A.}~\bibnamefont {Klypin}},
  \bibinfo {author} {\bibfnamefont {K.}~\bibnamefont {Abazajian}}, \bibinfo
  {author} {\bibfnamefont {M.~S.}\ \bibnamefont {Warren}},  \emph {et~al.},\
  }\href {\doibase 10.1086/591439} {\bibfield  {journal} {\bibinfo  {journal}
  {Astrophys.J.}\ }\textbf {\bibinfo {volume} {688}},\ \bibinfo {pages} {709}
  (\bibinfo {year} {2008})},\ \Eprint {http://arxiv.org/abs/0803.2706}
  {arXiv:0803.2706 [astro-ph]} \BibitemShut {NoStop}%
\bibitem [{\citenamefont {Tinker}\ \emph {et~al.}(2010)\citenamefont {Tinker},
  \citenamefont {Robertson}, \citenamefont {Kravtsov}, \citenamefont {Klypin},
  \citenamefont {Warren} \emph {et~al.}}]{Tinker:2010my}%
  \BibitemOpen
  \bibfield  {author} {\bibinfo {author} {\bibfnamefont {J.~L.}\ \bibnamefont
  {Tinker}}, \bibinfo {author} {\bibfnamefont {B.~E.}\ \bibnamefont
  {Robertson}}, \bibinfo {author} {\bibfnamefont {A.~V.}\ \bibnamefont
  {Kravtsov}}, \bibinfo {author} {\bibfnamefont {A.}~\bibnamefont {Klypin}},
  \bibinfo {author} {\bibfnamefont {M.~S.}\ \bibnamefont {Warren}},  \emph
  {et~al.},\ }\href {\doibase 10.1088/0004-637X/724/2/878} {\bibfield
  {journal} {\bibinfo  {journal} {Astrophys.J.}\ }\textbf {\bibinfo {volume}
  {724}},\ \bibinfo {pages} {878} (\bibinfo {year} {2010})},\ \Eprint
  {http://arxiv.org/abs/1001.3162} {arXiv:1001.3162 [astro-ph.CO]} \BibitemShut
  {NoStop}%
\bibitem [{\citenamefont {Navarro}\ \emph {et~al.}(1997)\citenamefont
  {Navarro}, \citenamefont {Frenk},\ and\ \citenamefont
  {White}}]{Navarro:1996gj}%
  \BibitemOpen
  \bibfield  {author} {\bibinfo {author} {\bibfnamefont {J.~F.}\ \bibnamefont
  {Navarro}}, \bibinfo {author} {\bibfnamefont {C.~S.}\ \bibnamefont {Frenk}},
  \ and\ \bibinfo {author} {\bibfnamefont {S.~D.}\ \bibnamefont {White}},\
  }\href {\doibase 10.1086/304888} {\bibfield  {journal} {\bibinfo  {journal}
  {Astrophys.J.}\ }\textbf {\bibinfo {volume} {490}},\ \bibinfo {pages} {493}
  (\bibinfo {year} {1997})},\ \Eprint {http://arxiv.org/abs/astro-ph/9611107}
  {arXiv:astro-ph/9611107 [astro-ph]} \BibitemShut {NoStop}%
\bibitem [{\citenamefont {Prada}\ \emph {et~al.}(2012)\citenamefont {Prada},
  \citenamefont {Klypin}, \citenamefont {Cuesta}, \citenamefont
  {Betancort-Rijo},\ and\ \citenamefont {Primack}}]{Prada:2011jf}%
  \BibitemOpen
  \bibfield  {author} {\bibinfo {author} {\bibfnamefont {F.}~\bibnamefont
  {Prada}}, \bibinfo {author} {\bibfnamefont {A.~A.}\ \bibnamefont {Klypin}},
  \bibinfo {author} {\bibfnamefont {A.~J.}\ \bibnamefont {Cuesta}}, \bibinfo
  {author} {\bibfnamefont {J.~E.}\ \bibnamefont {Betancort-Rijo}}, \ and\
  \bibinfo {author} {\bibfnamefont {J.}~\bibnamefont {Primack}},\ }\href
  {\doibase 10.1111/j.1365-2966.2012.21007.x} {\bibfield  {journal} {\bibinfo
  {journal} {Mon. Not. Roy. Astron. Soc.}\ }\textbf {\bibinfo {volume} {428}},\
  \bibinfo {pages} {3018} (\bibinfo {year} {2012})},\ \Eprint
  {http://arxiv.org/abs/1104.5130} {arXiv:1104.5130 [astro-ph.CO]} \BibitemShut
  {NoStop}%
\bibitem [{\citenamefont {Gao}\ \emph {et~al.}(2012)\citenamefont {Gao},
  \citenamefont {Frenk}, \citenamefont {Jenkins}, \citenamefont {Springel},\
  and\ \citenamefont {White}}]{Gao:2011rf}%
  \BibitemOpen
  \bibfield  {author} {\bibinfo {author} {\bibfnamefont {L.}~\bibnamefont
  {Gao}}, \bibinfo {author} {\bibfnamefont {C.}~\bibnamefont {Frenk}}, \bibinfo
  {author} {\bibfnamefont {A.}~\bibnamefont {Jenkins}}, \bibinfo {author}
  {\bibfnamefont {V.}~\bibnamefont {Springel}}, \ and\ \bibinfo {author}
  {\bibfnamefont {S.}~\bibnamefont {White}},\ }\href {\doibase
  10.1111/j.1365-2966.2011.19836.x} {\bibfield  {journal} {\bibinfo  {journal}
  {Mon.Not.Roy.Astron.Soc.}\ }\textbf {\bibinfo {volume} {419}},\ \bibinfo
  {pages} {1721} (\bibinfo {year} {2012})},\ \Eprint
  {http://arxiv.org/abs/1107.1916} {arXiv:1107.1916 [astro-ph.CO]} \BibitemShut
  {NoStop}%
\bibitem [{\citenamefont {S{\'a}nchez-Conde}\ and\ \citenamefont
  {Prada}(2014)}]{Sanchez-Conde:2013yxa}%
  \BibitemOpen
  \bibfield  {author} {\bibinfo {author} {\bibfnamefont {M.~A.}\ \bibnamefont
  {S{\'a}nchez-Conde}}\ and\ \bibinfo {author} {\bibfnamefont {F.}~\bibnamefont
  {Prada}},\ }\href {\doibase 10.1093/mnras/stu1014} {\bibfield  {journal}
  {\bibinfo  {journal} {Mon. Not. Roy. Astron. Soc.}\ }\textbf {\bibinfo
  {volume} {442}},\ \bibinfo {pages} {2271} (\bibinfo {year} {2014})},\ \Eprint
  {http://arxiv.org/abs/1312.1729} {arXiv:1312.1729 [astro-ph.CO]} \BibitemShut
  {NoStop}%
\bibitem [{\citenamefont {Bartels}\ and\ \citenamefont
  {Ando}(2015)}]{Bartels:2015uba}%
  \BibitemOpen
  \bibfield  {author} {\bibinfo {author} {\bibfnamefont {R.}~\bibnamefont
  {Bartels}}\ and\ \bibinfo {author} {\bibfnamefont {S.}~\bibnamefont {Ando}},\
  }\href {\doibase 10.1103/PhysRevD.92.123508} {\bibfield  {journal} {\bibinfo
  {journal} {Phys. Rev.}\ }\textbf {\bibinfo {volume} {D92}},\ \bibinfo {pages}
  {123508} (\bibinfo {year} {2015})},\ \Eprint
  {http://arxiv.org/abs/1507.08656} {arXiv:1507.08656 [astro-ph.CO]}
  \BibitemShut {NoStop}%
\bibitem [{\citenamefont {{Atwood}}\ \emph {et~al.}(2009)\citenamefont
  {{Atwood}}, \citenamefont {{Abdo}}, \citenamefont {{Ackermann}},
  \citenamefont {{Althouse}}, \citenamefont {{Anderson}}, \citenamefont
  {{Axelsson}}, \citenamefont {{Baldini}}, \citenamefont {{Ballet}},
  \citenamefont {{Band}}, \citenamefont {{Barbiellini}},\ and\ \citenamefont
  {et~al.}}]{2009ApJ...697.1071A}%
  \BibitemOpen
  \bibfield  {author} {\bibinfo {author} {\bibfnamefont {W.~B.}\ \bibnamefont
  {{Atwood}}}, \bibinfo {author} {\bibfnamefont {A.~A.}\ \bibnamefont
  {{Abdo}}}, \bibinfo {author} {\bibfnamefont {M.}~\bibnamefont {{Ackermann}}},
  \bibinfo {author} {\bibfnamefont {W.}~\bibnamefont {{Althouse}}}, \bibinfo
  {author} {\bibfnamefont {B.}~\bibnamefont {{Anderson}}}, \bibinfo {author}
  {\bibfnamefont {M.}~\bibnamefont {{Axelsson}}}, \bibinfo {author}
  {\bibfnamefont {L.}~\bibnamefont {{Baldini}}}, \bibinfo {author}
  {\bibfnamefont {J.}~\bibnamefont {{Ballet}}}, \bibinfo {author}
  {\bibfnamefont {D.~L.}\ \bibnamefont {{Band}}}, \bibinfo {author}
  {\bibfnamefont {G.}~\bibnamefont {{Barbiellini}}}, \ and\ \bibinfo {author}
  {\bibnamefont {et~al.}},\ }\href {\doibase 10.1088/0004-637X/697/2/1071}
  {\bibfield  {journal} {\bibinfo  {journal} {Astrophys. J.}\ }\textbf
  {\bibinfo {volume} {697}},\ \bibinfo {pages} {1071} (\bibinfo {year}
  {2009})},\ \Eprint {http://arxiv.org/abs/0902.1089} {arXiv:0902.1089
  [astro-ph.IM]} \BibitemShut {NoStop}%
\bibitem [{\citenamefont {Takahashi}\ \emph {et~al.}(2012)\citenamefont
  {Takahashi}, \citenamefont {Sato}, \citenamefont {Nishimichi}, \citenamefont
  {Taruya},\ and\ \citenamefont {Oguri}}]{Takahashi:2012em}%
  \BibitemOpen
  \bibfield  {author} {\bibinfo {author} {\bibfnamefont {R.}~\bibnamefont
  {Takahashi}}, \bibinfo {author} {\bibfnamefont {M.}~\bibnamefont {Sato}},
  \bibinfo {author} {\bibfnamefont {T.}~\bibnamefont {Nishimichi}}, \bibinfo
  {author} {\bibfnamefont {A.}~\bibnamefont {Taruya}}, \ and\ \bibinfo {author}
  {\bibfnamefont {M.}~\bibnamefont {Oguri}},\ }\href {\doibase
  10.1088/0004-637X/761/2/152} {\bibfield  {journal} {\bibinfo  {journal}
  {Astrophys. J.}\ }\textbf {\bibinfo {volume} {761}},\ \bibinfo {pages} {152}
  (\bibinfo {year} {2012})},\ \Eprint {http://arxiv.org/abs/1208.2701}
  {arXiv:1208.2701 [astro-ph.CO]} \BibitemShut {NoStop}%
\bibitem [{Note1()}]{Note1}%
  \BibitemOpen
  \bibinfo {note} {ULTRACLEANVETO events are the recommended class of photons
  for investigation of the IGRB at intermediate to high latitudes. The reader
  is referred to the Cicerone (\protect \url
  {http://fermi.gsfc.nasa.gov/ssc/data/analysis/documentation/Pass8_usage.html})
  for further details}\BibitemShut {NoStop}%
\bibitem [{\citenamefont {{Acero}}\ \emph {et~al.}(2015)\citenamefont
  {{Acero}}, \citenamefont {{Ackermann}}, \citenamefont {{Ajello}},
  \citenamefont {{Albert}} \emph {et~al.}}]{3FGL}%
  \BibitemOpen
  \bibfield  {author} {\bibinfo {author} {\bibfnamefont {F.}~\bibnamefont
  {{Acero}}}, \bibinfo {author} {\bibfnamefont {M.}~\bibnamefont
  {{Ackermann}}}, \bibinfo {author} {\bibfnamefont {M.}~\bibnamefont
  {{Ajello}}}, \bibinfo {author} {\bibfnamefont {A.}~\bibnamefont {{Albert}}},
  \emph {et~al.} (\bibinfo {collaboration} {Fermi-LAT}),\ }\href {\doibase
  10.1088/0067-0049/218/2/23} {\bibfield  {journal} {\bibinfo  {journal}
  {Astrophys.J.Suppl.}\ }\textbf {\bibinfo {volume} {218}},\ \bibinfo {eid}
  {23} (\bibinfo {year} {2015})},\ \Eprint {http://arxiv.org/abs/1501.02003}
  {arXiv:1501.02003 [astro-ph.HE]} \BibitemShut {NoStop}%
\bibitem [{Note2()}]{Note2}%
  \BibitemOpen
  \bibinfo {note} {\protect \url
  {http://fermi.gsfc.nasa.gov/ssc/data/analysis/documentation/}}\BibitemShut
  {NoStop}%
\bibitem [{Note3()}]{Note3}%
  \BibitemOpen
  \bibinfo {note} {Http://fermi.gsfc.nasa.gov/ssc/data/analysis/}\BibitemShut
  {NoStop}%
\bibitem [{Note4()}]{Note4}%
  \BibitemOpen
  \bibinfo {note} {\protect \url
  {http://fermi.gsfc.nasa.gov/ssc/data/analysis/scitools/solar_template.html}}\BibitemShut
  {NoStop}%
\bibitem [{\citenamefont {Macias}\ and\ \citenamefont
  {Gordon}(2014)}]{Macias:2013vya}%
  \BibitemOpen
  \bibfield  {author} {\bibinfo {author} {\bibfnamefont {O.}~\bibnamefont
  {Macias}}\ and\ \bibinfo {author} {\bibfnamefont {C.}~\bibnamefont
  {Gordon}},\ }\href {\doibase 10.1103/PhysRevD.89.063515} {\bibfield
  {journal} {\bibinfo  {journal} {Phys. Rev.}\ }\textbf {\bibinfo {volume}
  {D89}},\ \bibinfo {pages} {063515} (\bibinfo {year} {2014})},\ \Eprint
  {http://arxiv.org/abs/1312.6671} {arXiv:1312.6671 [astro-ph.HE]} \BibitemShut
  {NoStop}%
\bibitem [{\citenamefont {Abazajian}\ \emph {et~al.}(2014)\citenamefont
  {Abazajian}, \citenamefont {Canac}, \citenamefont {Horiuchi},\ and\
  \citenamefont {Kaplinghat}}]{Abazajian:2014fta}%
  \BibitemOpen
  \bibfield  {author} {\bibinfo {author} {\bibfnamefont {K.~N.}\ \bibnamefont
  {Abazajian}}, \bibinfo {author} {\bibfnamefont {N.}~\bibnamefont {Canac}},
  \bibinfo {author} {\bibfnamefont {S.}~\bibnamefont {Horiuchi}}, \ and\
  \bibinfo {author} {\bibfnamefont {M.}~\bibnamefont {Kaplinghat}},\ }\href
  {\doibase 10.1103/PhysRevD.90.023526} {\bibfield  {journal} {\bibinfo
  {journal} {Phys. Rev.}\ }\textbf {\bibinfo {volume} {D90}},\ \bibinfo {pages}
  {023526} (\bibinfo {year} {2014})},\ \Eprint {http://arxiv.org/abs/1402.4090}
  {arXiv:1402.4090 [astro-ph.HE]} \BibitemShut {NoStop}%
\bibitem [{\citenamefont {Ackermann}\ \emph {et~al.}(2015)\citenamefont
  {Ackermann} \emph {et~al.}}]{Ackermann:2014usa}%
  \BibitemOpen
  \bibfield  {author} {\bibinfo {author} {\bibfnamefont {M.}~\bibnamefont
  {Ackermann}} \emph {et~al.} (\bibinfo {collaboration} {Fermi-LAT}),\ }\href
  {\doibase 10.1088/0004-637X/799/1/86} {\bibfield  {journal} {\bibinfo
  {journal} {Astrophys. J.}\ }\textbf {\bibinfo {volume} {799}},\ \bibinfo
  {pages} {86} (\bibinfo {year} {2015})},\ \Eprint
  {http://arxiv.org/abs/1410.3696} {arXiv:1410.3696 [astro-ph.HE]} \BibitemShut
  {NoStop}%
\bibitem [{\citenamefont {Heymans}\ \emph {et~al.}(2012)\citenamefont
  {Heymans}, \citenamefont {Van~Waerbeke}, \citenamefont {Miller},
  \citenamefont {Erben}, \citenamefont {Hildebrandt} \emph
  {et~al.}}]{Heymans:2012gg}%
  \BibitemOpen
  \bibfield  {author} {\bibinfo {author} {\bibfnamefont {C.}~\bibnamefont
  {Heymans}}, \bibinfo {author} {\bibfnamefont {L.}~\bibnamefont
  {Van~Waerbeke}}, \bibinfo {author} {\bibfnamefont {L.}~\bibnamefont
  {Miller}}, \bibinfo {author} {\bibfnamefont {T.}~\bibnamefont {Erben}},
  \bibinfo {author} {\bibfnamefont {H.}~\bibnamefont {Hildebrandt}},  \emph
  {et~al.},\ }\href@noop {} {\bibfield  {journal} {\bibinfo  {journal}
  {Mon.Not.Roy.Astron.Soc.}\ }\textbf {\bibinfo {volume} {427}},\ \bibinfo
  {pages} {146} (\bibinfo {year} {2012})},\ \Eprint
  {http://arxiv.org/abs/1210.0032} {arXiv:1210.0032 [astro-ph.CO]} \BibitemShut
  {NoStop}%
\bibitem [{\citenamefont {Hildebrandt}\ \emph {et~al.}(2016)\citenamefont
  {Hildebrandt} \emph {et~al.}}]{Hildebrandt:2016wyi}%
  \BibitemOpen
  \bibfield  {author} {\bibinfo {author} {\bibfnamefont {H.}~\bibnamefont
  {Hildebrandt}} \emph {et~al.},\ }\href@noop {} {\  (\bibinfo {year}
  {2016})},\ \Eprint {http://arxiv.org/abs/1603.07722} {arXiv:1603.07722
  [astro-ph.CO]} \BibitemShut {NoStop}%
\bibitem [{\citenamefont {Harnois-D{\'e}raps}\ \emph
  {et~al.}(2016)\citenamefont {Harnois-D{\'e}raps} \emph
  {et~al.}}]{Harnois-Deraps:2016huu}%
  \BibitemOpen
  \bibfield  {author} {\bibinfo {author} {\bibfnamefont {J.}~\bibnamefont
  {Harnois-D{\'e}raps}} \emph {et~al.},\ }\href {\doibase 10.1093/mnras/stw947}
  {\  (\bibinfo {year} {2016}),\ 10.1093/mnras/stw947},\ \Eprint
  {http://arxiv.org/abs/1603.07723} {arXiv:1603.07723 [astro-ph.CO]}
  \BibitemShut {NoStop}%
\bibitem [{\citenamefont {Hildebrandt}\ \emph {et~al.}(2012)\citenamefont
  {Hildebrandt}, \citenamefont {Erben}, \citenamefont {Kuijken}, \citenamefont
  {van Waerbeke}, \citenamefont {Heymans} \emph {et~al.}}]{Hildebrandt:2011hb}%
  \BibitemOpen
  \bibfield  {author} {\bibinfo {author} {\bibfnamefont {H.}~\bibnamefont
  {Hildebrandt}}, \bibinfo {author} {\bibfnamefont {T.}~\bibnamefont {Erben}},
  \bibinfo {author} {\bibfnamefont {K.}~\bibnamefont {Kuijken}}, \bibinfo
  {author} {\bibfnamefont {L.}~\bibnamefont {van Waerbeke}}, \bibinfo {author}
  {\bibfnamefont {C.}~\bibnamefont {Heymans}},  \emph {et~al.},\ }\href@noop {}
  {\bibfield  {journal} {\bibinfo  {journal} {Mon.Not.Roy.Astron.Soc.}\
  }\textbf {\bibinfo {volume} {421}},\ \bibinfo {pages} {2355} (\bibinfo {year}
  {2012})},\ \Eprint {http://arxiv.org/abs/1111.4434} {arXiv:1111.4434
  [astro-ph.CO]} \BibitemShut {NoStop}%
\bibitem [{\citenamefont {Erben}\ \emph {et~al.}(2013)\citenamefont {Erben},
  \citenamefont {Hildebrandt}, \citenamefont {Miller}, \citenamefont {van
  Waerbeke}, \citenamefont {Heymans} \emph {et~al.}}]{Erben:2012zw}%
  \BibitemOpen
  \bibfield  {author} {\bibinfo {author} {\bibfnamefont {T.}~\bibnamefont
  {Erben}}, \bibinfo {author} {\bibfnamefont {H.}~\bibnamefont {Hildebrandt}},
  \bibinfo {author} {\bibfnamefont {L.}~\bibnamefont {Miller}}, \bibinfo
  {author} {\bibfnamefont {L.}~\bibnamefont {van Waerbeke}}, \bibinfo {author}
  {\bibfnamefont {C.}~\bibnamefont {Heymans}},  \emph {et~al.},\ }\href@noop {}
  {\bibfield  {journal} {\bibinfo  {journal} {Mon.Not.Roy.Astron.Soc.}\
  }\textbf {\bibinfo {volume} {433}},\ \bibinfo {pages} {2545} (\bibinfo {year}
  {2013})},\ \Eprint {http://arxiv.org/abs/1210.8156} {arXiv:1210.8156
  [astro-ph.CO]} \BibitemShut {NoStop}%
\bibitem [{\citenamefont {Miller}\ \emph {et~al.}(2013)\citenamefont {Miller},
  \citenamefont {Heymans}, \citenamefont {Kitching}, \citenamefont
  {Van~Waerbeke}, \citenamefont {Erben} \emph {et~al.}}]{Miller:2012am}%
  \BibitemOpen
  \bibfield  {author} {\bibinfo {author} {\bibfnamefont {L.}~\bibnamefont
  {Miller}}, \bibinfo {author} {\bibfnamefont {C.}~\bibnamefont {Heymans}},
  \bibinfo {author} {\bibfnamefont {T.}~\bibnamefont {Kitching}}, \bibinfo
  {author} {\bibfnamefont {L.}~\bibnamefont {Van~Waerbeke}}, \bibinfo {author}
  {\bibfnamefont {T.}~\bibnamefont {Erben}},  \emph {et~al.},\ }\href {\doibase
  10.1093/mnras/sts454} {\bibfield  {journal} {\bibinfo  {journal}
  {Mon.Not.Roy.Astron.Soc.}\ }\textbf {\bibinfo {volume} {429}},\ \bibinfo
  {pages} {2858} (\bibinfo {year} {2013})},\ \Eprint
  {http://arxiv.org/abs/1210.8201} {arXiv:1210.8201 [astro-ph.CO]} \BibitemShut
  {NoStop}%
\bibitem [{\citenamefont {Benjamin}\ \emph {et~al.}(2013)\citenamefont
  {Benjamin}, \citenamefont {Van~Waerbeke}, \citenamefont {Heymans},
  \citenamefont {Kilbinger}, \citenamefont {Erben} \emph
  {et~al.}}]{Benjamin:2012qp}%
  \BibitemOpen
  \bibfield  {author} {\bibinfo {author} {\bibfnamefont {J.}~\bibnamefont
  {Benjamin}}, \bibinfo {author} {\bibfnamefont {L.}~\bibnamefont
  {Van~Waerbeke}}, \bibinfo {author} {\bibfnamefont {C.}~\bibnamefont
  {Heymans}}, \bibinfo {author} {\bibfnamefont {M.}~\bibnamefont {Kilbinger}},
  \bibinfo {author} {\bibfnamefont {T.}~\bibnamefont {Erben}},  \emph
  {et~al.},\ }\href@noop {} {\bibfield  {journal} {\bibinfo  {journal}
  {Mon.Not.Roy.Astron.Soc.}\ }\textbf {\bibinfo {volume} {431}},\ \bibinfo
  {pages} {1547} (\bibinfo {year} {2013})},\ \Eprint
  {http://arxiv.org/abs/1212.3327} {arXiv:1212.3327 [astro-ph.CO]} \BibitemShut
  {NoStop}%
\bibitem [{\citenamefont {Benitez}(2000)}]{Benitez:1998br}%
  \BibitemOpen
  \bibfield  {author} {\bibinfo {author} {\bibfnamefont {N.}~\bibnamefont
  {Benitez}},\ }\href {\doibase 10.1086/308947} {\bibfield  {journal} {\bibinfo
   {journal} {Astrophys.J.}\ }\textbf {\bibinfo {volume} {536}},\ \bibinfo
  {pages} {571} (\bibinfo {year} {2000})},\ \Eprint
  {http://arxiv.org/abs/astro-ph/9811189} {arXiv:astro-ph/9811189 [astro-ph]}
  \BibitemShut {NoStop}%
\bibitem [{\citenamefont {Gilbank}\ \emph {et~al.}(2011)\citenamefont
  {Gilbank}, \citenamefont {Gladders}, \citenamefont {Yee},\ and\ \citenamefont
  {Hsieh}}]{Gilbank:2010zv}%
  \BibitemOpen
  \bibfield  {author} {\bibinfo {author} {\bibfnamefont {D.~G.}\ \bibnamefont
  {Gilbank}}, \bibinfo {author} {\bibfnamefont {M.~D.}\ \bibnamefont
  {Gladders}}, \bibinfo {author} {\bibfnamefont {H.~K.~C.}\ \bibnamefont
  {Yee}}, \ and\ \bibinfo {author} {\bibfnamefont {B.~C.}\ \bibnamefont
  {Hsieh}},\ }\href {\doibase 10.1088/0004-6256/141/3/94} {\bibfield  {journal}
  {\bibinfo  {journal} {Astron. J.}\ }\textbf {\bibinfo {volume} {141}},\
  \bibinfo {pages} {94} (\bibinfo {year} {2011})},\ \Eprint
  {http://arxiv.org/abs/1012.3470} {arXiv:1012.3470 [astro-ph.CO]} \BibitemShut
  {NoStop}%
\bibitem [{\citenamefont {Coupon}\ \emph {et~al.}(2015)\citenamefont {Coupon}
  \emph {et~al.}}]{Coupon:2015rua}%
  \BibitemOpen
  \bibfield  {author} {\bibinfo {author} {\bibfnamefont {J.}~\bibnamefont
  {Coupon}} \emph {et~al.},\ }\href {\doibase 10.1093/mnras/stv276} {\bibfield
  {journal} {\bibinfo  {journal} {Mon. Not. Roy. Astron. Soc.}\ }\textbf
  {\bibinfo {volume} {449}},\ \bibinfo {pages} {1352} (\bibinfo {year}
  {2015})},\ \Eprint {http://arxiv.org/abs/1502.02867} {arXiv:1502.02867
  [astro-ph.CO]} \BibitemShut {NoStop}%
\bibitem [{\citenamefont {Shirasaki}\ and\ \citenamefont
  {Yoshida}(2014)}]{Shirasaki:2013zpa}%
  \BibitemOpen
  \bibfield  {author} {\bibinfo {author} {\bibfnamefont {M.}~\bibnamefont
  {Shirasaki}}\ and\ \bibinfo {author} {\bibfnamefont {N.}~\bibnamefont
  {Yoshida}},\ }\href@noop {} {\bibfield  {journal} {\bibinfo  {journal}
  {Astrophys.J.}\ }\textbf {\bibinfo {volume} {786}},\ \bibinfo {pages} {43}
  (\bibinfo {year} {2014})},\ \Eprint {http://arxiv.org/abs/1312.5032}
  {arXiv:1312.5032 [astro-ph.CO]} \BibitemShut {NoStop}%
\bibitem [{\citenamefont {Shirasaki}\ \emph
  {et~al.}(2015{\natexlab{a}})\citenamefont {Shirasaki}, \citenamefont
  {Hamana},\ and\ \citenamefont {Yoshida}}]{Shirasaki:2015dga}%
  \BibitemOpen
  \bibfield  {author} {\bibinfo {author} {\bibfnamefont {M.}~\bibnamefont
  {Shirasaki}}, \bibinfo {author} {\bibfnamefont {T.}~\bibnamefont {Hamana}}, \
  and\ \bibinfo {author} {\bibfnamefont {N.}~\bibnamefont {Yoshida}},\ }\href
  {\doibase 10.1093/mnras/stv1854} {\bibfield  {journal} {\bibinfo  {journal}
  {Mon. Not. Roy. Astron. Soc.}\ }\textbf {\bibinfo {volume} {453}},\ \bibinfo
  {pages} {3043} (\bibinfo {year} {2015}{\natexlab{a}})},\ \Eprint
  {http://arxiv.org/abs/1504.05672} {arXiv:1504.05672 [astro-ph.CO]}
  \BibitemShut {NoStop}%
\bibitem [{\citenamefont {Ackermann}\ \emph {et~al.}(2012)\citenamefont
  {Ackermann} \emph {et~al.}}]{ackermannajelloatwood2012}%
  \BibitemOpen
  \bibfield  {author} {\bibinfo {author} {\bibfnamefont {M.}~\bibnamefont
  {Ackermann}} \emph {et~al.},\ }\href
  {http://stacks.iop.org/0004-637X/750/i=1/a=3} {\bibfield  {journal} {\bibinfo
   {journal} {The Astrophysical Journal}\ }\textbf {\bibinfo {volume} {750}},\
  \bibinfo {pages} {3} (\bibinfo {year} {2012})}\BibitemShut {NoStop}%
\bibitem [{\citenamefont {{Strong}}\ \emph {et~al.}()\citenamefont {{Strong}}
  \emph {et~al.}}]{Galpropsupplementary}%
  \BibitemOpen
  \bibfield  {author} {\bibinfo {author} {\bibfnamefont {A.~W.}\ \bibnamefont
  {{Strong}}} \emph {et~al.},\ }\href@noop {} {\enquote {\bibinfo {title}
  {Galprop version 54: Explanatory supplement},}\ }\bibinfo {howpublished}
  {\protect\url{http://galprop.stanford.edu}},\ \bibinfo {note} {accessed:
  2014-02-20}\BibitemShut {NoStop}%
\bibitem [{\citenamefont {Shirasaki}\ \emph
  {et~al.}(2015{\natexlab{b}})\citenamefont {Shirasaki}, \citenamefont
  {Horiuchi},\ and\ \citenamefont {Yoshida}}]{Shirasaki:2015nqp}%
  \BibitemOpen
  \bibfield  {author} {\bibinfo {author} {\bibfnamefont {M.}~\bibnamefont
  {Shirasaki}}, \bibinfo {author} {\bibfnamefont {S.}~\bibnamefont {Horiuchi}},
  \ and\ \bibinfo {author} {\bibfnamefont {N.}~\bibnamefont {Yoshida}},\ }\href
  {\doibase 10.1103/PhysRevD.92.123540} {\bibfield  {journal} {\bibinfo
  {journal} {Phys. Rev.}\ }\textbf {\bibinfo {volume} {D92}},\ \bibinfo {pages}
  {123540} (\bibinfo {year} {2015}{\natexlab{b}})},\ \Eprint
  {http://arxiv.org/abs/1511.07092} {arXiv:1511.07092 [astro-ph.CO]}
  \BibitemShut {NoStop}%
\bibitem [{\citenamefont {{Steigman}}\ \emph {et~al.}(2012)\citenamefont
  {{Steigman}}, \citenamefont {{Dasgupta}},\ and\ \citenamefont
  {{Beacom}}}]{2012PhRvD..86b3506S}%
  \BibitemOpen
  \bibfield  {author} {\bibinfo {author} {\bibfnamefont {G.}~\bibnamefont
  {{Steigman}}}, \bibinfo {author} {\bibfnamefont {B.}~\bibnamefont
  {{Dasgupta}}}, \ and\ \bibinfo {author} {\bibfnamefont {J.~F.}\ \bibnamefont
  {{Beacom}}},\ }\href {\doibase 10.1103/PhysRevD.86.023506} {\bibfield
  {journal} {\bibinfo  {journal} {\prd}\ }\textbf {\bibinfo {volume} {86}},\
  \bibinfo {eid} {023506} (\bibinfo {year} {2012})},\ \Eprint
  {http://arxiv.org/abs/1204.3622} {arXiv:1204.3622 [hep-ph]} \BibitemShut
  {NoStop}%
\bibitem [{\citenamefont {Hryczuk}\ and\ \citenamefont
  {Iengo}(2012)}]{Hryczuk:2011vi}%
  \BibitemOpen
  \bibfield  {author} {\bibinfo {author} {\bibfnamefont {A.}~\bibnamefont
  {Hryczuk}}\ and\ \bibinfo {author} {\bibfnamefont {R.}~\bibnamefont
  {Iengo}},\ }\href {\doibase 10.1007/JHEP01(2012)163, 10.1007/JHEP06(2012)137}
  {\bibfield  {journal} {\bibinfo  {journal} {JHEP}\ }\textbf {\bibinfo
  {volume} {01}},\ \bibinfo {pages} {163} (\bibinfo {year} {2012})},\ \bibinfo
  {note} {[Erratum: JHEP06,137(2012)]},\ \Eprint
  {http://arxiv.org/abs/1111.2916} {arXiv:1111.2916 [hep-ph]} \BibitemShut
  {NoStop}%
\bibitem [{\citenamefont {Randall}\ and\ \citenamefont
  {Sundrum}(1999)}]{Randall:1998uk}%
  \BibitemOpen
  \bibfield  {author} {\bibinfo {author} {\bibfnamefont {L.}~\bibnamefont
  {Randall}}\ and\ \bibinfo {author} {\bibfnamefont {R.}~\bibnamefont
  {Sundrum}},\ }\href {\doibase 10.1016/S0550-3213(99)00359-4} {\bibfield
  {journal} {\bibinfo  {journal} {Nucl. Phys.}\ }\textbf {\bibinfo {volume}
  {B557}},\ \bibinfo {pages} {79} (\bibinfo {year} {1999})},\ \Eprint
  {http://arxiv.org/abs/hep-th/9810155} {arXiv:hep-th/9810155 [hep-th]}
  \BibitemShut {NoStop}%
\bibitem [{\citenamefont {Giudice}\ \emph {et~al.}(1998)\citenamefont
  {Giudice}, \citenamefont {Luty}, \citenamefont {Murayama},\ and\
  \citenamefont {Rattazzi}}]{Giudice:1998xp}%
  \BibitemOpen
  \bibfield  {author} {\bibinfo {author} {\bibfnamefont {G.~F.}\ \bibnamefont
  {Giudice}}, \bibinfo {author} {\bibfnamefont {M.~A.}\ \bibnamefont {Luty}},
  \bibinfo {author} {\bibfnamefont {H.}~\bibnamefont {Murayama}}, \ and\
  \bibinfo {author} {\bibfnamefont {R.}~\bibnamefont {Rattazzi}},\ }\href
  {\doibase 10.1088/1126-6708/1998/12/027} {\bibfield  {journal} {\bibinfo
  {journal} {JHEP}\ }\textbf {\bibinfo {volume} {12}},\ \bibinfo {pages} {027}
  (\bibinfo {year} {1998})},\ \Eprint {http://arxiv.org/abs/hep-ph/9810442}
  {arXiv:hep-ph/9810442 [hep-ph]} \BibitemShut {NoStop}%
\bibitem [{\citenamefont {Aad}\ \emph {et~al.}(2012)\citenamefont {Aad} \emph
  {et~al.}}]{Aad:2012tfa}%
  \BibitemOpen
  \bibfield  {author} {\bibinfo {author} {\bibfnamefont {G.}~\bibnamefont
  {Aad}} \emph {et~al.} (\bibinfo {collaboration} {ATLAS}),\ }\href {\doibase
  10.1016/j.physletb.2012.08.020} {\bibfield  {journal} {\bibinfo  {journal}
  {Phys. Lett.}\ }\textbf {\bibinfo {volume} {B716}},\ \bibinfo {pages} {1}
  (\bibinfo {year} {2012})},\ \Eprint {http://arxiv.org/abs/1207.7214}
  {arXiv:1207.7214 [hep-ex]} \BibitemShut {NoStop}%
\bibitem [{\citenamefont {Chatrchyan}\ \emph {et~al.}(2012)\citenamefont
  {Chatrchyan} \emph {et~al.}}]{Chatrchyan:2012xdj}%
  \BibitemOpen
  \bibfield  {author} {\bibinfo {author} {\bibfnamefont {S.}~\bibnamefont
  {Chatrchyan}} \emph {et~al.} (\bibinfo {collaboration} {CMS}),\ }\href
  {\doibase 10.1016/j.physletb.2012.08.021} {\bibfield  {journal} {\bibinfo
  {journal} {Phys. Lett.}\ }\textbf {\bibinfo {volume} {B716}},\ \bibinfo
  {pages} {30} (\bibinfo {year} {2012})},\ \Eprint
  {http://arxiv.org/abs/1207.7235} {arXiv:1207.7235 [hep-ex]} \BibitemShut
  {NoStop}%
\bibitem [{\citenamefont {Hall}\ and\ \citenamefont
  {Nomura}(2012)}]{Hall:2011jd}%
  \BibitemOpen
  \bibfield  {author} {\bibinfo {author} {\bibfnamefont {L.~J.}\ \bibnamefont
  {Hall}}\ and\ \bibinfo {author} {\bibfnamefont {Y.}~\bibnamefont {Nomura}},\
  }\href {\doibase 10.1007/JHEP01(2012)082} {\bibfield  {journal} {\bibinfo
  {journal} {JHEP}\ }\textbf {\bibinfo {volume} {01}},\ \bibinfo {pages} {082}
  (\bibinfo {year} {2012})},\ \Eprint {http://arxiv.org/abs/1111.4519}
  {arXiv:1111.4519 [hep-ph]} \BibitemShut {NoStop}%
\bibitem [{\citenamefont {Hall}\ \emph {et~al.}(2013)\citenamefont {Hall},
  \citenamefont {Nomura},\ and\ \citenamefont {Shirai}}]{Hall:2012zp}%
  \BibitemOpen
  \bibfield  {author} {\bibinfo {author} {\bibfnamefont {L.~J.}\ \bibnamefont
  {Hall}}, \bibinfo {author} {\bibfnamefont {Y.}~\bibnamefont {Nomura}}, \ and\
  \bibinfo {author} {\bibfnamefont {S.}~\bibnamefont {Shirai}},\ }\href
  {\doibase 10.1007/JHEP01(2013)036} {\bibfield  {journal} {\bibinfo  {journal}
  {JHEP}\ }\textbf {\bibinfo {volume} {01}},\ \bibinfo {pages} {036} (\bibinfo
  {year} {2013})},\ \Eprint {http://arxiv.org/abs/1210.2395} {arXiv:1210.2395
  [hep-ph]} \BibitemShut {NoStop}%
\bibitem [{\citenamefont {Ibe}\ and\ \citenamefont
  {Yanagida}(2012)}]{Ibe:2011aa}%
  \BibitemOpen
  \bibfield  {author} {\bibinfo {author} {\bibfnamefont {M.}~\bibnamefont
  {Ibe}}\ and\ \bibinfo {author} {\bibfnamefont {T.~T.}\ \bibnamefont
  {Yanagida}},\ }\href {\doibase 10.1016/j.physletb.2012.02.034} {\bibfield
  {journal} {\bibinfo  {journal} {Phys. Lett.}\ }\textbf {\bibinfo {volume}
  {B709}},\ \bibinfo {pages} {374} (\bibinfo {year} {2012})},\ \Eprint
  {http://arxiv.org/abs/1112.2462} {arXiv:1112.2462 [hep-ph]} \BibitemShut
  {NoStop}%
\bibitem [{\citenamefont {Ibe}\ \emph {et~al.}(2012)\citenamefont {Ibe},
  \citenamefont {Matsumoto},\ and\ \citenamefont {Yanagida}}]{Ibe:2012hu}%
  \BibitemOpen
  \bibfield  {author} {\bibinfo {author} {\bibfnamefont {M.}~\bibnamefont
  {Ibe}}, \bibinfo {author} {\bibfnamefont {S.}~\bibnamefont {Matsumoto}}, \
  and\ \bibinfo {author} {\bibfnamefont {T.~T.}\ \bibnamefont {Yanagida}},\
  }\href {\doibase 10.1103/PhysRevD.85.095011} {\bibfield  {journal} {\bibinfo
  {journal} {Phys. Rev.}\ }\textbf {\bibinfo {volume} {D85}},\ \bibinfo {pages}
  {095011} (\bibinfo {year} {2012})},\ \Eprint {http://arxiv.org/abs/1202.2253}
  {arXiv:1202.2253 [hep-ph]} \BibitemShut {NoStop}%
\bibitem [{\citenamefont {Arvanitaki}\ \emph {et~al.}(2013)\citenamefont
  {Arvanitaki}, \citenamefont {Craig}, \citenamefont {Dimopoulos},\ and\
  \citenamefont {Villadoro}}]{Arvanitaki:2012ps}%
  \BibitemOpen
  \bibfield  {author} {\bibinfo {author} {\bibfnamefont {A.}~\bibnamefont
  {Arvanitaki}}, \bibinfo {author} {\bibfnamefont {N.}~\bibnamefont {Craig}},
  \bibinfo {author} {\bibfnamefont {S.}~\bibnamefont {Dimopoulos}}, \ and\
  \bibinfo {author} {\bibfnamefont {G.}~\bibnamefont {Villadoro}},\ }\href
  {\doibase 10.1007/JHEP02(2013)126} {\bibfield  {journal} {\bibinfo  {journal}
  {JHEP}\ }\textbf {\bibinfo {volume} {02}},\ \bibinfo {pages} {126} (\bibinfo
  {year} {2013})},\ \Eprint {http://arxiv.org/abs/1210.0555} {arXiv:1210.0555
  [hep-ph]} \BibitemShut {NoStop}%
\bibitem [{\citenamefont {Arkani-Hamed}\ \emph {et~al.}(2012)\citenamefont
  {Arkani-Hamed}, \citenamefont {Gupta}, \citenamefont {Kaplan}, \citenamefont
  {Weiner},\ and\ \citenamefont {Zorawski}}]{ArkaniHamed:2012gw}%
  \BibitemOpen
  \bibfield  {author} {\bibinfo {author} {\bibfnamefont {N.}~\bibnamefont
  {Arkani-Hamed}}, \bibinfo {author} {\bibfnamefont {A.}~\bibnamefont {Gupta}},
  \bibinfo {author} {\bibfnamefont {D.~E.}\ \bibnamefont {Kaplan}}, \bibinfo
  {author} {\bibfnamefont {N.}~\bibnamefont {Weiner}}, \ and\ \bibinfo {author}
  {\bibfnamefont {T.}~\bibnamefont {Zorawski}},\ }\href@noop {} {\  (\bibinfo
  {year} {2012})},\ \Eprint {http://arxiv.org/abs/1212.6971} {arXiv:1212.6971
  [hep-ph]} \BibitemShut {NoStop}%
\bibitem [{\citenamefont {Nomura}\ and\ \citenamefont
  {Shirai}(2014)}]{Nomura:2014asa}%
  \BibitemOpen
  \bibfield  {author} {\bibinfo {author} {\bibfnamefont {Y.}~\bibnamefont
  {Nomura}}\ and\ \bibinfo {author} {\bibfnamefont {S.}~\bibnamefont
  {Shirai}},\ }\href {\doibase 10.1103/PhysRevLett.113.111801} {\bibfield
  {journal} {\bibinfo  {journal} {Phys. Rev. Lett.}\ }\textbf {\bibinfo
  {volume} {113}},\ \bibinfo {pages} {111801} (\bibinfo {year} {2014})},\
  \Eprint {http://arxiv.org/abs/1407.3785} {arXiv:1407.3785 [hep-ph]}
  \BibitemShut {NoStop}%
\bibitem [{\citenamefont {Okada}\ \emph
  {et~al.}(1991{\natexlab{a}})\citenamefont {Okada}, \citenamefont
  {Yamaguchi},\ and\ \citenamefont {Yanagida}}]{Okada:1990vk}%
  \BibitemOpen
  \bibfield  {author} {\bibinfo {author} {\bibfnamefont {Y.}~\bibnamefont
  {Okada}}, \bibinfo {author} {\bibfnamefont {M.}~\bibnamefont {Yamaguchi}}, \
  and\ \bibinfo {author} {\bibfnamefont {T.}~\bibnamefont {Yanagida}},\ }\href
  {\doibase 10.1143/PTP.85.1} {\bibfield  {journal} {\bibinfo  {journal} {Prog.
  Theor. Phys.}\ }\textbf {\bibinfo {volume} {85}},\ \bibinfo {pages} {1}
  (\bibinfo {year} {1991}{\natexlab{a}})}\BibitemShut {NoStop}%
\bibitem [{\citenamefont {Okada}\ \emph
  {et~al.}(1991{\natexlab{b}})\citenamefont {Okada}, \citenamefont
  {Yamaguchi},\ and\ \citenamefont {Yanagida}}]{Okada:1990gg}%
  \BibitemOpen
  \bibfield  {author} {\bibinfo {author} {\bibfnamefont {Y.}~\bibnamefont
  {Okada}}, \bibinfo {author} {\bibfnamefont {M.}~\bibnamefont {Yamaguchi}}, \
  and\ \bibinfo {author} {\bibfnamefont {T.}~\bibnamefont {Yanagida}},\ }\href
  {\doibase 10.1016/0370-2693(91)90642-4} {\bibfield  {journal} {\bibinfo
  {journal} {Phys. Lett.}\ }\textbf {\bibinfo {volume} {B262}},\ \bibinfo
  {pages} {54} (\bibinfo {year} {1991}{\natexlab{b}})}\BibitemShut {NoStop}%
\bibitem [{\citenamefont {Ellis}\ \emph
  {et~al.}(1991{\natexlab{a}})\citenamefont {Ellis}, \citenamefont {Ridolfi},\
  and\ \citenamefont {Zwirner}}]{Ellis:1990nz}%
  \BibitemOpen
  \bibfield  {author} {\bibinfo {author} {\bibfnamefont {J.~R.}\ \bibnamefont
  {Ellis}}, \bibinfo {author} {\bibfnamefont {G.}~\bibnamefont {Ridolfi}}, \
  and\ \bibinfo {author} {\bibfnamefont {F.}~\bibnamefont {Zwirner}},\ }\href
  {\doibase 10.1016/0370-2693(91)90863-L} {\bibfield  {journal} {\bibinfo
  {journal} {Phys. Lett.}\ }\textbf {\bibinfo {volume} {B257}},\ \bibinfo
  {pages} {83} (\bibinfo {year} {1991}{\natexlab{a}})}\BibitemShut {NoStop}%
\bibitem [{\citenamefont {Haber}\ and\ \citenamefont
  {Hempfling}(1991)}]{Haber:1990aw}%
  \BibitemOpen
  \bibfield  {author} {\bibinfo {author} {\bibfnamefont {H.~E.}\ \bibnamefont
  {Haber}}\ and\ \bibinfo {author} {\bibfnamefont {R.}~\bibnamefont
  {Hempfling}},\ }\href {\doibase 10.1103/PhysRevLett.66.1815} {\bibfield
  {journal} {\bibinfo  {journal} {Phys. Rev. Lett.}\ }\textbf {\bibinfo
  {volume} {66}},\ \bibinfo {pages} {1815} (\bibinfo {year}
  {1991})}\BibitemShut {NoStop}%
\bibitem [{\citenamefont {Ellis}\ \emph
  {et~al.}(1991{\natexlab{b}})\citenamefont {Ellis}, \citenamefont {Ridolfi},\
  and\ \citenamefont {Zwirner}}]{Ellis:1991zd}%
  \BibitemOpen
  \bibfield  {author} {\bibinfo {author} {\bibfnamefont {J.~R.}\ \bibnamefont
  {Ellis}}, \bibinfo {author} {\bibfnamefont {G.}~\bibnamefont {Ridolfi}}, \
  and\ \bibinfo {author} {\bibfnamefont {F.}~\bibnamefont {Zwirner}},\ }\href
  {\doibase 10.1016/0370-2693(91)90626-2} {\bibfield  {journal} {\bibinfo
  {journal} {Phys. Lett.}\ }\textbf {\bibinfo {volume} {B262}},\ \bibinfo
  {pages} {477} (\bibinfo {year} {1991}{\natexlab{b}})}\BibitemShut {NoStop}%
\bibitem [{\citenamefont {Hisano}\ \emph {et~al.}(2007)\citenamefont {Hisano},
  \citenamefont {Matsumoto}, \citenamefont {Nagai}, \citenamefont {Saito},\
  and\ \citenamefont {Senami}}]{Hisano:2006nn}%
  \BibitemOpen
  \bibfield  {author} {\bibinfo {author} {\bibfnamefont {J.}~\bibnamefont
  {Hisano}}, \bibinfo {author} {\bibfnamefont {S.}~\bibnamefont {Matsumoto}},
  \bibinfo {author} {\bibfnamefont {M.}~\bibnamefont {Nagai}}, \bibinfo
  {author} {\bibfnamefont {O.}~\bibnamefont {Saito}}, \ and\ \bibinfo {author}
  {\bibfnamefont {M.}~\bibnamefont {Senami}},\ }\href {\doibase
  10.1016/j.physletb.2007.01.012} {\bibfield  {journal} {\bibinfo  {journal}
  {Phys. Lett.}\ }\textbf {\bibinfo {volume} {B646}},\ \bibinfo {pages} {34}
  (\bibinfo {year} {2007})},\ \Eprint {http://arxiv.org/abs/hep-ph/0610249}
  {arXiv:hep-ph/0610249 [hep-ph]} \BibitemShut {NoStop}%
\bibitem [{\citenamefont {Aad}\ \emph {et~al.}(2013)\citenamefont {Aad} \emph
  {et~al.}}]{Aad:2013yna}%
  \BibitemOpen
  \bibfield  {author} {\bibinfo {author} {\bibfnamefont {G.}~\bibnamefont
  {Aad}} \emph {et~al.} (\bibinfo {collaboration} {ATLAS}),\ }\href {\doibase
  10.1103/PhysRevD.88.112006} {\bibfield  {journal} {\bibinfo  {journal} {Phys.
  Rev.}\ }\textbf {\bibinfo {volume} {D88}},\ \bibinfo {pages} {112006}
  (\bibinfo {year} {2013})},\ \Eprint {http://arxiv.org/abs/1310.3675}
  {arXiv:1310.3675 [hep-ex]} \BibitemShut {NoStop}%
\bibitem [{\citenamefont {Khachatryan}\ \emph {et~al.}(2015)\citenamefont
  {Khachatryan} \emph {et~al.}}]{CMS:2014gxa}%
  \BibitemOpen
  \bibfield  {author} {\bibinfo {author} {\bibfnamefont {V.}~\bibnamefont
  {Khachatryan}} \emph {et~al.} (\bibinfo {collaboration} {CMS}),\ }\href
  {\doibase 10.1007/JHEP01(2015)096} {\bibfield  {journal} {\bibinfo  {journal}
  {JHEP}\ }\textbf {\bibinfo {volume} {01}},\ \bibinfo {pages} {096} (\bibinfo
  {year} {2015})},\ \Eprint {http://arxiv.org/abs/1411.6006} {arXiv:1411.6006
  [hep-ex]} \BibitemShut {NoStop}%
\bibitem [{\citenamefont {Hisano}\ \emph {et~al.}(2010)\citenamefont {Hisano},
  \citenamefont {Ishiwata},\ and\ \citenamefont {Nagata}}]{Hisano:2010fy}%
  \BibitemOpen
  \bibfield  {author} {\bibinfo {author} {\bibfnamefont {J.}~\bibnamefont
  {Hisano}}, \bibinfo {author} {\bibfnamefont {K.}~\bibnamefont {Ishiwata}}, \
  and\ \bibinfo {author} {\bibfnamefont {N.}~\bibnamefont {Nagata}},\ }\href
  {\doibase 10.1016/j.physletb.2010.05.047} {\bibfield  {journal} {\bibinfo
  {journal} {Phys. Lett.}\ }\textbf {\bibinfo {volume} {B690}},\ \bibinfo
  {pages} {311} (\bibinfo {year} {2010})},\ \Eprint
  {http://arxiv.org/abs/1004.4090} {arXiv:1004.4090 [hep-ph]} \BibitemShut
  {NoStop}%
\bibitem [{\citenamefont {Hisano}\ \emph {et~al.}(2013)\citenamefont {Hisano},
  \citenamefont {Ishiwata},\ and\ \citenamefont {Nagata}}]{Hisano:2012wm}%
  \BibitemOpen
  \bibfield  {author} {\bibinfo {author} {\bibfnamefont {J.}~\bibnamefont
  {Hisano}}, \bibinfo {author} {\bibfnamefont {K.}~\bibnamefont {Ishiwata}}, \
  and\ \bibinfo {author} {\bibfnamefont {N.}~\bibnamefont {Nagata}},\ }\href
  {\doibase 10.1103/PhysRevD.87.035020} {\bibfield  {journal} {\bibinfo
  {journal} {Phys. Rev.}\ }\textbf {\bibinfo {volume} {D87}},\ \bibinfo {pages}
  {035020} (\bibinfo {year} {2013})},\ \Eprint {http://arxiv.org/abs/1210.5985}
  {arXiv:1210.5985 [hep-ph]} \BibitemShut {NoStop}%
\bibitem [{\citenamefont {Hisano}\ \emph {et~al.}(2015)\citenamefont {Hisano},
  \citenamefont {Ishiwata},\ and\ \citenamefont {Nagata}}]{Hisano:2015rsa}%
  \BibitemOpen
  \bibfield  {author} {\bibinfo {author} {\bibfnamefont {J.}~\bibnamefont
  {Hisano}}, \bibinfo {author} {\bibfnamefont {K.}~\bibnamefont {Ishiwata}}, \
  and\ \bibinfo {author} {\bibfnamefont {N.}~\bibnamefont {Nagata}},\ }\href
  {\doibase 10.1007/JHEP06(2015)097} {\bibfield  {journal} {\bibinfo  {journal}
  {JHEP}\ }\textbf {\bibinfo {volume} {06}},\ \bibinfo {pages} {097} (\bibinfo
  {year} {2015})},\ \Eprint {http://arxiv.org/abs/1504.00915} {arXiv:1504.00915
  [hep-ph]} \BibitemShut {NoStop}%
\bibitem [{\citenamefont {Hisano}\ \emph {et~al.}(2004)\citenamefont {Hisano},
  \citenamefont {Matsumoto},\ and\ \citenamefont {Nojiri}}]{Hisano:2003ec}%
  \BibitemOpen
  \bibfield  {author} {\bibinfo {author} {\bibfnamefont {J.}~\bibnamefont
  {Hisano}}, \bibinfo {author} {\bibfnamefont {S.}~\bibnamefont {Matsumoto}}, \
  and\ \bibinfo {author} {\bibfnamefont {M.~M.}\ \bibnamefont {Nojiri}},\
  }\href {\doibase 10.1103/PhysRevLett.92.031303} {\bibfield  {journal}
  {\bibinfo  {journal} {Phys. Rev. Lett.}\ }\textbf {\bibinfo {volume} {92}},\
  \bibinfo {pages} {031303} (\bibinfo {year} {2004})},\ \Eprint
  {http://arxiv.org/abs/hep-ph/0307216} {arXiv:hep-ph/0307216 [hep-ph]}
  \BibitemShut {NoStop}%
\bibitem [{\citenamefont {Hisano}\ \emph {et~al.}(2005)\citenamefont {Hisano},
  \citenamefont {Matsumoto}, \citenamefont {Nojiri},\ and\ \citenamefont
  {Saito}}]{Hisano:2004ds}%
  \BibitemOpen
  \bibfield  {author} {\bibinfo {author} {\bibfnamefont {J.}~\bibnamefont
  {Hisano}}, \bibinfo {author} {\bibfnamefont {S.}~\bibnamefont {Matsumoto}},
  \bibinfo {author} {\bibfnamefont {M.~M.}\ \bibnamefont {Nojiri}}, \ and\
  \bibinfo {author} {\bibfnamefont {O.}~\bibnamefont {Saito}},\ }\href
  {\doibase 10.1103/PhysRevD.71.063528} {\bibfield  {journal} {\bibinfo
  {journal} {Phys. Rev.}\ }\textbf {\bibinfo {volume} {D71}},\ \bibinfo {pages}
  {063528} (\bibinfo {year} {2005})},\ \Eprint
  {http://arxiv.org/abs/hep-ph/0412403} {arXiv:hep-ph/0412403 [hep-ph]}
  \BibitemShut {NoStop}%
\bibitem [{\citenamefont {Cohen}\ \emph {et~al.}(2013)\citenamefont {Cohen},
  \citenamefont {Lisanti}, \citenamefont {Pierce},\ and\ \citenamefont
  {Slatyer}}]{Cohen:2013ama}%
  \BibitemOpen
  \bibfield  {author} {\bibinfo {author} {\bibfnamefont {T.}~\bibnamefont
  {Cohen}}, \bibinfo {author} {\bibfnamefont {M.}~\bibnamefont {Lisanti}},
  \bibinfo {author} {\bibfnamefont {A.}~\bibnamefont {Pierce}}, \ and\ \bibinfo
  {author} {\bibfnamefont {T.~R.}\ \bibnamefont {Slatyer}},\ }\href {\doibase
  10.1088/1475-7516/2013/10/061} {\bibfield  {journal} {\bibinfo  {journal}
  {JCAP}\ }\textbf {\bibinfo {volume} {1310}},\ \bibinfo {pages} {061}
  (\bibinfo {year} {2013})},\ \Eprint {http://arxiv.org/abs/1307.4082}
  {arXiv:1307.4082} \BibitemShut {NoStop}%
\bibitem [{\citenamefont {Fan}\ and\ \citenamefont
  {Reece}(2013)}]{Fan:2013faa}%
  \BibitemOpen
  \bibfield  {author} {\bibinfo {author} {\bibfnamefont {J.}~\bibnamefont
  {Fan}}\ and\ \bibinfo {author} {\bibfnamefont {M.}~\bibnamefont {Reece}},\
  }\href {\doibase 10.1007/JHEP10(2013)124} {\bibfield  {journal} {\bibinfo
  {journal} {JHEP}\ }\textbf {\bibinfo {volume} {10}},\ \bibinfo {pages} {124}
  (\bibinfo {year} {2013})},\ \Eprint {http://arxiv.org/abs/1307.4400}
  {arXiv:1307.4400 [hep-ph]} \BibitemShut {NoStop}%
\bibitem [{\citenamefont {Hryczuk}\ \emph {et~al.}(2014)\citenamefont
  {Hryczuk}, \citenamefont {Cholis}, \citenamefont {Iengo}, \citenamefont
  {Tavakoli},\ and\ \citenamefont {Ullio}}]{Hryczuk:2014hpa}%
  \BibitemOpen
  \bibfield  {author} {\bibinfo {author} {\bibfnamefont {A.}~\bibnamefont
  {Hryczuk}}, \bibinfo {author} {\bibfnamefont {I.}~\bibnamefont {Cholis}},
  \bibinfo {author} {\bibfnamefont {R.}~\bibnamefont {Iengo}}, \bibinfo
  {author} {\bibfnamefont {M.}~\bibnamefont {Tavakoli}}, \ and\ \bibinfo
  {author} {\bibfnamefont {P.}~\bibnamefont {Ullio}},\ }\href {\doibase
  10.1088/1475-7516/2014/07/031} {\bibfield  {journal} {\bibinfo  {journal}
  {JCAP}\ }\textbf {\bibinfo {volume} {1407}},\ \bibinfo {pages} {031}
  (\bibinfo {year} {2014})},\ \Eprint {http://arxiv.org/abs/1401.6212}
  {arXiv:1401.6212 [astro-ph.HE]} \BibitemShut {NoStop}%
\bibitem [{\citenamefont {Bhattacherjee}\ \emph {et~al.}(2014)\citenamefont
  {Bhattacherjee}, \citenamefont {Ibe}, \citenamefont {Ichikawa}, \citenamefont
  {Matsumoto},\ and\ \citenamefont {Nishiyama}}]{Bhattacherjee:2014dya}%
  \BibitemOpen
  \bibfield  {author} {\bibinfo {author} {\bibfnamefont {B.}~\bibnamefont
  {Bhattacherjee}}, \bibinfo {author} {\bibfnamefont {M.}~\bibnamefont {Ibe}},
  \bibinfo {author} {\bibfnamefont {K.}~\bibnamefont {Ichikawa}}, \bibinfo
  {author} {\bibfnamefont {S.}~\bibnamefont {Matsumoto}}, \ and\ \bibinfo
  {author} {\bibfnamefont {K.}~\bibnamefont {Nishiyama}},\ }\href {\doibase
  10.1007/JHEP07(2014)080} {\bibfield  {journal} {\bibinfo  {journal} {JHEP}\
  }\textbf {\bibinfo {volume} {07}},\ \bibinfo {pages} {080} (\bibinfo {year}
  {2014})},\ \Eprint {http://arxiv.org/abs/1405.4914} {arXiv:1405.4914
  [hep-ph]} \BibitemShut {NoStop}%
\bibitem [{\citenamefont {Ibe}\ \emph {et~al.}(2015)\citenamefont {Ibe},
  \citenamefont {Matsumoto}, \citenamefont {Shirai},\ and\ \citenamefont
  {Yanagida}}]{Ibe:2015tma}%
  \BibitemOpen
  \bibfield  {author} {\bibinfo {author} {\bibfnamefont {M.}~\bibnamefont
  {Ibe}}, \bibinfo {author} {\bibfnamefont {S.}~\bibnamefont {Matsumoto}},
  \bibinfo {author} {\bibfnamefont {S.}~\bibnamefont {Shirai}}, \ and\ \bibinfo
  {author} {\bibfnamefont {T.~T.}\ \bibnamefont {Yanagida}},\ }\href {\doibase
  10.1103/PhysRevD.91.111701} {\bibfield  {journal} {\bibinfo  {journal} {Phys.
  Rev.}\ }\textbf {\bibinfo {volume} {D91}},\ \bibinfo {pages} {111701}
  (\bibinfo {year} {2015})},\ \Eprint {http://arxiv.org/abs/1504.05554}
  {arXiv:1504.05554 [hep-ph]} \BibitemShut {NoStop}%
\bibitem [{Note5()}]{Note5}%
  \BibitemOpen
  \bibinfo {note} {If we adopted the Planck results \cite {Ade:2015xua}, we
  would have derived slightly tighter constraints because of the difference in
  $\Omega _{\protect \rm m0}$ and $\sigma _{8}$}\BibitemShut {NoStop}%
\bibitem [{\citenamefont {Su}\ \emph {et~al.}(2010)\citenamefont {Su},
  \citenamefont {Slatyer},\ and\ \citenamefont {Finkbeiner}}]{Su:2010qj}%
  \BibitemOpen
  \bibfield  {author} {\bibinfo {author} {\bibfnamefont {M.}~\bibnamefont
  {Su}}, \bibinfo {author} {\bibfnamefont {T.~R.}\ \bibnamefont {Slatyer}}, \
  and\ \bibinfo {author} {\bibfnamefont {D.~P.}\ \bibnamefont {Finkbeiner}},\
  }\href {\doibase 10.1088/0004-637X/724/2/1044} {\bibfield  {journal}
  {\bibinfo  {journal} {Astrophys. J.}\ }\textbf {\bibinfo {volume} {724}},\
  \bibinfo {pages} {1044} (\bibinfo {year} {2010})},\ \Eprint
  {http://arxiv.org/abs/1005.5480} {arXiv:1005.5480 [astro-ph.HE]} \BibitemShut
  {NoStop}%
\bibitem [{\citenamefont {Goodenough}\ and\ \citenamefont
  {Hooper}(2009)}]{Goodenough:2009gk}%
  \BibitemOpen
  \bibfield  {author} {\bibinfo {author} {\bibfnamefont {L.}~\bibnamefont
  {Goodenough}}\ and\ \bibinfo {author} {\bibfnamefont {D.}~\bibnamefont
  {Hooper}},\ }\href@noop {} {\  (\bibinfo {year} {2009})},\ \Eprint
  {http://arxiv.org/abs/0910.2998} {arXiv:0910.2998 [hep-ph]} \BibitemShut
  {NoStop}%
\bibitem [{\citenamefont {Hooper}\ and\ \citenamefont
  {Goodenough}(2011)}]{Hooper:2010mq}%
  \BibitemOpen
  \bibfield  {author} {\bibinfo {author} {\bibfnamefont {D.}~\bibnamefont
  {Hooper}}\ and\ \bibinfo {author} {\bibfnamefont {L.}~\bibnamefont
  {Goodenough}},\ }\href {\doibase 10.1016/j.physletb.2011.02.029} {\bibfield
  {journal} {\bibinfo  {journal} {Phys.Lett.}\ }\textbf {\bibinfo {volume}
  {B697}},\ \bibinfo {pages} {412} (\bibinfo {year} {2011})},\ \Eprint
  {http://arxiv.org/abs/1010.2752} {arXiv:1010.2752 [hep-ph]} \BibitemShut
  {NoStop}%
\bibitem [{\citenamefont {Boyarsky}\ \emph {et~al.}(2011)\citenamefont
  {Boyarsky}, \citenamefont {Malyshev},\ and\ \citenamefont
  {Ruchayskiy}}]{Boyarsky:2010dr}%
  \BibitemOpen
  \bibfield  {author} {\bibinfo {author} {\bibfnamefont {A.}~\bibnamefont
  {Boyarsky}}, \bibinfo {author} {\bibfnamefont {D.}~\bibnamefont {Malyshev}},
  \ and\ \bibinfo {author} {\bibfnamefont {O.}~\bibnamefont {Ruchayskiy}},\
  }\href {\doibase 10.1016/j.physletb.2011.10.014} {\bibfield  {journal}
  {\bibinfo  {journal} {Phys.Lett.}\ }\textbf {\bibinfo {volume} {B705}},\
  \bibinfo {pages} {165} (\bibinfo {year} {2011})},\ \Eprint
  {http://arxiv.org/abs/1012.5839} {arXiv:1012.5839 [hep-ph]} \BibitemShut
  {NoStop}%
\bibitem [{\citenamefont {Hooper}\ and\ \citenamefont
  {Linden}(2011)}]{Hooper:2011ti}%
  \BibitemOpen
  \bibfield  {author} {\bibinfo {author} {\bibfnamefont {D.}~\bibnamefont
  {Hooper}}\ and\ \bibinfo {author} {\bibfnamefont {T.}~\bibnamefont
  {Linden}},\ }\href {\doibase 10.1103/PhysRevD.84.123005} {\bibfield
  {journal} {\bibinfo  {journal} {Phys.Rev.}\ }\textbf {\bibinfo {volume}
  {D84}},\ \bibinfo {pages} {123005} (\bibinfo {year} {2011})},\ \Eprint
  {http://arxiv.org/abs/1110.0006} {arXiv:1110.0006 [astro-ph.HE]} \BibitemShut
  {NoStop}%
\bibitem [{\citenamefont {Abazajian}\ and\ \citenamefont
  {Kaplinghat}(2012)}]{Abazajian:2012pn}%
  \BibitemOpen
  \bibfield  {author} {\bibinfo {author} {\bibfnamefont {K.~N.}\ \bibnamefont
  {Abazajian}}\ and\ \bibinfo {author} {\bibfnamefont {M.}~\bibnamefont
  {Kaplinghat}},\ }\href {\doibase 10.1103/PhysRevD.86.083511} {\bibfield
  {journal} {\bibinfo  {journal} {Phys.Rev.}\ }\textbf {\bibinfo {volume}
  {D86}},\ \bibinfo {pages} {083511} (\bibinfo {year} {2012})},\ \Eprint
  {http://arxiv.org/abs/1207.6047} {arXiv:1207.6047 [astro-ph.HE]} \BibitemShut
  {NoStop}%
\bibitem [{\citenamefont {Gordon}\ and\ \citenamefont
  {Macias}(2013)}]{Gordon:2013vta}%
  \BibitemOpen
  \bibfield  {author} {\bibinfo {author} {\bibfnamefont {C.}~\bibnamefont
  {Gordon}}\ and\ \bibinfo {author} {\bibfnamefont {O.}~\bibnamefont
  {Macias}},\ }\href {\doibase 10.1103/PhysRevD.88.083521} {\bibfield
  {journal} {\bibinfo  {journal} {Phys.Rev.}\ }\textbf {\bibinfo {volume}
  {D88}},\ \bibinfo {pages} {083521} (\bibinfo {year} {2013})},\ \Eprint
  {http://arxiv.org/abs/1306.5725} {arXiv:1306.5725 [astro-ph.HE]} \BibitemShut
  {NoStop}%
\bibitem [{\citenamefont {Calore}\ \emph
  {et~al.}(2015{\natexlab{a}})\citenamefont {Calore}, \citenamefont {Cholis},
  \citenamefont {McCabe},\ and\ \citenamefont {Weniger}}]{Calore:2014nla}%
  \BibitemOpen
  \bibfield  {author} {\bibinfo {author} {\bibfnamefont {F.}~\bibnamefont
  {Calore}}, \bibinfo {author} {\bibfnamefont {I.}~\bibnamefont {Cholis}},
  \bibinfo {author} {\bibfnamefont {C.}~\bibnamefont {McCabe}}, \ and\ \bibinfo
  {author} {\bibfnamefont {C.}~\bibnamefont {Weniger}},\ }\href {\doibase
  10.1103/PhysRevD.91.063003} {\bibfield  {journal} {\bibinfo  {journal} {Phys.
  Rev.}\ }\textbf {\bibinfo {volume} {D91}},\ \bibinfo {pages} {063003}
  (\bibinfo {year} {2015}{\natexlab{a}})},\ \Eprint
  {http://arxiv.org/abs/1411.4647} {arXiv:1411.4647 [hep-ph]} \BibitemShut
  {NoStop}%
\bibitem [{\citenamefont {Calore}\ \emph
  {et~al.}(2015{\natexlab{b}})\citenamefont {Calore}, \citenamefont {Cholis},\
  and\ \citenamefont {Weniger}}]{Calore:2014xka}%
  \BibitemOpen
  \bibfield  {author} {\bibinfo {author} {\bibfnamefont {F.}~\bibnamefont
  {Calore}}, \bibinfo {author} {\bibfnamefont {I.}~\bibnamefont {Cholis}}, \
  and\ \bibinfo {author} {\bibfnamefont {C.}~\bibnamefont {Weniger}},\ }\href
  {\doibase 10.1088/1475-7516/2015/03/038} {\bibfield  {journal} {\bibinfo
  {journal} {JCAP}\ }\textbf {\bibinfo {volume} {1503}},\ \bibinfo {pages}
  {038} (\bibinfo {year} {2015}{\natexlab{b}})},\ \Eprint
  {http://arxiv.org/abs/1409.0042} {arXiv:1409.0042 [astro-ph.CO]} \BibitemShut
  {NoStop}%
\bibitem [{\citenamefont {Daylan}\ \emph {et~al.}(2016)\citenamefont {Daylan},
  \citenamefont {Finkbeiner}, \citenamefont {Hooper}, \citenamefont {Linden},
  \citenamefont {Portillo}, \citenamefont {Rodd},\ and\ \citenamefont
  {Slatyer}}]{Daylan:2014rsa}%
  \BibitemOpen
  \bibfield  {author} {\bibinfo {author} {\bibfnamefont {T.}~\bibnamefont
  {Daylan}}, \bibinfo {author} {\bibfnamefont {D.~P.}\ \bibnamefont
  {Finkbeiner}}, \bibinfo {author} {\bibfnamefont {D.}~\bibnamefont {Hooper}},
  \bibinfo {author} {\bibfnamefont {T.}~\bibnamefont {Linden}}, \bibinfo
  {author} {\bibfnamefont {S.~K.~N.}\ \bibnamefont {Portillo}}, \bibinfo
  {author} {\bibfnamefont {N.~L.}\ \bibnamefont {Rodd}}, \ and\ \bibinfo
  {author} {\bibfnamefont {T.~R.}\ \bibnamefont {Slatyer}},\ }\href {\doibase
  10.1016/j.dark.2015.12.005} {\bibfield  {journal} {\bibinfo  {journal} {Phys.
  Dark Univ.}\ }\textbf {\bibinfo {volume} {12}},\ \bibinfo {pages} {1}
  (\bibinfo {year} {2016})},\ \Eprint {http://arxiv.org/abs/1402.6703}
  {arXiv:1402.6703 [astro-ph.HE]} \BibitemShut {NoStop}%
\bibitem [{\citenamefont {Macias}\ \emph {et~al.}(2015)\citenamefont {Macias},
  \citenamefont {Crocker}, \citenamefont {Gordon},\ and\ \citenamefont
  {Profumo}}]{Macias:2014sta}%
  \BibitemOpen
  \bibfield  {author} {\bibinfo {author} {\bibfnamefont {O.}~\bibnamefont
  {Macias}}, \bibinfo {author} {\bibfnamefont {R.}~\bibnamefont {Crocker}},
  \bibinfo {author} {\bibfnamefont {C.}~\bibnamefont {Gordon}}, \ and\ \bibinfo
  {author} {\bibfnamefont {S.}~\bibnamefont {Profumo}},\ }\href {\doibase
  10.1093/mnras/stv1002} {\bibfield  {journal} {\bibinfo  {journal} {Mon. Not.
  Roy. Astron. Soc.}\ }\textbf {\bibinfo {volume} {451}},\ \bibinfo {pages}
  {1833} (\bibinfo {year} {2015})},\ \Eprint {http://arxiv.org/abs/1410.1678}
  {arXiv:1410.1678 [astro-ph.HE]} \BibitemShut {NoStop}%
\bibitem [{\citenamefont {Ajello}\ \emph {et~al.}(2016)\citenamefont {Ajello}
  \emph {et~al.}}]{TheFermi-LAT:2015kwa}%
  \BibitemOpen
  \bibfield  {author} {\bibinfo {author} {\bibfnamefont {M.}~\bibnamefont
  {Ajello}} \emph {et~al.} (\bibinfo {collaboration} {Fermi-LAT}),\ }\href
  {\doibase 10.3847/0004-637X/819/1/44} {\bibfield  {journal} {\bibinfo
  {journal} {Astrophys. J.}\ }\textbf {\bibinfo {volume} {819}},\ \bibinfo
  {pages} {44} (\bibinfo {year} {2016})},\ \Eprint
  {http://arxiv.org/abs/1511.02938} {arXiv:1511.02938 [astro-ph.HE]}
  \BibitemShut {NoStop}%
\bibitem [{\citenamefont {Lacroix}\ \emph {et~al.}(2016)\citenamefont
  {Lacroix}, \citenamefont {Macias}, \citenamefont {Gordon}, \citenamefont
  {Panci}, \citenamefont {B{\oe}hm},\ and\ \citenamefont
  {Silk}}]{Lacroix:2015wfx}%
  \BibitemOpen
  \bibfield  {author} {\bibinfo {author} {\bibfnamefont {T.}~\bibnamefont
  {Lacroix}}, \bibinfo {author} {\bibfnamefont {O.}~\bibnamefont {Macias}},
  \bibinfo {author} {\bibfnamefont {C.}~\bibnamefont {Gordon}}, \bibinfo
  {author} {\bibfnamefont {P.}~\bibnamefont {Panci}}, \bibinfo {author}
  {\bibfnamefont {C.}~\bibnamefont {B{\oe}hm}}, \ and\ \bibinfo {author}
  {\bibfnamefont {J.}~\bibnamefont {Silk}},\ }\href {\doibase
  10.1103/PhysRevD.93.103004} {\bibfield  {journal} {\bibinfo  {journal} {Phys.
  Rev.}\ }\textbf {\bibinfo {volume} {D93}},\ \bibinfo {pages} {103004}
  (\bibinfo {year} {2016})},\ \Eprint {http://arxiv.org/abs/1512.01846}
  {arXiv:1512.01846 [astro-ph.HE]} \BibitemShut {NoStop}%
\bibitem [{\citenamefont {Horiuchi}\ \emph
  {et~al.}(2016{\natexlab{a}})\citenamefont {Horiuchi}, \citenamefont
  {Kaplinghat},\ and\ \citenamefont {Kwa}}]{Horiuchi:2016zwu}%
  \BibitemOpen
  \bibfield  {author} {\bibinfo {author} {\bibfnamefont {S.}~\bibnamefont
  {Horiuchi}}, \bibinfo {author} {\bibfnamefont {M.}~\bibnamefont
  {Kaplinghat}}, \ and\ \bibinfo {author} {\bibfnamefont {A.}~\bibnamefont
  {Kwa}},\ }\href@noop {} {\  (\bibinfo {year} {2016}{\natexlab{a}})},\ \Eprint
  {http://arxiv.org/abs/1604.01402} {arXiv:1604.01402 [astro-ph.HE]}
  \BibitemShut {NoStop}%
\bibitem [{\citenamefont {Horiuchi}\ \emph
  {et~al.}(2016{\natexlab{b}})\citenamefont {Horiuchi}, \citenamefont {Macias},
  \citenamefont {Restrepo}, \citenamefont {Rivera}, \citenamefont {Zapata},\
  and\ \citenamefont {Silverwood}}]{Horiuchi:2016tqw}%
  \BibitemOpen
  \bibfield  {author} {\bibinfo {author} {\bibfnamefont {S.}~\bibnamefont
  {Horiuchi}}, \bibinfo {author} {\bibfnamefont {O.}~\bibnamefont {Macias}},
  \bibinfo {author} {\bibfnamefont {D.}~\bibnamefont {Restrepo}}, \bibinfo
  {author} {\bibfnamefont {A.}~\bibnamefont {Rivera}}, \bibinfo {author}
  {\bibfnamefont {O.}~\bibnamefont {Zapata}}, \ and\ \bibinfo {author}
  {\bibfnamefont {H.}~\bibnamefont {Silverwood}},\ }\href {\doibase
  10.1088/1475-7516/2016/03/048} {\bibfield  {journal} {\bibinfo  {journal}
  {JCAP}\ }\textbf {\bibinfo {volume} {1603}},\ \bibinfo {pages} {048}
  (\bibinfo {year} {2016}{\natexlab{b}})},\ \Eprint
  {http://arxiv.org/abs/1602.04788} {arXiv:1602.04788 [hep-ph]} \BibitemShut
  {NoStop}%
\bibitem [{\citenamefont {Maturi}\ \emph {et~al.}(2005)\citenamefont {Maturi},
  \citenamefont {Meneghetti}, \citenamefont {Bartelmann}, \citenamefont
  {Dolag},\ and\ \citenamefont {Moscardini}}]{Maturi:2004rn}%
  \BibitemOpen
  \bibfield  {author} {\bibinfo {author} {\bibfnamefont {M.}~\bibnamefont
  {Maturi}}, \bibinfo {author} {\bibfnamefont {M.}~\bibnamefont {Meneghetti}},
  \bibinfo {author} {\bibfnamefont {M.}~\bibnamefont {Bartelmann}}, \bibinfo
  {author} {\bibfnamefont {K.}~\bibnamefont {Dolag}}, \ and\ \bibinfo {author}
  {\bibfnamefont {L.}~\bibnamefont {Moscardini}},\ }\href {\doibase
  10.1051/0004-6361:20042600} {\bibfield  {journal} {\bibinfo  {journal}
  {Astron. Astrophys.}\ }\textbf {\bibinfo {volume} {442}},\ \bibinfo {pages}
  {851} (\bibinfo {year} {2005})},\ \Eprint
  {http://arxiv.org/abs/astro-ph/0412604} {arXiv:astro-ph/0412604 [astro-ph]}
  \BibitemShut {NoStop}%
\bibitem [{\citenamefont {Hennawi}\ and\ \citenamefont
  {Spergel}(2005)}]{Hennawi:2004ai}%
  \BibitemOpen
  \bibfield  {author} {\bibinfo {author} {\bibfnamefont {J.~F.}\ \bibnamefont
  {Hennawi}}\ and\ \bibinfo {author} {\bibfnamefont {D.~N.}\ \bibnamefont
  {Spergel}},\ }\href {\doibase 10.1086/428749} {\bibfield  {journal} {\bibinfo
   {journal} {Astrophys. J.}\ }\textbf {\bibinfo {volume} {624}},\ \bibinfo
  {pages} {59} (\bibinfo {year} {2005})},\ \Eprint
  {http://arxiv.org/abs/astro-ph/0404349} {arXiv:astro-ph/0404349 [astro-ph]}
  \BibitemShut {NoStop}%
\bibitem [{\citenamefont {Ade}\ \emph {et~al.}(2015)\citenamefont {Ade} \emph
  {et~al.}}]{Ade:2015xua}%
  \BibitemOpen
  \bibfield  {author} {\bibinfo {author} {\bibfnamefont {P.~A.~R.}\
  \bibnamefont {Ade}} \emph {et~al.} (\bibinfo {collaboration} {Planck}),\
  }\href@noop {} {\  (\bibinfo {year} {2015})},\ \Eprint
  {http://arxiv.org/abs/1502.01589} {arXiv:1502.01589 [astro-ph.CO]}
  \BibitemShut {NoStop}%
\end{thebibliography}%
\end{document}